\documentclass[aps,prl,floatfix,twocolumn,superscriptaddress]{revtex4-2}
\usepackage[latin9]{inputenc}
\usepackage{amsmath}
\usepackage{amssymb}
\usepackage{graphicx}
\usepackage{esint}
\usepackage{epstopdf}
\usepackage{braket}
\usepackage{color}
\usepackage[colorlinks=true,citecolor=blue,linkcolor=blue,urlcolor=blue]{hyperref}
\usepackage{graphicx}
\usepackage{stackengine}
\usepackage{appendix}
\usepackage{leftindex}

\makeatletter
\@ifundefined{textcolor}{}
{
 \definecolor{BLACK}{gray}{0}
 \definecolor{WHITE}{gray}{1}
 \definecolor{RED}{rgb}{1,0,0}
 \definecolor{GREEN}{rgb}{0,1,0}
 \definecolor{BLUE}{rgb}{0,0,1}
 \definecolor{CYAN}{cmyk}{1,0,0,0}
 \definecolor{MAGENTA}{cmyk}{0,1,0,0}
 \definecolor{YELLOW}{cmyk}{0,0,1,0}
}

\makeatother

\begin{document}

% \title{Hawking time crystals}

\title{Time Crystal from Self-Amplification of Spontaneous Analog Hawking Radiation}

\author{Juan Ram\'on Mu\~noz de Nova}
\affiliation{Instituto de Estructura de la Materia, IEM-CSIC, Serrano, 123 E-28006 Madrid, Spain}
\affiliation{Departamento de F\'isica de Materiales, Universidad Complutense de
Madrid, E-28040 Madrid, Spain}

\author{Fernando Sols}
\affiliation{Departamento de F\'isica de Materiales, Universidad Complutense de Madrid, E-28040 Madrid, Spain}

\begin{abstract}
We propose a time crystal based on a quantum black-hole laser, where the genuinely spontaneous character of the symmetry breaking stems from the self-amplification of spontaneous Hawking radiation. The resulting Hawking time crystal (HTC) is characterized by the periodic dependence of the out-of-time density-density correlation function, while equal-time observables are time independent because they embody averages over different realizations with a random oscillation phase. The HTC provides a nonlinear periodic analog of the Andreev-Hawking effect, exhibiting anticorrelation bands resulting from the spontaneous, quantum emission of pairs of dispersive waves and solitons into the upstream and downstream regions. Remarkably, we prove that any parametric amplifier has associated a time operator, which leads to a unique characterization of the time-crystal formation in terms of two time operators: one associated with the initial black-hole laser and another associated with the final spontaneous Floquet state.
\end{abstract}

\date{\today}

\maketitle

\textit{Introduction.---} Analog gravity \cite{Unruh1981} investigates the simulation of otherwise inaccessible gravitational phenomena by using a variety of tabletop experiments, including atomic condensates \cite{Garay2000,Lahav2010}, nonlinear optical fibers \cite{Belgiorno2010,Drori2019}, ion rings \cite{Horstmann2010,Wittemer2019}, water waves \cite{Weinfurtner2011,Euve2016}, quantum fluids of light \cite{Carusotto2013,Falque2025}, superconducting qubits \cite{Shi2023}, Fermi gases \cite{FulgadoClaudio2023}, or superfluid helium \cite{Vsvanvcara2024}. This has resulted in observations of the dynamical Casimir effect \cite{Jaskula2012}, Sakharov oscillations \cite{Hung2013}, superradiance \cite{Torres2017}, inflation \cite{Eckel2018}, Unruh effect \cite{Hu2019}, quasinormal ringdown \cite{Torres2020}, backreaction \cite{Patrick2021}, and cosmological particle creation \cite{Steinhauer2022a,Viermann2022}. A central topic is the observation of spontaneous Hawking radiation \cite{Leonhardt2003a,Balbinot2008,Carusotto2008,Macher2009a,Recati2009,Zapata2011,Larre2012,deNova2014,Finazzi2014,Busch2014,deNova2015,Michel2016,Isoard2021,Ribeiro2022,Ribeiro2023,Ciliberto2024}, finally achieved in atomic condensates \cite{Steinhauer2016,deNova2019,Kolobov2021}, manifested as the correlated quantum emission of Bogoliubov quasiparticles from a subsonic/supersonic interface playing the role of the event horizon. Since scattering within the supersonic region also provides a bosonic analog of the Andreev effect \cite{Zapata2009a}, by borrowing concepts from quantum optics \cite{Schleich2001,Walls2008} the whole process can be understood as a joint Andreev-Hawking effect \cite{deNova2024}.

A major remaining challenge is the observation of a black-hole laser (BHL) \cite{Corley1999,Leonhardt2003,Barcelo2006,Jain2007,Coutant2010,Finazzi2010,Faccio_2012,Michel2013,Michel2015,deNova2016,Peloquin2016,Bermudez2018,Burkle2018,deNova2021,RinconEstrada2021,Katayama2021,Steinhauer2022,deNova2023}, where Hawking radiation is self-amplified by successive reflections between a pair of horizons. Of particular interest is the late-time behavior of a BHL, after the initial instability has saturated, where it may exhibit a periodic regime of continuous emission of solitons (CES) \cite{deNova2016,deNova2021}, representing the \textit{bona-fide} BHL \cite{deNova2016}. Indeed, the CES state is a universal feature of flowing condensates, providing one of the simplest realizations of the more general concept of spontaneous Floquet state: a state of a time-independent Hamiltonian which oscillates like a Floquet state due to interactions \cite{deNova2022}. Among other intriguing features, spontaneous Floquet states display a temporal Floquet-Nambu-Goldstone (FNG) mode with zero quasifrequency, whose quantum amplitude represents a rare tangible realization of time operator in Quantum Mechanics \cite{deNova2025}. This connects with the physics of time crystals \cite{Wilczek2012,Sacha2017,Sacha2020}, which are out-of-equilibrium phases spontaneously breaking time-translation symmetry, and are robust against external perturbations. Both discrete \cite{Sacha2015,Else2016,Pizzi2021,Ye2021,Bhowmick2023} and continuous \cite{Syrwid2017,Iemini2018,Buca2019,Booker2020,Daviet2024} time crystals have been achieved in a wide range of systems \cite{Choi2017,Zhang2017,Autti2018,Smits2018,Rovny2018,Kyprianidis2021,Randall2021,Google2022,Frey2022,Kongkhambut2022,Dreon2022,Liu2023,Greilich2024,Carraro2024,Wang2025}. 

Spontaneous Floquet states are highly-excited states, shifted above the ground state by a macroscopic energy \cite{deNova2025}, thus evading the no-go theorem for time crystals \cite{Watanabe2015}, in analogy to excited eigenstates \cite{Syrwid2017}. In particular, the CES state belongs in a dynamical phase diagram, and exhibits typical time-crystalline robustness \cite{deNova2022}. However, these states have been described so far only within mean-field descriptions imposing a global time origin, so, even though they possess a temporal FNG mode, the symmetry breaking is not truly spontaneous \cite{deNova2025}. 

Here we show that a \textit{bona-fide} continuous time crystal can be achieved in a spontaneous Floquet state by leveraging the quantum nature of the spontaneous Hawking radiation self-amplified in a BHL, thus denoted as Hawking time crystal (HTC). We also prove that the initial BHL has an associated time operator by exploiting its description as a parametric amplifier (PA) \cite{Finazzi2010,Ribeiro2022,deNova2024}, whose behavior determines the asymptotic time crystallization, in turn characterized by another time operator. Furthermore, an HTC provides a nonlinear periodic counterpart of the Andreev-Hawking effect.

\begin{figure}[!tb]
\begin{tabular}{@{}cc@{}}
\stackinset{l}{0pt}{t}{0pt}{(a)}{\includegraphics[width=0.5\columnwidth]{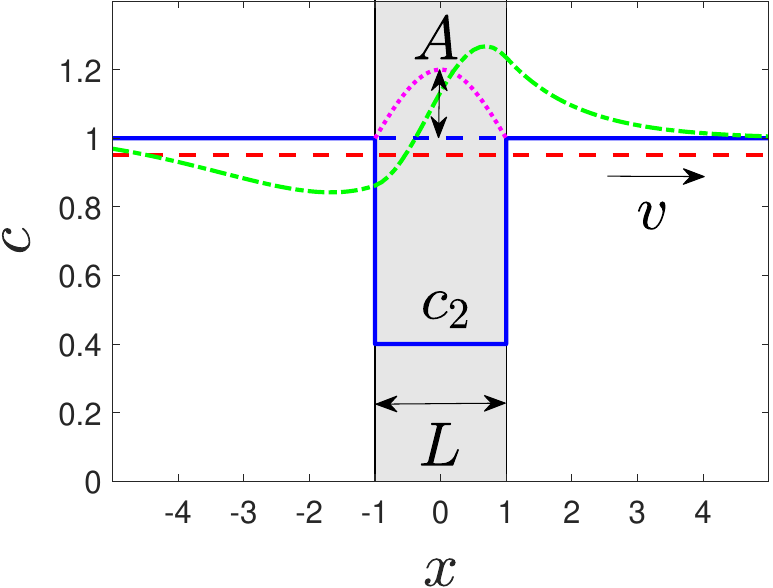}} & \stackinset{l}{0pt}{t}{0pt}{(b)}{\includegraphics[width=0.5\columnwidth]{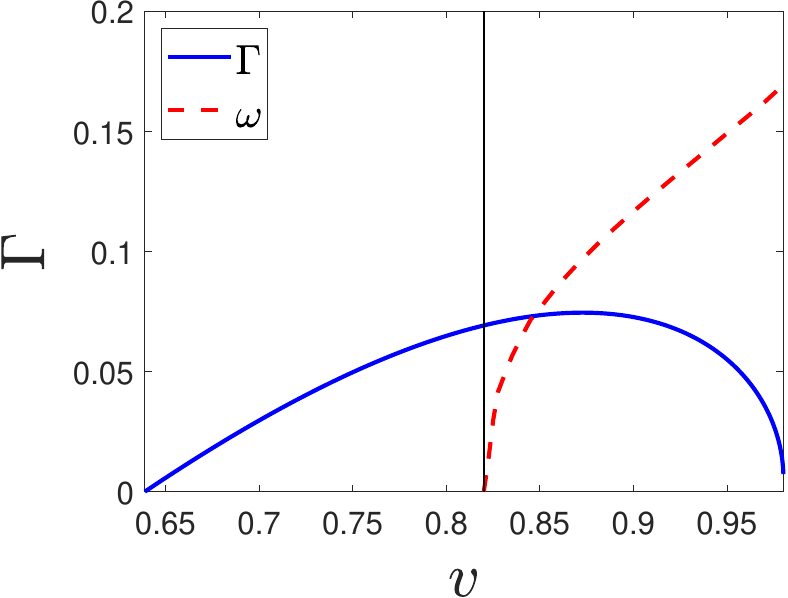}}\\~~~~~
\stackinset{l}{-18pt}{t}{0pt}{(c)}{\includegraphics[width=0.42\columnwidth]{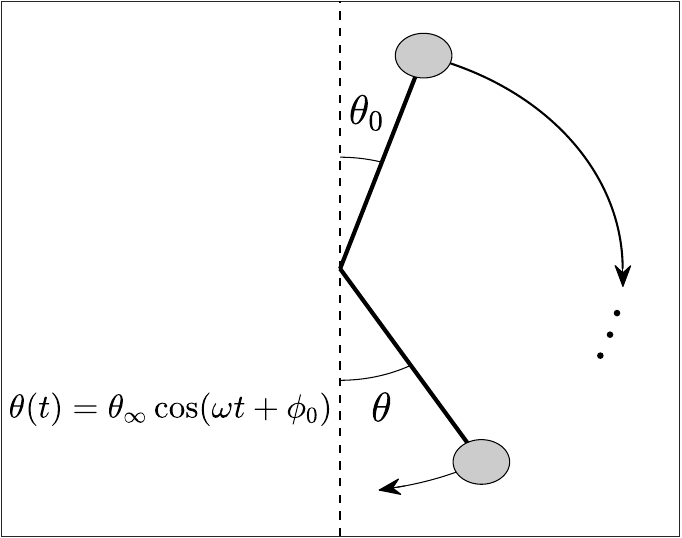}} & ~~~~~~\stackinset{l}{-18pt}{t}{0pt}{(d)}{\includegraphics[width=0.42\columnwidth]{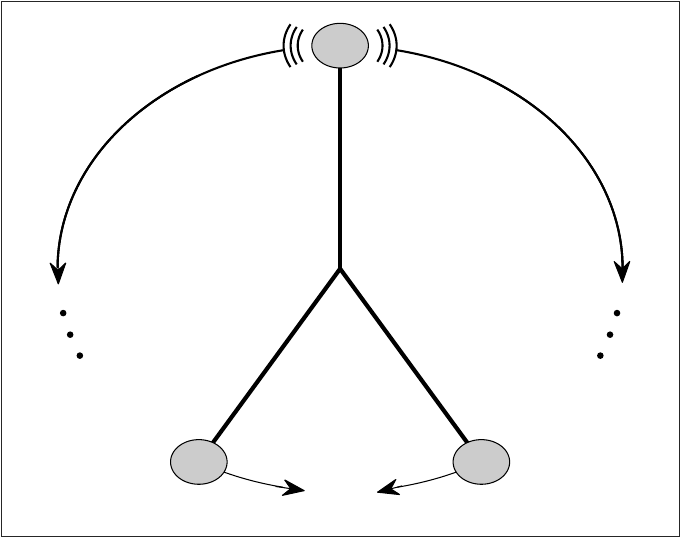}}
\end{tabular}
\caption{(a) Sound (solid blue) and flow (dashed red) velocity profile of the FPBHL at $t=0$. Horizontal dashed blue line represents the initial homogeneous condensate $\Psi_0(x)=e^{ivx}$. The shaded area indicates the lasing cavity of length $L$, where the coupling constant is locally quenched for $t\geq 0$ so its sound speed becomes $c_2<v$. A perturbation of amplitude $A$ (dotted magenta) is added to the initial GP condition as classical seed for the unstable mode (dashed-dotted green). (b) Velocity dependence of $\Gamma$ (solid blue) and $\omega$ (dashed red) for $c_2=0.4,L=2$. Vertical line indicates $v=v_c$. (c) Classical trajectory of a pendulum initially separated from its unstable position by an angle $\theta_0$, playing the role of the classical amplitude $A$ in the FPBHL model. After some transient, the pendulum ends up oscillating around its true equilibrium position with a well-defined oscillation phase $\phi_0$, similar to the mean-field CES state $\tilde{\Psi}_0(x,\phi_0+\omega t)$. (d) A pendulum in its unstable position collapses at the quantum level, leading to a superposition of trajectories with different $\phi_0$.}
\label{fig:Model}
\end{figure}

\textit{BHL model.---} We take the flat-profile BHL (FPBHL) model as test ground since, due to its simplicity, it neatly captures the physics at play in more realistic scenarios \cite{Carusotto2008,Recati2009,Michel2013,deNova2016,deNova2023} (see Appendix A for a comprehensive characterization of the FPBHL). For $t<0$, a 1D homogeneous atomic quasicondensate \cite{Menotti2002} flows from left to right, described by a stationary Gross-Pitaevskii (GP) wavefunction $\Psi_0(x)=\sqrt{n_0}e^{iqx}$, whose flow and sound speeds are $v=\hbar q/m<c_0=\sqrt{gn_0/m}$, where $g$ is the coupling constant describing the short-ranged interactions between atoms and $m$ is their mass. Hereafter, $\hbar=m=c_0=1$, and the GP wavefunction is rescaled as $\Psi(x,t) \to \sqrt{n_0}\Psi(x,t)$. For $t\geq 0$, the external potential and the coupling constant are piecewise quenched so that $\Psi_0(x)$ remains stationary, and all the dynamics is driven by fluctuations. Specifically, we choose $g(x)$ such that the condensate is supersonic for $|x|<L/2$, with sound speed $c_2<v$, so there are two sonic horizons at $x=\pm L/2$, resulting in a stationary BHL,  Fig. \ref{fig:Model}a.

A BHL is characterized by a finite Bogoliubov spectrum of complex frequencies. We focus on short cavities containing only one degenerate unstable mode with purely imaginary frequency $i\Gamma$, $\Gamma>0$. This optimal cavity choice both maximizes $\Gamma$ and minimizes the transient towards the final state \cite{Michel2013,deNova2016,deNova2023}. 

Once excited, the lasing mode is exponentially amplified up to the nonlinear saturation regime \cite{Michel2013,Michel2015,deNova2016,deNova2017a}. At long times, the final state of the system exhibits a dynamical phase diagram \cite{Moeckel2008,Sciolla2010,Lang2018}, displaying only two phases \cite{deNova2016,deNova2021}: the stationary nonlinear ground state and the periodic CES state. In general, any condensate flowing over some obstacle with velocity $v$ above certain critical value $v_c$ asymptotically approaches the CES state \cite{deNova2022},
\begin{equation}\label{eq:CES}
    \Psi(x,t)\xrightarrow[t\to\infty]{} e^{-i\mu t}\Psi_0(x,t),
\end{equation}
$\mu$ being the quasichemical potential and $\Psi_0(x,t)$ a periodic wavefunction, $\Psi_0(x,t)=\tilde{\Psi}_0(x,\phi_0+\omega t)$, with $\tilde{\Psi}_0(x,\phi+2\pi)=\tilde{\Psi}_0(x,\phi)$ and $\omega=2\pi/T$. The velocity dependence of $\Gamma,\omega$ for a FPBHL is shown in Fig. \ref{fig:Model}b. Due to the time independence of the underlying Hamiltonian, the CES state is a spontaneous Floquet state, where the oscillation phase-shift $\phi_0$ is not \textit{a priori} fixed, in contrast to conventional Floquet systems. This leads to the emergence of a temporal FNG mode in the Bogoliubov spectrum, whose quantum amplitude $\hat{t}_0$ is the time operator describing the quantum fluctuations of the global time-shift $t_0=-\phi_0/\omega$ \cite{deNova2025}. However, at the mean-field level, $\phi_0$ is still fixed by the deterministic GP equation once an initial condition is set.

\begin{figure*}
\begin{tabular}{@{}cccc@{}}
    \stackinset{l}{0pt}{t}{0pt}{(a)}{\includegraphics[width=0.25\textwidth]{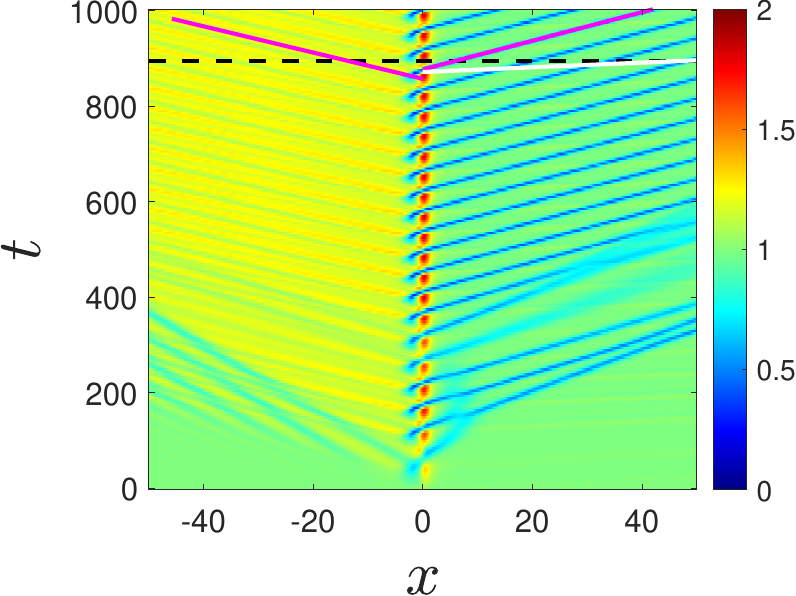}} 
    & \stackinset{l}{0pt}{t}{0pt}{(b)}{\includegraphics[width=0.25\textwidth]{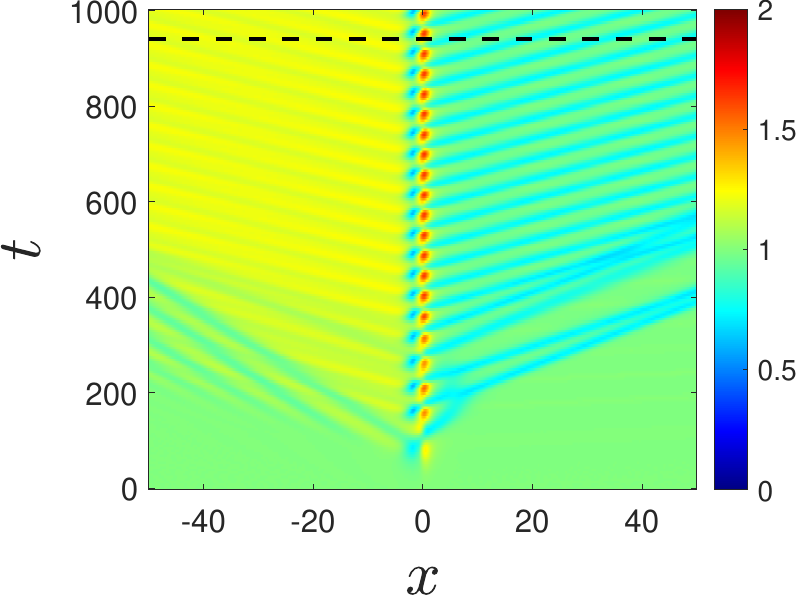}}& 
    \stackinset{l}{0pt}{t}{0pt}{(c)}{\includegraphics[width=0.25\textwidth]{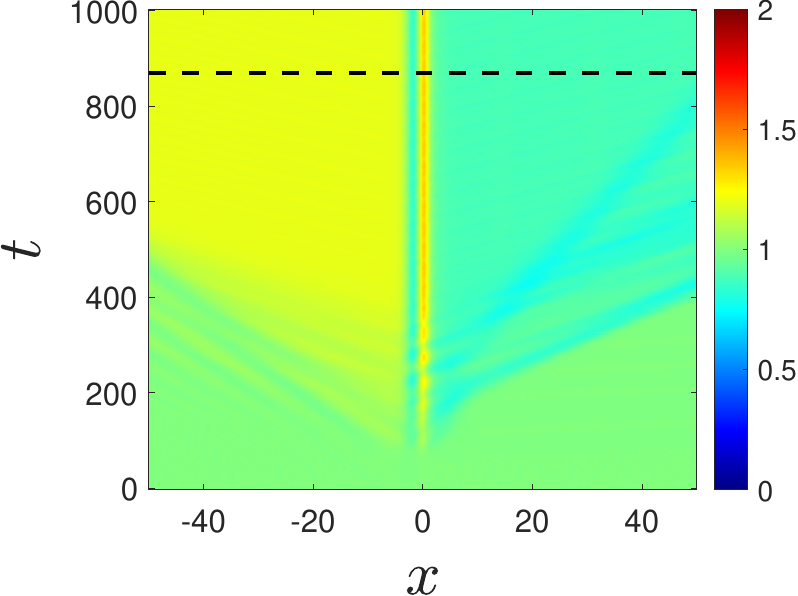}} &
    \stackinset{l}{0pt}{t}{0pt}{(d)}{\includegraphics[width=0.24\textwidth]{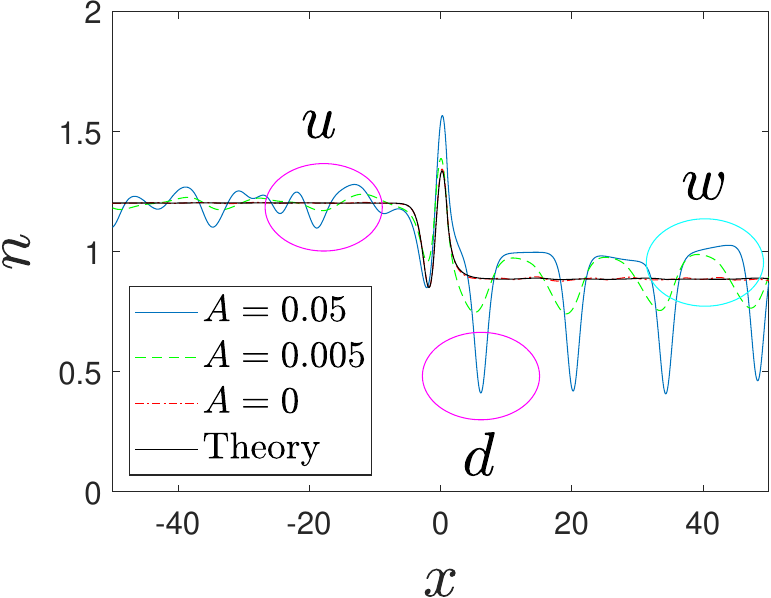}}
    \\
    \stackinset{l}{0pt}{t}{0pt}{(e)}{\includegraphics[width=0.25\textwidth]{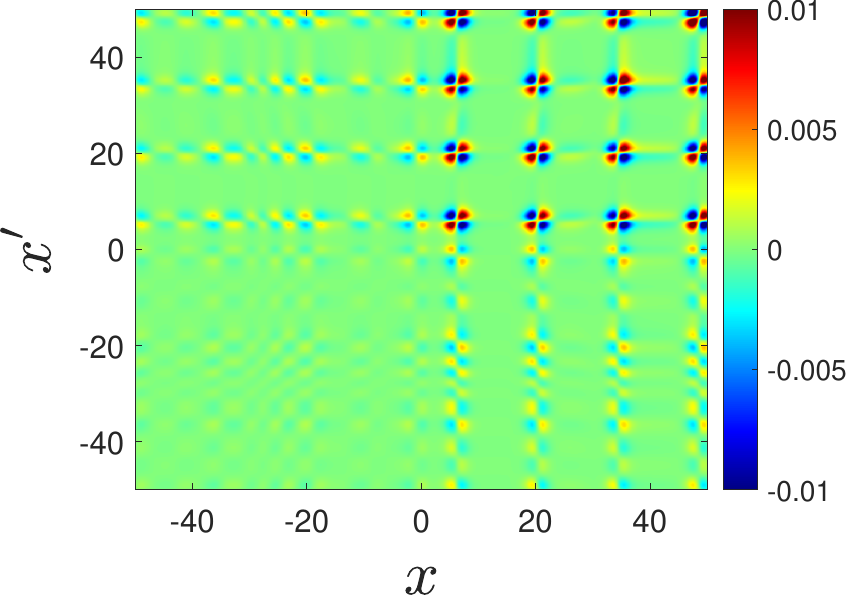}} 
    & \stackinset{l}{0pt}{t}{0pt}{(f)}{\includegraphics[width=0.25\textwidth]{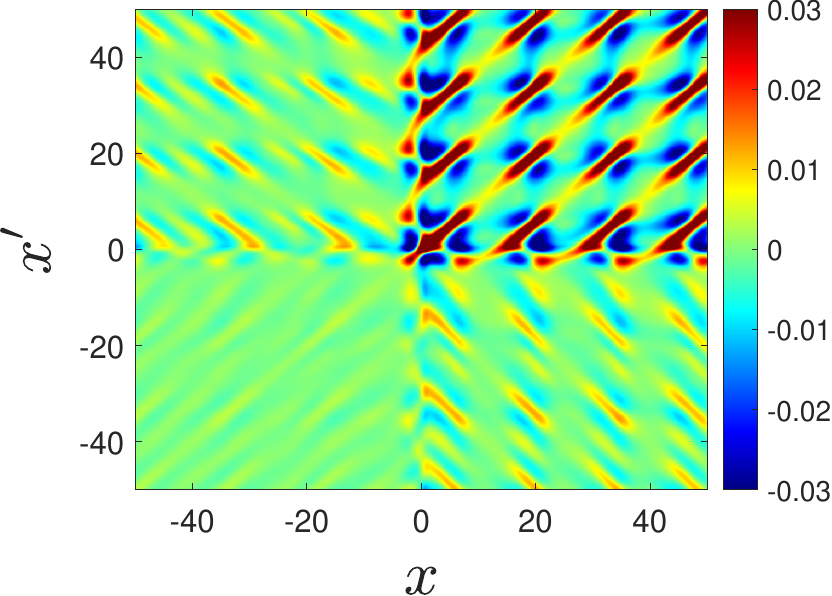}}& 
    \stackinset{l}{0pt}{t}{0pt}{(g)}{\includegraphics[width=0.25\textwidth]{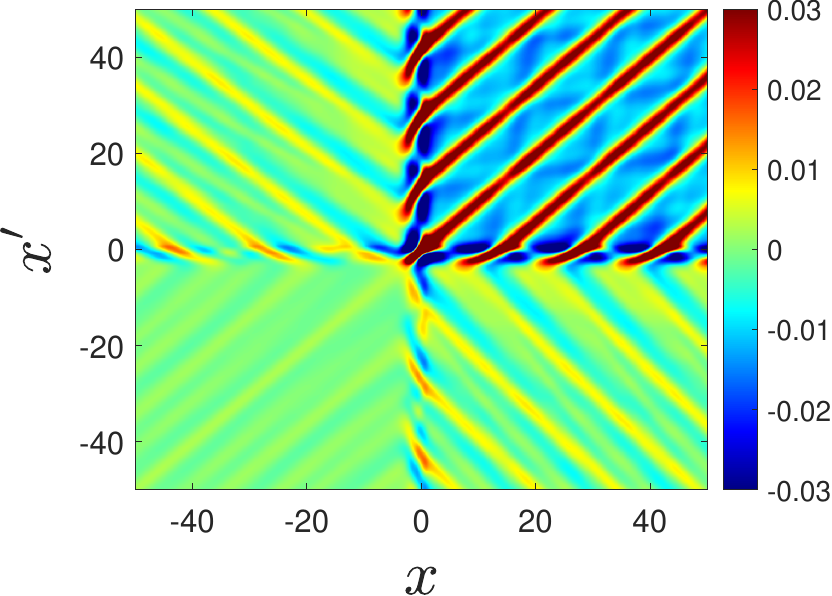}} &
    \stackinset{l}{0pt}{t}{0pt}{(h)}{\includegraphics[width=0.25\textwidth]{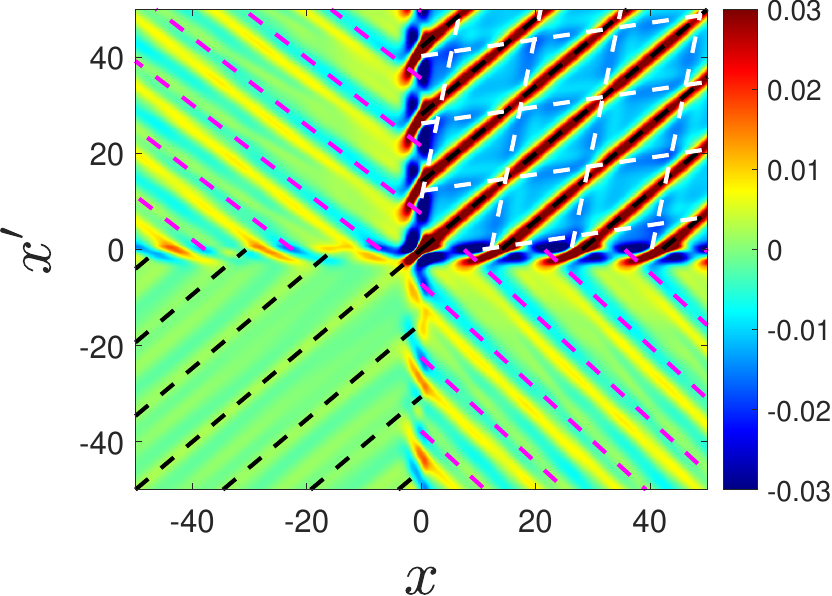}}
    \\ 
     \stackinset{l}{0pt}{t}{0pt}{(i)}{\includegraphics[width=0.25\textwidth]{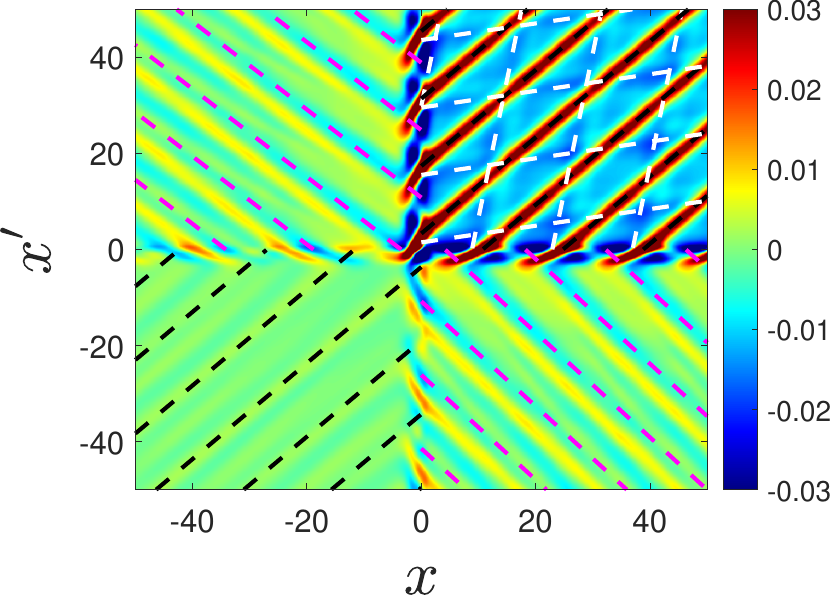}} 
    & \stackinset{l}{0pt}{t}{0pt}{(j)}{\includegraphics[width=0.25\textwidth]{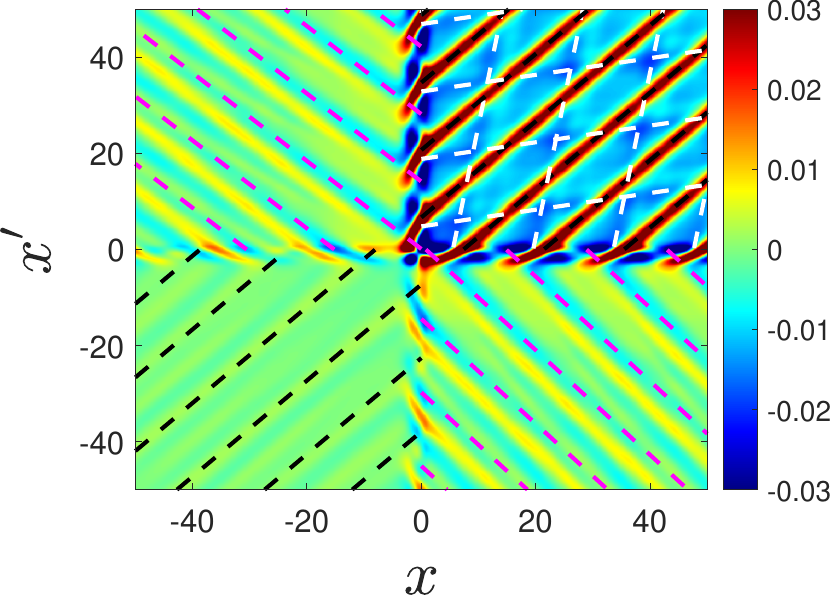}}& 
    \stackinset{l}{0pt}{t}{0pt}{(k)}{\includegraphics[width=0.25\textwidth]{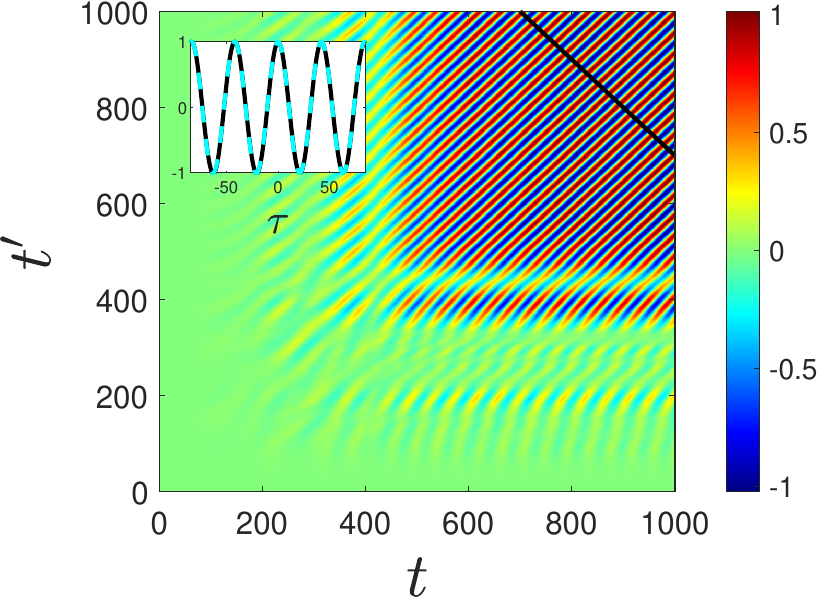}} &
    \stackinset{l}{0pt}{t}{0pt}{(l)}{\includegraphics[width=0.25\textwidth]{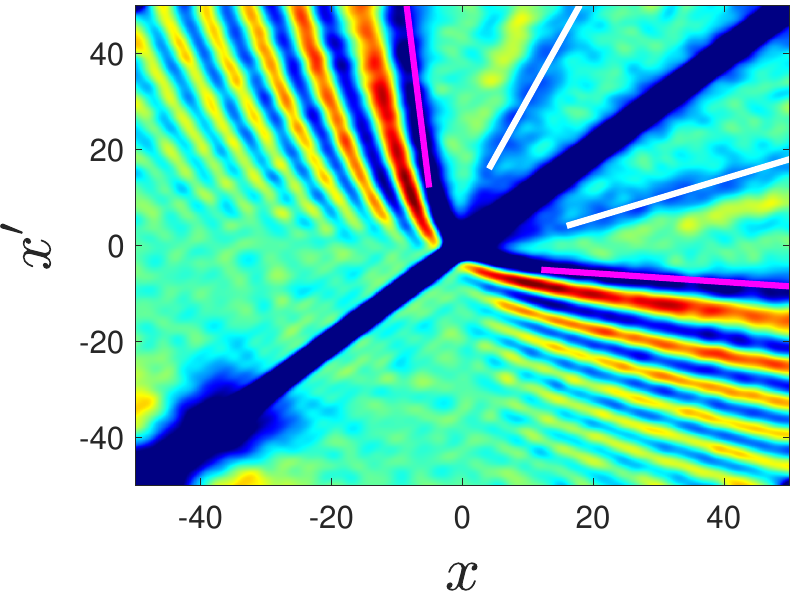}}
\end{tabular}
\caption{Expectation values for a FPBHL with  $v=0.95,c_2=0.4,L=2$, computed via TW method. (a)-(c) Ensemble-averaged density $n(x,t)$ for initial classical amplitudes $A=0.05,0.005,0$, respectively. (d) Snapshots of $n(x,t)$ at the times indicated by horizontal dashed line in (a)-(c). Black line is the theoretical prediction $n_{\rm{HTC}}(x)$, Eq. (\ref{eq:Hteo}). (e)-(g) ETCF $g^{(2)}(x,x',t)$ for (a)-(c), evaluated at the times shown in (d). (h) Theoretical ETCF $g_{\rm{HTC}}^{(2)}(x,x',\tau=0)$, Eq. (\ref{eq:Hteo}). Dashed lines indicate the expected correlation bands, Eqs. (\ref{eq:HawkingParallelOOT}), (\ref{eq:NonLinearAndreedHawkingOOT}). (i)-(j) OTCF $g^{(2)}(x,x',\tau;t)$ for $A=0$, evaluated at $\tau=10,20$ and fixed $t=870$. (k) $\textrm{Re}\, \mathcal{G}(t,t')$ for $A=0$. Inset: 1D profile along  black line in main plot. Dashed cyan is the theoretical prediction $\textrm{Re}\, \mathcal{G}_{\rm{HTC}}(\tau)=\cos\omega\tau$. l) Spatial structure of the ETCF for a flat-profile black hole with $v=0.95,c_2=0.4$. Magenta (white) solid lines indicate the Hawking (Andreev) correlation bands.}
\label{fig:ClassicalQuantum}
\end{figure*}

%go beyond mean-field by exploring

%provides a coherent

\textit{Hawking time crystal.---} In order to achieve a genuine spontaneous temporal symmetry breaking, we explore the classical-quantum crossover of a BHL \cite{deNova2023}. Specifically, we add a perturbation within the lasing cavity (dotted magenta in Fig. \ref{fig:Model}a), so the initial GP wavefunction at $t=0$ now reads $\Psi(x,0)=\left[1+A\delta\Psi_C(x)\right]e^{ivx}$, with $\delta\Psi_C(x)=\cos(\pi x/L)\theta\left(x+L/2\right)\theta\left(L/2-x\right)$ and $\theta$ the Heaviside function. This classical seed coherently stimulates the lasing mode (dashed-dotted green), mimicking the effect of Bogoliubov-Cherenkov-Landau radiation \cite{Carusotto2006} in real experiments \cite{Wang2016,Wang2017,Kolobov2021,Steinhauer2022}. As initial quantum state, we take the quasiparticle vacuum. Hence, there is a competition between the classical amplitude and vacuum quantum fluctuations, with $A$ acting as a control parameter of the classical-quantum crossover: for large $A$, the dynamics is described by a Bogoliubov approximation of small quantum fluctuations around the mean-field GP trajectory generated by the amplification of the initial classical seed; for small $A$, the dynamics is governed by the squeezing of the initial vacuum, representing the self-amplification of spontaneous Hawking radiation. Qualitatively, we can understand our FPBHL model as an unstable pendulum, where $A$ is akin to the initial separation from the unstable equilibrium position, and the CES state is akin to the periodic oscillations around the true equilibrium position, Figs. \ref{fig:Model}c,d.

The time evolution of the system is computed via Truncated Wigner (TW) method \cite{Sinatra2002}, a quantum Monte Carlo technique used in both analog gravity \cite{Carusotto2008,deNova2019,Jacquet2022,Butera2023,deNova2023} and time crystals \cite{Kessler2019,Kongkhambut2022} that goes beyond the standard Bogoliubov approximation; here we consider ensembles of $1000$ simulations (see Supplemental Material \cite{SuppMatTimeCrystal}). We evaluate the ensemble-averaged density 
\begin{equation}
     n(x,t)\equiv \braket{\hat{n}(x,t)}=\braket{\hat{\Psi}^\dagger(x,t)\hat{\Psi}(x,t)},
\end{equation}
and the second-order correlation function
\begin{eqnarray}\label{eq:RelativeCorrelations}
   \nonumber g^{(2)}(x,x',t,t')&\equiv&\braket{\hat{\Psi}^\dagger(x,t)\hat{\Psi}^\dagger(x',t')\hat{\Psi}(x',t')\hat{\Psi}(x,t)}\\
   &-&n(x,t)n(x',t').
\end{eqnarray}
We distinguish between the \textit{equal-time} correlation function (ETCF) $g^{(2)}(x,x',t)\equiv g^{(2)}(x,x',t,t)$ and the \textit{out-of-time} correlation function (OTCF) $g^{(2)}(x,x',\tau;t)\equiv g^{(2)}(x,x',t,t+\tau)$.

Figures \ref{fig:ClassicalQuantum}a-c, e-g show $n(x,t)$ and $g^{(2)}(x,x',t)$ for decreasing $A=0.05,0.005,0$, respectively. For large $A$, Figs. \ref{fig:ClassicalQuantum}a,e, a good agreement is found with the Bogoliubov prediction \cite{deNova2025}: 
\begin{eqnarray}
    n(x,t)&\simeq &n_0(x,t)\equiv |\Psi_0(x,t)|^2,\\
    \nonumber g^{(2)}(x,x',t)&\simeq &  \partial_t n_0(x,t)\partial_t n_0(x',t)C(t).
\end{eqnarray}
The quadratic function $C(t)=\braket{\hat{t}^2_0(t)}$ is the CES time-operator variance, increasing in time due to frequency fluctuations (these are found negligible in the present simulations). Spontaneous symmetry breaking requires going beyond the Bogoliubov regime of small quantum fluctuations, $\Delta \phi_0=\sqrt{\braket{\hat{\phi}^2_0}-\braket{\hat{\phi}_0}^2}\gtrsim 1$, with $\hat{\phi}_0=-\omega\hat{t}_0$ the phase-shift operator. This is achieved for a purely quantum BHL (i.e., $A=0$), Figs. \ref{fig:ClassicalQuantum}c,g, where there is no mean-field evolution and all the dynamics is due to quantum fluctuations, leading to a random $\phi_0$ in each TW realization. If we assume a uniform distribution, ensemble averages then represent averages over $t_0\in[0,T)$, resulting in time-independent values $n(x,t)=n_{\rm{HTC}}(x)$ and
$g^{(2)}(x,x',t)=g_{\rm{HTC}}^{(2)}(x,x',\tau=0)$, with
\begin{eqnarray}\label{eq:Hteo}
n_{\rm{HTC}}(x)&=&\frac{1}{T}\int^{T}_{0}\mathrm{d}t_0~n_0(x,t_0),\\
\nonumber g_{\rm{HTC}}^{(2)}(x,x',\tau)&=&\frac{1}{T}\int^{T}_{0}\mathrm{d}t_0~n_0(x,t_0)n_0(x',t_0+\tau)\\
\nonumber&-&n_{\rm{HTC}}(x)n_{\rm{HTC}}(x').
\end{eqnarray}
An excellent agreement with this theoretical prediction is found at late times for both density (Fig. \ref{fig:ClassicalQuantum}d) and ETCF (Figs. \ref{fig:ClassicalQuantum}g,h).

\begin{figure*}[!tb]
\begin{tabular}{@{}cccc@{}}   
\stackinset{l}{0pt}{t}{0pt}{(a)}{\includegraphics[width=0.25\textwidth]{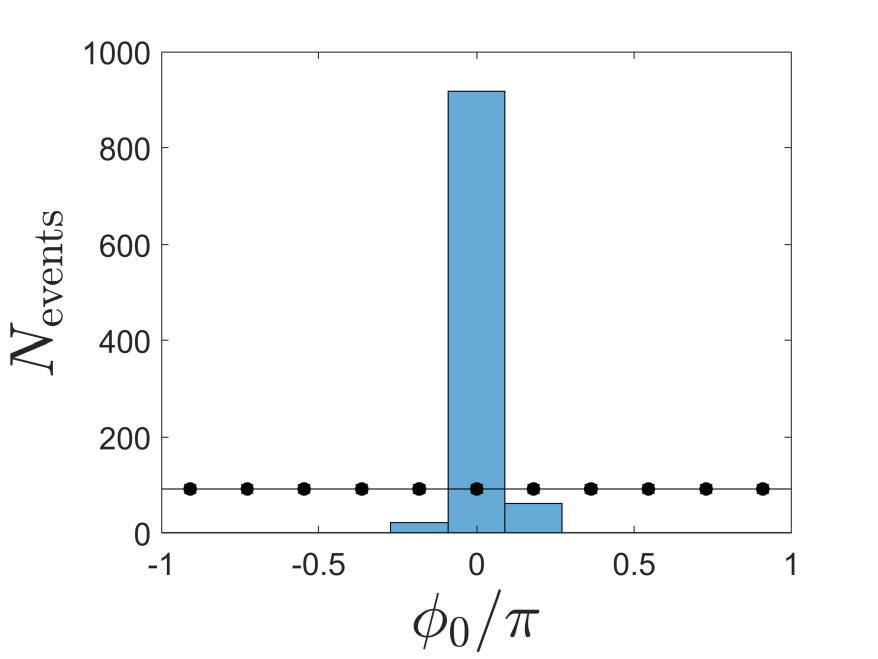}} 
& \stackinset{l}{0pt}{t}{0pt}{(b)}{\includegraphics[width=0.25\textwidth]{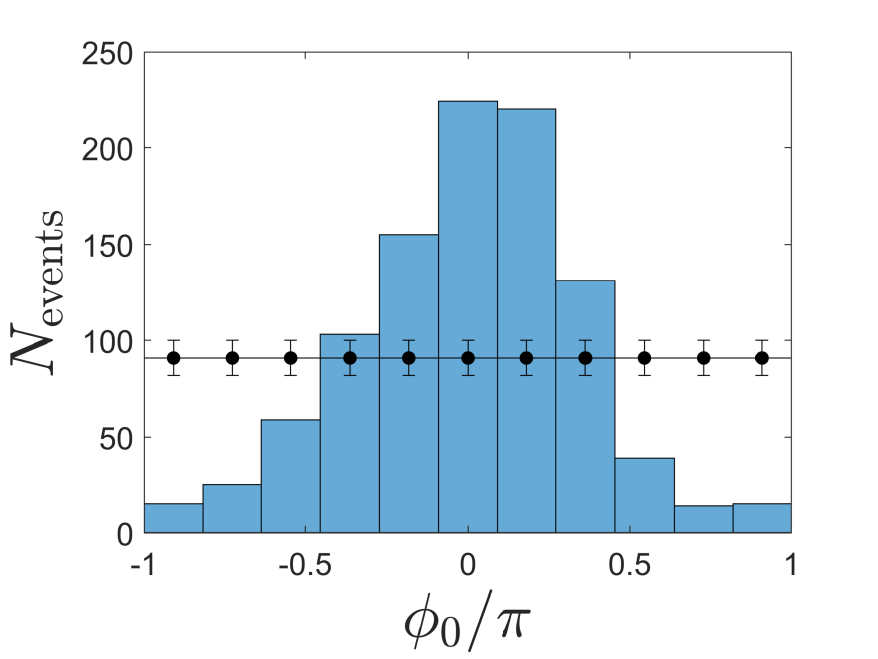}}& 
\stackinset{l}{0pt}{t}{0pt}{(c)}{\includegraphics[width=0.25\textwidth]{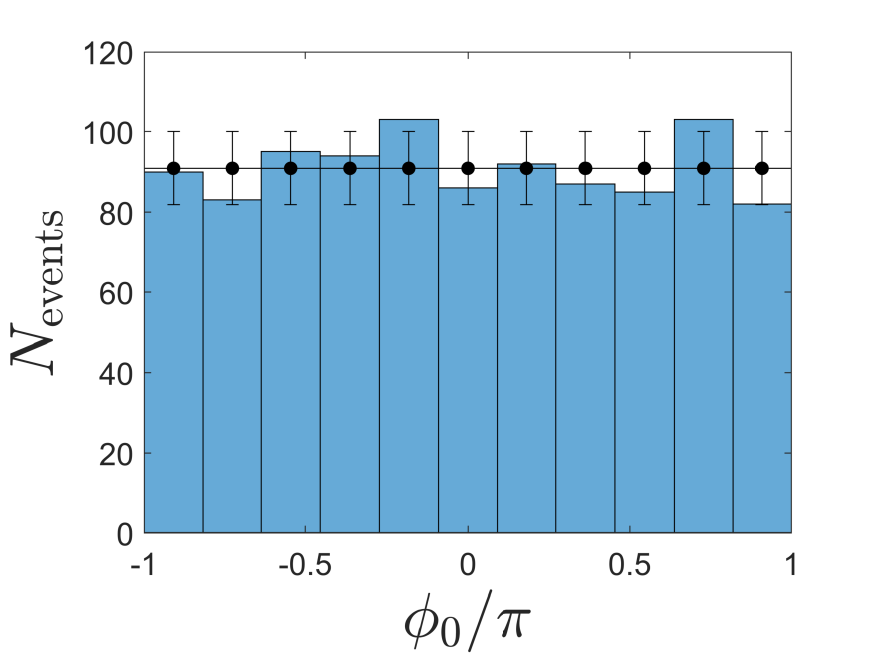}} &
\stackinset{l}{0pt}{t}{0pt}{(d)}{\includegraphics[width=0.24\textwidth]{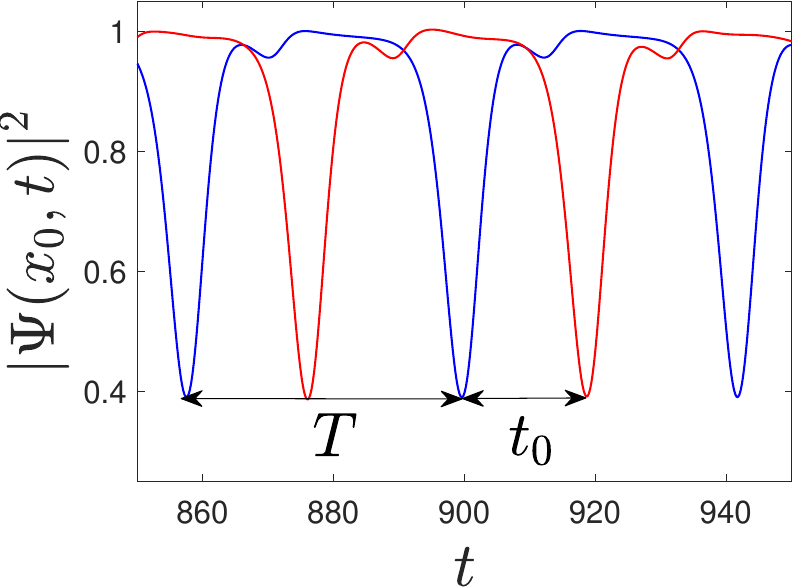}}
\\
\stackinset{l}{0pt}{t}{0pt}{(e)}{\includegraphics[width=0.25\textwidth]{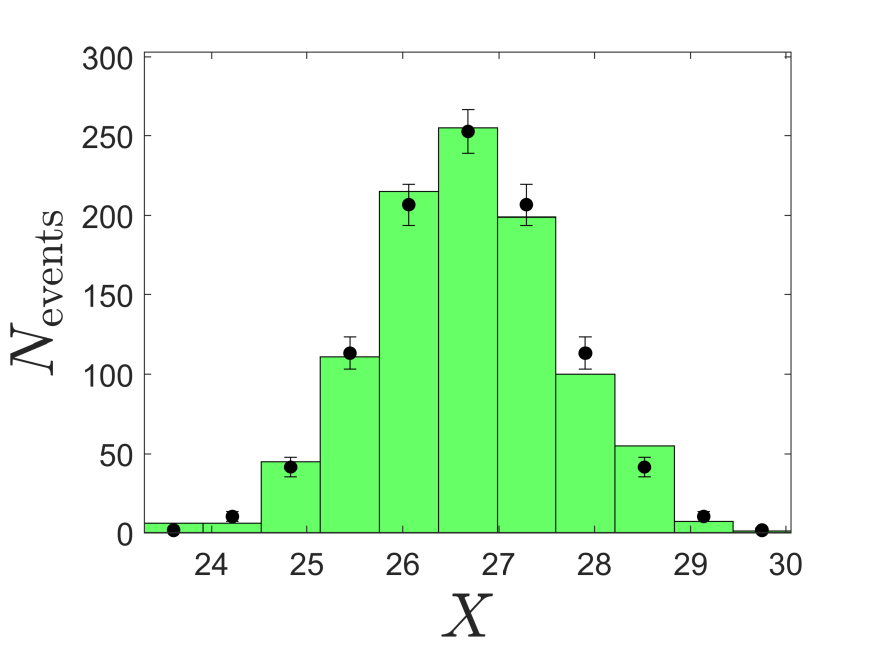}} 
& \stackinset{l}{0pt}{t}{0pt}{(f)}{\includegraphics[width=0.25\textwidth]{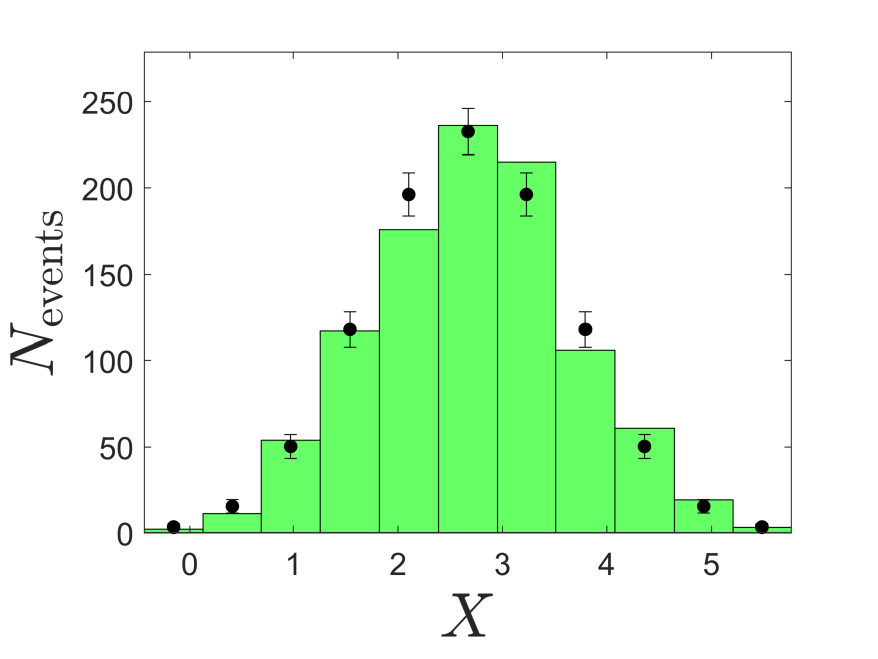}}& 
\stackinset{l}{0pt}{t}{0pt}{(g)}{\includegraphics[width=0.25\textwidth]{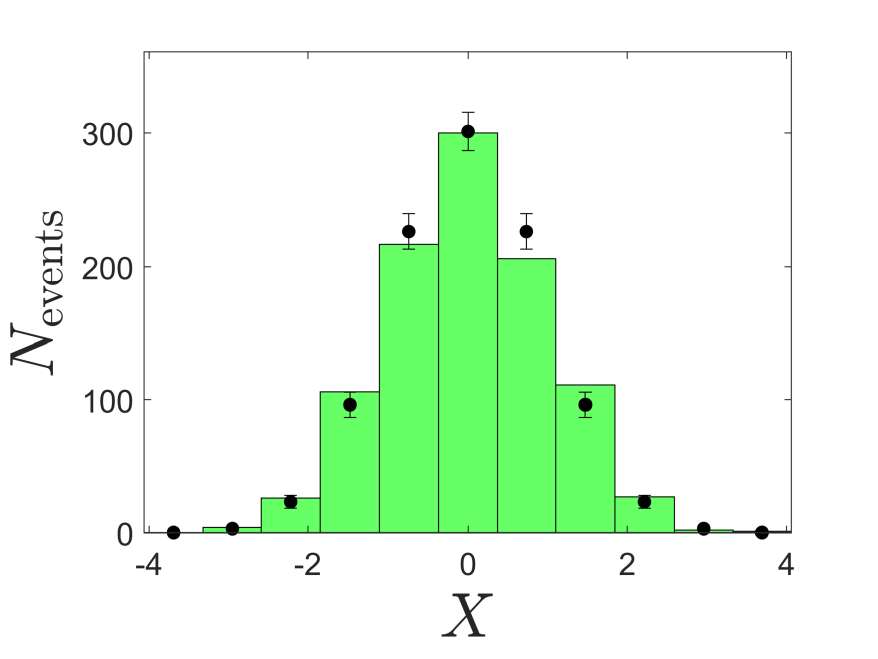}} &
\stackinset{l}{-1pt}{t}{0pt}{(h)}{\includegraphics[width=0.25\textwidth]{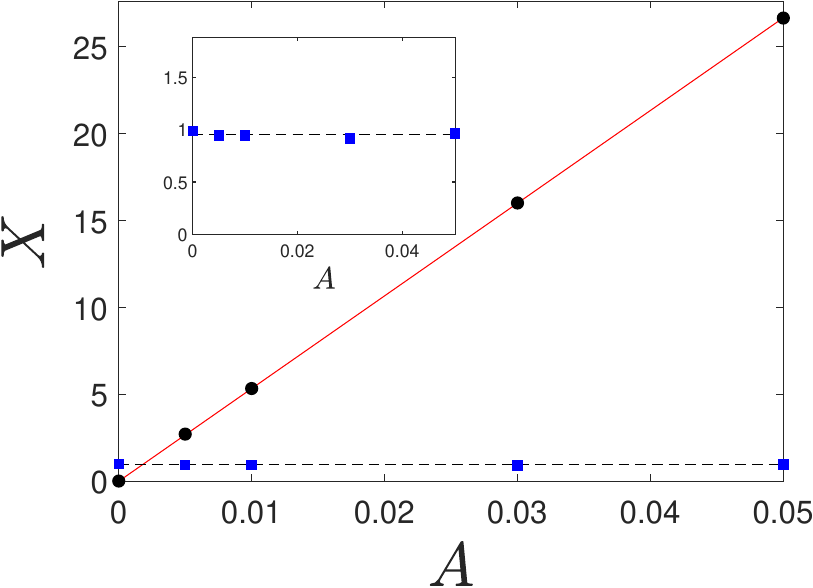}}
\end{tabular}
\caption{(a)-(c) Histogram: phase-shift $\phi_0$ for the TW ensembles of Figs. \ref{fig:ClassicalQuantum}a-c. Horizontal line with error bars: uniform distribution and its statistical uncertainty. (d) Schematic obtention of $\phi_0=-\omega t_0$. Blue line: mean-field trajectory. Red line: particular TW realization. (e)-(g) Histogram: lasing amplitude $X$ at $t=0$ for (a)-(c). Dots with error bars: expected Gaussian distribution (\ref{eq:GaussianWignerLaser}) and its statistical uncertainty. (h) Mean $X_C$ (black dots) and statistical deviation $\Delta X_Q$ (blue squares) as a function of $A$, and their theoretical predictions (solid red and horizontal dashed black lines, respectively). Inset: Zoom of $\Delta X_Q$.}
\label{fig:Histogram}
\end{figure*}

The system periodicity is manifested through the OTCF, $g^{(2)}(x,x',\tau;t)=g_{\rm{HTC}}^{(2)}(x,x',\tau)=g_{\rm{HTC}}^{(2)}(x,x',\tau+T)$, displayed in Figs. \ref{fig:ClassicalQuantum}i,j for $\tau=10,20$ and fixed $t=870$. As figure of merit for the OTCF periodicity, we choose the peak of its spatial Fourier spectrum within the downstream-upstream quadrant, $\mathcal{G}(t,t')$ (see Appendix B). The real part of $\mathcal{G}(t,t')$ is shown in Fig. \ref{fig:ClassicalQuantum}k; the imaginary part displays a similar behavior \cite{SuppMatTimeCrystal}. At late times, $\mathcal{G}(t,t')$ becomes a function solely of $\tau=t'-t$ (diagonal fringes). The inset shows the 1D profile along the perpendicular cut in the main plot, in great agreement with the HTC prediction $\mathcal{G}_{\rm{HTC}}(\tau)=e^{i\omega\tau}$ \cite{SuppMatTimeCrystal} (dashed cyan).

The suppression of the time dependence of the density and the ETCF, as well as the periodic nature of the OTCF, are the characteristic hallmarks of a \textit{bona-fide} time crystal \cite{Sacha2020}, here revealing the formation of an HTC, i.e., a continuous time crystal whose spontaneous symmetry breaking results from the self-amplification of spontaneous Hawking radiation.

\textit{Nonlinear Andreev-Hawking effect.---} The CES state results from the periodic upstream ($x<0$) emission of a soliton, which is dragged back to the downstream ($x>0$) region by the background flow (magenta circle labeled as $d$ in Fig. \ref{fig:ClassicalQuantum}d). The passage of the soliton through the lasing cavity restarts the cycle, resulting in the additional emission of upstream and downstream waves \cite{deNova2016,deNova2022} (magenta, cyan circles labeled as $u,w$ around blue solid line in Fig. \ref{fig:ClassicalQuantum}d, respectively). Their trajectories are well fitted by a ballistic trajectory $x_{i}(t)=x^{(0)}_{i}+v_{i}t$ \cite{SuppMatTimeCrystal} (magenta lines for $i=u,d$, white line for $i=w$ in Fig. \ref{fig:ClassicalQuantum}a).

These features are imprinted on the spatial structure of the correlations in Fig. \ref{fig:ClassicalQuantum}. From Eq. (\ref{eq:Hteo}), strong self-correlations between the upstream waves/downstream solitons and their past and future counterparts are predicted along the diagonals \cite{SuppMatTimeCrystal}
\begin{equation}\label{eq:HawkingParallelOOT}
    \frac{x'-x}{v_{u,d}}= \tau+nT,~n\in\mathbb{Z},
\end{equation}
dashed black lines in Figs. \ref{fig:ClassicalQuantum}h-j. In turn, the density defect carried by a soliton is correlated with the positive amplitude of the upstream and downstream waves, spawning the anticorrelation bands \cite{SuppMatTimeCrystal}
\begin{eqnarray}\label{eq:NonLinearAndreedHawkingOOT}
    \dfrac{x'}{v_{u}}-\dfrac{x}{v_{d}}&=&\dfrac{x^{(0)}_u}{v_{u}}-\dfrac{x^{(0)}_d}{v_{d}}+\tau+nT,~n\in\mathbb{Z}, \\
    \nonumber \frac{x'}{v_{w}}-\frac{x}{v_{d}}&=&\dfrac{x^{(0)}_w}{v_{w}}-\dfrac{x^{(0)}_d}{v_{d}}+\tau+nT,~n\in\mathbb{Z}.
\end{eqnarray}
The remaining features are obtained by exchanging $x\longleftrightarrow x'$, $\tau\to -\tau$. Upper (lower) Eq. (\ref{eq:NonLinearAndreedHawkingOOT}) is indicated by dashed magenta (white) lines in Figs. \ref{fig:ClassicalQuantum}h-j, matching the observed patterns. 

%, computed by Eq. (\ref{eq:QuantumAmplitudeTeo})

Remarkably, these anticorrelation bands are a nonlinear analog of those arising from the spontaneous Andreev-Hawking effect \cite{Recati2009}, Fig. \ref{fig:ClassicalQuantum}l (see Appendix C). The celebrated Hawking moustache (magenta line) and the Andreev band (white line) result, respectively, from the anticorrelation of the  Hawking (upstream, subsonic region) and copropagating (downstream, supersonic region) modes with the  downstream partner modes. In an HTC, the partner role is played by the soliton (also confined downstream by the flow), synchronously emitted along with one wave upstream (Hawking) and another downstream (Andreev). Their correlation patterns become spatially periodic as a result of the HTC time periodicity.

\textit{Time operators.---} The spontaneous nature of the symmetry breaking is further revealed by the phase-shift distribution within the TW ensemble, Figs. \ref{fig:Histogram}a-c, obtained from the time-shifts in the soliton passage in the late-time density $|\Psi(x_0, t)|^2$ at fixed $x=x_0>0$, Fig. \ref{fig:Histogram}d. To fix a time origin, we set $\phi_0=0$ for the mean-field wavefunction; for a purely quantum BHL, $\phi_0=0$ corresponds to the CES wavefunction resulting from a mean-field simulation with $A=0.003$ (there is no mean-field CES background for $A=0$). As expected, the distribution broadens in the purely quantum limit, Fig. \ref{fig:Histogram}c, approaching a uniform distribution. 

The late-time shifts are determined by the initial distribution of the quantum amplitude $\hat{X}$ of the lasing mode, which behaves as the position operator in a degenerate PA. The numerical results, Figs. \ref{fig:Histogram}e-g, are in excellent agreement with the expected Gaussian distribution (black dots with error bars; see Appendix D):
\begin{equation}\label{eq:GaussianWignerLaser}
W_0(X)=\frac{e^{-\frac{(X-X_C)^2}{2\Delta X^2_Q}}}{\sqrt{2\pi \Delta X^2_Q}}. 
\end{equation}
The mean $X_C\sim \sqrt{n_0}A$ is the coherent component of the lasing amplitude, while $\Delta X_Q\sim 1$ controls its quantum fluctuations; their values are shown in Fig. \ref{fig:Histogram}h as a function of $A$. 

In a BHL, the quantization procedure leads to a spontaneous breaking of time-translation symmetry due to the absence of a well-defined instantaneous vacuum \cite{Ribeiro2022}. Using this observation, we prove that a time operator can be defined for any PA \cite{SuppMatTimeCrystal}, 
\begin{equation}
    \hat{t}\equiv-\frac{1}{\Gamma}\ln \left|\dfrac{\hat{X}}{X_C}\right|,
\end{equation}
which allows to associate a time-shift $t_0$ with each lasing amplitude by $X=X_Ce^{-\Gamma t_0}$. This early time-shift is inherited by the asymptotic CES state, becoming a phase-shift $\phi_0=-\omega t_0$. Late-time expectation values then take the form
\begin{eqnarray}
\braket{O}&\simeq&\int^\pi_{-\pi}\mathrm{d} \phi_0~W(\phi_0)O[\phi_0],\\
\nonumber W(\phi_0)&=&\left|\frac{dX}{d\phi_0}\right|W_0(X),~\phi_0(X)=\frac{\omega}{\Gamma}\ln\frac{X}{X_C},
\end{eqnarray}
where $O[\phi_0]$ results from evaluating the observable $O$ using $\hat{\Psi}(x,t)= \tilde{\Psi}_0(x,\phi_0+\omega t)$. Thus, $W(\phi_0)$ is the statistical distribution for the phase-shift operator $\hat{\phi}_0$, whose width $\Delta \phi_0$ is in turn controlled by $\Delta X_Q$ and $\omega/\Gamma$. The HTC formation requires $\Delta\phi_0\gtrsim 1$, which implies $\Delta X_Q\gtrsim X_C$, $\omega/\Gamma\gtrsim  1$ \cite{SuppMatTimeCrystal}. Finite temperature and experimental noise do not spoil but rather enhance the phase-shift randomness as they increase $\Delta \phi_0$ \cite{SuppMatTimeCrystal}. 

\textit{Conclusions.}---We formulate the concept of Hawking time crystal, i.e., a  time crystal whose spontaneous symmetry breaking results from the self-amplification of spontaneous Hawking radiation. The HTC is signaled by a time-independent density and  ETCF, coexisting with a periodic OTCF. Interestingly, an HTC provides a nonlinear periodic analog of the Andreev-Hawking effect, and its formation is understood in terms of two different time operators. Although we focus on a particular model, the emergence of an HTC in a BHL is expected to be universal since its basic ingredients are quite general: strong lasing fluctuations and the long-time achievement of the CES state \cite{deNova2021}.

The formation of an HTC further demonstrates how analog concepts can inspire applications in tabletop experiments, in addition to quantum amplifiers \cite{deNova2023} or low-pass filters \cite{deNova2014a,deNova2017b}. Conversely, an HTC and the associated nonlinear Andreev-Hawking effect represent novel concepts in the analog field, enhancing the intrinsic interest of a potential BHL realization. Moreover, this model provides an ideal test ground for the study of backreaction \cite{Schutzhold2005,Balbinot2005a,Baak2022,Butera2023,Ciliberto2025}. Our results can be translated into other systems governed by similar nonlinear equations of motion, such as magnonic condensates or nonlinear photonics \cite{Makinen2024,Carusotto2013,Drummond2014,Drori2019,Nguyen2015,Falque2025}. Future work should address the thermodynamical characterization of the HTC by using the tools of Floquet thermodynamics \cite{deNova2025}. Another intriguing line of research is the behavior as quantum clocks \cite{Singh2026,Viotti2026} of the two time operators here considered. From a broader perspective, any realization of a time operator within a gravitational analog scenario is of fundamental interest, since it provides a novel and tangible platform for foundational research on the quantum nature of time, including potential connections to quantum gravity.

\acknowledgments

\textit{Acknowledgments.}---This work has received funding from European Union's Horizon 2020 research and innovation programme (Marie Sk\l{}odowska-Curie Grant Agreement No. 847635), and from Spain's MICIU/AEI through Proyectos de Generaci\'on de Conocimiento (Grant No. PID2022-139288NB-I00) and through Ram\'on y Cajal program (Grant No. RYC2024-050437-I). 

\bibliographystyle{apsrev4-1}
\bibliography{Hawking}

\section{Appendix}

\textit{Appendix A: FPBHL dynamics}. The condensate dynamics for $t\geq 0$ in our FPBHL model is governed by the GP equation 
\begin{eqnarray}\label{eq:GPEquationFPBHL}
\nonumber i\partial_t\Psi(x,t)&=&\left[-\frac{1}{2}\partial_x^2+g(x)\left[|\Psi(x,t)|^2-1\right]+1\right]\Psi(x,t),\\
g(x)&=&1+(c^2_2-1)\theta\left(x+L/2\right)\theta\left(L/2-x\right).
\end{eqnarray}
For times $t<0$ before the quench, $g(x)=1$. Hence, $\Psi(x,t)=e^{i(vx-\mu_0 t)}$, with $\mu_0=1+v^2/2$, is a stationary GP solution at all times. However, the field fluctuations 
\begin{equation}
    \hat{\Psi}(x,t)=\left[1+\frac{\hat{\varphi}(x,t)}{\sqrt{n_0}}\right]e^{i(vx-\mu_0 t)} %$\hat{\Psi}(x,t)=[1+\hat{\varphi}(x,t)]e^{i(vx-\mu_0 t)}$
\end{equation}
do experience nontrivial dynamics after the quench, described at the linear level by the Bogoliubov-de Gennes (BdG) equations 
\begin{eqnarray}\label{eq:FPBHLBdG}
i\partial_t\hat{\Phi}&=&M_0\hat{\Phi},~\hat{\Phi}=\left[\begin{array}{c}\hat{\varphi}\\ \hat{\varphi}^{\dagger}\end{array}\right],\\
\nonumber~M_0&=&\left[\begin{array}{cc}-\frac{1}{2}\partial^2_x-iv\partial_x+g(x) & g(x)\\ -g(x) & \frac{1}{2}\partial^2_x-iv\partial_x-g(x)\end{array}\right].
\end{eqnarray}
In general, BdG solutions can be expanded in terms of a complete set of eigenmodes $M_0z_n=\omega_n z_n$, which are orthogonal under the inner product 
\begin{equation}
    (z_n|z_m)\equiv \int\mathrm{d}x~z_n^\dagger(x) \sigma_z z_m(x),
\end{equation}
$\sigma_i$ being the Pauli matrices. This is because $M_0$ is pseudo-Hermitian, i.e., $(z_n|M_0 z_m)=(M_0 z_n|z_m)$. Nevertheless, both $M_0$ and the inner product are not positive semidefinite; the conjugate mode $\bar{z}_n\equiv \sigma_x z^*_n$ satisfies $M_0 \bar{z}_n=-\omega^*_n\bar{z}_n,~(z_n|z_m)=-(\bar{z}_m|\bar{z}_n)$.

For $t<0$, $M_0$ is a homogeneous operator, diagonal in the Fourier basis:
\begin{align}
    M_0z_k&=\omega_k z_k,~z_k(x)=\frac{e^{ikx}}{\sqrt{L}}s_k,~s_k=\left[\begin{array}{c}u_{k}\\ v_{k}\end{array}\right]\\
    \nonumber u_k&=\frac{\frac{k^2}{2}+(\omega_k-vk)}{\sqrt{2k^2|\omega_k-vk|}},~
v_k=\frac{\frac{k^2}{2}-(\omega_k-vk)}{\sqrt{2k^2|\omega_k-vk|}},
\end{align}
where the dispersion relation is 
\begin{equation}\label{eq:DispersionRelation}
(\omega_k-vk)^2=c^2k^2+\frac{k^4}{4},
\end{equation}
the speed of sound $c=\sqrt{g}$ being $c=1$ in this case. The field spinor is expanded for $t<0$ as
\begin{equation}
    \hat{\Phi}(x,t)=\sum_k\hat{\alpha}_kz_k(x)e^{-i\omega_k t}+ \hat{\alpha}^\dagger_k\bar{z}_k(x)e^{i\omega_k t},
\end{equation}
with $\hat{\alpha}_k$ the annihilation operator of the mode $z_k$. 

For $t\geq 0$, apart from the real spectrum of scattering states \cite{Finazzi2010} (irrelevant for our discussion and neglected in the following), a dynamical instability emerges above critical velocities $v_n$ such that \cite{Michel2013}
\begin{equation}\label{eq:Criticalvs}
    L=\frac{n\pi +\arctan\sqrt{\dfrac{1-v_n^2}{v_n^2-c_2^2}}}{\sqrt{v_n^2-c_2^2}},~n=0,1\ldots
\end{equation}
The $n$-th mode is degenerate for $v_n\leq v\leq v_{n+1/2}$, developing a nonvanishing real part of the frequency for $v>v_{n+1/2}$. In this work, we restrict ourselves to short cavities satisfying $v_0<v<v_{1/2}$, so that the FPBHL only displays a single, degenerate unstable mode, with purely imaginary frequency $\Omega=-\Omega^*\equiv i\Gamma$, $\Gamma>0$. Specifically, there are two complex BdG modes, 
\begin{equation}
M_0z_I=\Omega z_I,~ M_0z_S=\Omega^* z_S,~z_{I,S}=\left[\begin{array}{c}u_{I,S}\\ v_{I,S}\end{array}\right].
\end{equation}
It can be seen that one can choose both $z_I=\bar{z}_I$ and $z_S=\bar{z}_S$, satisfying the normalization condition $(z_I|z_S)=i$, with $(z_I|z_I)=(z_S|z_S)=0$. Moreover, due to the even parity of $g(x)$, $g(x)=g(-x)$, the BdG equations are $PT$ invariant, which implies that one can take $z_S(x)=-z^*_I(-x)$. Their contribution to the field spinor reads as
\begin{equation}
    \hat{\Phi}(x,t)\simeq \hat{X}z_I(x)e^{\Gamma t}+\hat{P}z_S(x)e^{-\Gamma t},
\end{equation}
since their quantum amplitudes precisely behave as conjugate position-momentum operators because 
\begin{equation}
    \hat{X}= i(z_{S}|\hat{\Phi}),~\hat{P}= -i(z_{I}|\hat{\Phi})\Longrightarrow [\hat{X},\hat{P}]=(z_I|z_S)=i.
\end{equation}
A proper annihilation operator is then constructed as
\begin{equation}  
\hat{a}=\frac{\hat{X}+i\hat{P}}{\sqrt{2}}.
\end{equation}
In terms of this amplitude, the BHL behaves as a degenerate PA since, after neglecting $c$-number contributions, the grand-canonical Hamiltonian governing the BdG dynamics takes the form \cite{Finazzi2010,deNova2024} 
\begin{equation}\label{eq:DPA}
\hat{K}=\hat{H}-\mu_0\hat{N}\simeq\hat{K}_{\rm{DPA}}\equiv  i\Gamma\frac{(\hat{a}^\dagger)^2-\hat{a}^2}{2}=\Gamma\frac{\hat{X}\hat{P}+\hat{P}\hat{X}}{2}.
\end{equation}
The density fluctuations in the BdG approximation read $\hat{\Psi}^\dagger(x,t)\hat{\Psi}(x,t)\simeq 1+\delta\hat{n}(x,t)$, with 
\begin{equation}
    \delta\hat{n}(x,t)\equiv\frac{\hat{\varphi}(x,t)+\hat{\varphi}^\dagger(x,t)}{\sqrt{n_0}}=\frac{\hat{X}r_I(x)e^{\Gamma t}+\hat{P}r_S(x)e^{-\Gamma t}}{\sqrt{n_0}},
\end{equation}
$r_{I,S}(x)=u_{I,S}(x)+v_{I,S}(x)=2\,\textrm{Re}\, u_{I,S}(x)$ being the corresponding density wavefunction; $\textrm{Re}\, u_I(x)$ is depicted in Fig. \ref{fig:Model}a in dashed-dotted green. The nonlinear saturation regime, where the BdG approximation ceases to be valid, is reached when $\delta\hat{n}(x,t)\sim 1$, i.e., $\hat{X}e^{\Gamma t}\sim \sqrt{n_0}\gg 1$.

%where we have invoked the canonical commutation relations, $[\hat{\varphi}(x,t),\hat{\varphi}^\dagger(x',t)]=\delta(x-x')$. 

%Notice that the first term in the r.h.s. is the Hamiltonian of a , or that of an unstable harmonic oscillator (apart from some trivial phase transformation). 

%+\sum_\omega \omega \hat{\alpha}^\dagger_\omega\hat{\alpha}_\omega

% \begin{eqnarray}\label{eq:FieldExpansionBHL}
%     \hat{\Phi}(x,t)&=&\hat{X}z_I(x)e^{\Gamma t}+\hat{P}z_S(x)e^{-\Gamma t}\\
%     \nonumber &+&\sum_\omega\hat{\alpha}_\omega z_\omega(x)e^{-i\omega t}+\hat{\alpha}^\dagger_\omega \bar{z}_\omega(x) e^{i\omega t}.
% \end{eqnarray}

%Moreover, due to the even parity of $g(x)$, $g(x)=g(-x)$, the BdG equations are $PT$ invariant, which implies that one can take $z_S(x)=-z^*_I(-x)$.

\textit{Appendix B: HTC figure of merit}. We compute the spatial Fourier transform $g^{(2)}(k,k',t,t')$ of the OTCF $g^{(2)}(x,x',t,t')$ in the downstream-upstream quadrant ($x>0$, $x'<0$), which characterizes the Hawking-like correlations between the upstream wave and the downstream soliton. Due to their ballistic nature, we expect Fourier peaks in the upstream/downstream regions at $k_{u,d}=\omega/|v_{u,d}|$, respectively \cite{SuppMatTimeCrystal}. Thus, we take 
\begin{equation}
    \mathcal{G}(t,t')\equiv \frac{g^{(2)}(k_d,k_u,t,t')}{g_{\rm{HTC}}^{(2)}(k_d,k_u,\tau=0)}
\end{equation}
as figure of merit. The normalization is chosen such that the theoretical prediction is simply $\mathcal{G}_{\rm{HTC}}(\tau)=e^{i\omega\tau}$ \cite{SuppMatTimeCrystal}, which results from replacing the numerator above by $g_{\rm{HTC}}^{(2)}(k_d,k_u,\tau)$, so $\mathcal{G}(\tau; t)\equiv \mathcal{G}(t,t+\tau)=\mathcal{G}_{\rm{HTC}}(\tau)$.

\begin{figure}[!tb]
\begin{tabular}{@{}cc@{}}
\stackinset{l}{-4pt}{t}{-7pt}{(a)}{\includegraphics[width=0.5\columnwidth]{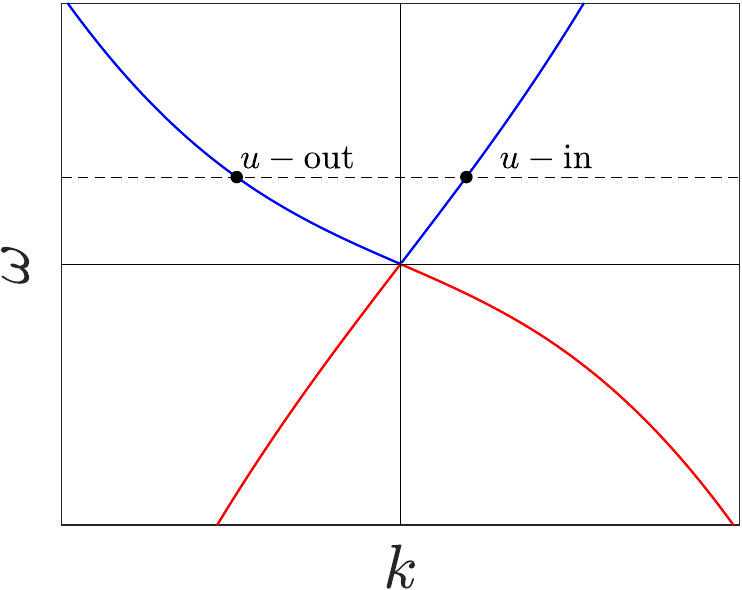}} & \stackinset{l}{-4pt}{t}{-7pt}{(b)}{\includegraphics[width=0.5\columnwidth]{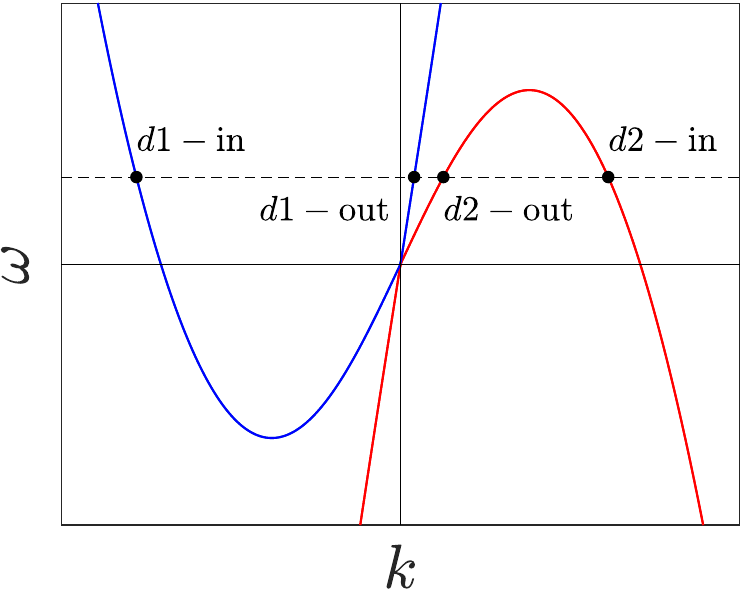}} \\
\stackinset{l}{-4pt}{t}{-7pt}{(c)}{\includegraphics[width=0.5\columnwidth]{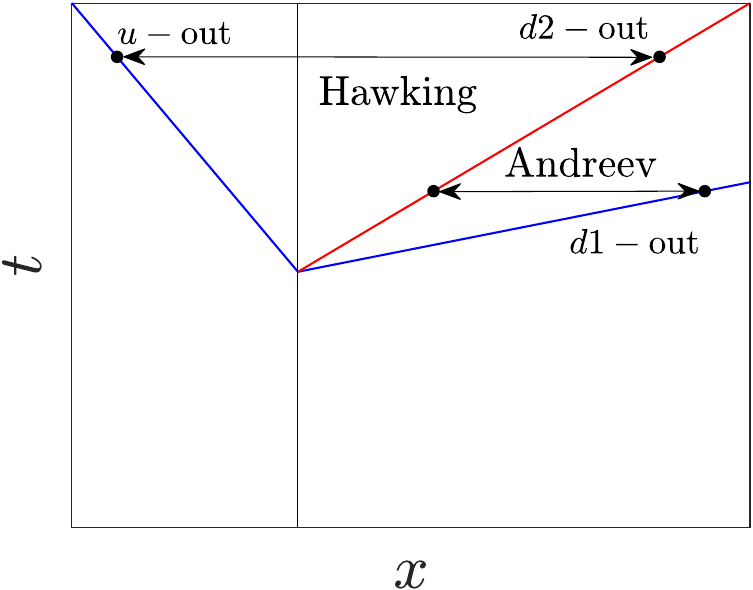}} & \stackinset{l}{-4pt}{t}{-7pt}{(d)}{\includegraphics[width=0.5\columnwidth]{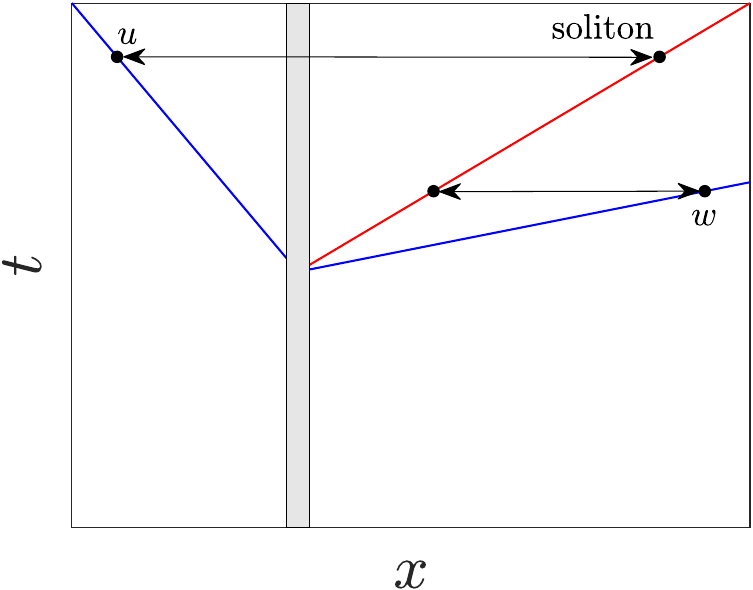}} 
\end{tabular}
\caption{(a)-(b) Schematic representation of the dispersion relation (\ref{eq:DispersionRelation}) for a subsonic ($c>v$) and supersonic ($c<v$) flow. (c) Spacetime diagram of the outgoing modes in an analog black hole. Vertical line indicates the horizon at $x=0$. (d) Same as (c) but for an HTC, where the vertical gray band now represents the finite lasing cavity. }
\label{fig:DispersionLaser}
\end{figure}

\textit{Appendix C: Andreev-Hawking vs. HTC}. Figure \ref{fig:ClassicalQuantum}l shows the ETCF at $t=500$ for a TW simulation of a black hole, computed in the same way as for the FPBHL case (\ref{eq:GPEquationFPBHL}) but now $g(x)=1+(c^2_2-1)\theta(x)$ for $t\geq 0$, with $c_2<v$. Thus, the upstream/downstream regions are subsonic/supersonic, with their dispersion relation schematically depicted in Figs. \ref{fig:DispersionLaser}a-b, where ``in'' modes travel towards the horizon while ``out'' modes travel outwards the horizon, here exactly placed at $x=0$. Blue line represents $\omega_k$, Eq. (\ref{eq:DispersionRelation}), and red line is the conjugate $-\omega_{-k}$. 

The Andreev-Hawking effect is characterized by the spontaneous, correlated emission of outgoing modes from the horizon (see spacetime diagram in Fig. \ref{fig:DispersionLaser}c). Specifically, Hawking correlations involve the normal $u$-out (denoted as the Hawking mode in this context) and the anomalous $d2$-out (partner) modes, while Andreev correlations involve the normal $d1$-out (copropagating) and the anomalous $d2$-out modes. This establishes a direct analogy with the traveling features of an HTC, whose spacetime diagram is shown in Fig. \ref{fig:DispersionLaser}d. The nonlinear counterpart of the anomalous partner $d2$ modes are the solitons, carrying a density defect, while the upstream/downstream waves with positive amplitude are akin to the normal $u$/$d1$ modes. The role of the horizon (vertical black line in Fig. \ref{fig:DispersionLaser}c) is here played by the localized lasing cavity (vertical gray band in Fig. \ref{fig:DispersionLaser}d).

The Hawking moustache and the Andreev band in Fig. \ref{fig:ClassicalQuantum}l are predicted in the long-wavelength hydrodynamic limit, where $\omega_k\simeq v_ik$, with $v_u=v-1$, $v_{d1,d2}=v\pm c_2$ \cite{Recati2009}. Their trajectory is obtained by replacing $v_u\to v_u$, $v_{w}\to v_{d1}$, $v_{d}\to v_{d2}$, and setting $\tau=T=0$ in Eq. (\ref{eq:NonLinearAndreedHawkingOOT}), reinforcing the physical picture of the HTC as a nonlinear periodic analog of the Andreev-Hawking effect. Dispersive effects also manifest similarly: i) in the HTC, as parallel fringes to the dashed magenta lines in Figs. \ref{fig:ClassicalQuantum}h-j, representing the positive correlation between the soliton and the upstream wave minima (see magenta circles in Fig. \ref{fig:ClassicalQuantum}d); ii) in the Andreev-Hawking effect, as parallel fringes to the main correlation bands in Fig. \ref{fig:ClassicalQuantum}l.

\textit{Appendix D: Lasing and phase-shift distributions.} 
For $t<0$, the quantum state is the quasiparticle vacuum $\ket{0}$, $\hat{\alpha}_k\ket{0}=0$. The coherent perturbation can be described as a classical contribution $\Phi_C(x)=[\delta\Psi_C(x) ~\,\delta\Psi^*_C(x)]^{\rm{T}}$ to the initial field fluctuations:
\begin{equation}\label{eq:InitialFieldSpinor}
    \hat{\Phi}(x,0)=\sqrt{n_0}A\Phi_C(x)+\sum_k\hat{\alpha}_kz_k(x)+ \hat{\alpha}^\dagger_k\bar{z}_k(x).
\end{equation}
The lasing amplitude at $t=0$ then reads as
\begin{eqnarray}\label{eq:BetaAmplitudeLasing} \nonumber \hat{X}&=&i(z_S|\hat{\Phi}(0))=X_C+\frac{1}{\sqrt{L}}\sum_k\hat{\alpha}_k\beta_k+ \hat{\alpha}^\dagger_k\beta^*_k,\\
X_C&\equiv& i(z_S|\Phi_C)\sqrt{n_0}A,~\beta_k\equiv i\sqrt{L}(z_S|z_k).
\end{eqnarray}

Since the initial Wigner distribution for the quasiparticle amplitudes $\alpha_k$ is Gaussian \cite{SuppMatTimeCrystal}, the initial Wigner distribution for $X$ is also Gaussian, Eq. (\ref{eq:GaussianWignerLaser}), determined by its mean and variance:
\begin{eqnarray}
\braket{\hat{X}}&=&X_C,\\
\nonumber \Delta X^2_Q&\equiv&\braket{\hat{X}^2}-\braket{\hat{X}}^2=\frac{1}{L}\sum_k|\beta_k|^2=\frac{1}{2\pi }\int\mathrm{d}k~|\beta_k|^2.
\end{eqnarray}
Hence, $\beta_k$ controls the quantum fluctuations of the lasing mode, in analogy to the usual Hawking coefficient $\beta_\omega$ \cite{Macher2009a}.

For $t\geq0$, the degenerate PA (\ref{eq:DPA}) exponentially amplifies position eigenstates ($\hat{X}\ket{X}=X\ket{X}$) as $e^{-i\hat{K}t}\ket{X}=e^{\frac{\Gamma t}{2}}\ket{Xe^{\Gamma t}}$. This allows to define a time operator by \cite{SuppMatTimeCrystal}
\begin{equation}
    \hat{t}\equiv-\frac{1}{\Gamma}\ln \left|\dfrac{\hat{X}}{X_C}\right|,~[\hat{t},\hat{K}]=-i.
\end{equation}
Thus, the evolution of any lasing amplitude $X=X_Ce^{-\Gamma t_0}$ is obtained by shifting the mean-field trajectory a time $t_0$, yielding an asymptotic oscillation phase-shift
\begin{equation}
    \phi_0(X)=-\omega t_0=\frac{\omega}{\Gamma}\ln\frac{X}{X_C}.
\end{equation}
This derivation assumes that $X$ and $X_C$ have the same sign, valid for weak-to-moderate quantum fluctuations. The case of arbitrary $X$ is also understood in terms of the time operator in a similar fashion, with a minor technical correction \cite{SuppMatTimeCrystal}.

\begin{appendices}

\onecolumngrid

\setcounter{equation}{0}
\setcounter{figure}{0}
\setcounter{table}{0}

\renewcommand{\theequation}{S\arabic{equation}}
\renewcommand{\thefigure}{S\arabic{figure}}
%\renewcommand{\bibnumfmt}[1]{[S#1]}
%\renewcommand{\citenumfont}[1]{S#1}

% IMPORTANTE para hyperref
\renewcommand{\theHequation}{S\arabic{equation}}
\renewcommand{\theHfigure}{S\arabic{figure}}

\section{Supplemental Material}

\subsection{Lasing mode computation}\label{SM:BHL}

We briefly sketch the computation of the complex modes $z_I,z_S$ with frequencies $\Omega,\Omega^*$ ($\textrm{Im}\,\Omega>0$) for a flat-profile black-hole laser (FPBHL). First, they can be determined from the time-independent Bogoliubov-de Gennes (BdG) equations (see Refs. \cite{Michel2013,Michel2015} for more details):
\begin{eqnarray}\label{eq:FPBHLBdG}
M_0 z_I&=&\Omega z_I,~M_0z_S=\Omega^* z_S,~z_{I,S}=\left[\begin{array}{c}u_{I,S}\\ v_{I,S}\end{array}\right],\\
\nonumber M_0&=&\left[\begin{array}{cc}-\frac{1}{2}\partial^2_x-iv\partial_x+g(x) & g(x)\\ -g(x) & \frac{1}{2}\partial^2_x-iv\partial_x-g(x)\end{array}\right],~g(x)=1+(c^2_2-1)\theta\left(x+L/2\right)\theta\left(L/2-x\right).
\end{eqnarray}
Since the problem is piecewise homogeneous, it can be solved by matching the different plane waves in each homogeneous region, which now have complex wavevectors, given by the zeros of the polynomial equation
\begin{equation}
    (\Omega-vk)^2=c^2k^2+\frac{k^4}{4},
\end{equation}
where $c$ is the corresponding speed of sound. The BdG plane-wave solutions associated with each wavevector read
\begin{equation}
    M_0z_k=\Omega z_k,~z_k(x)=e^{ikx}\left[\begin{array}{c}\frac{k^2}{2}+(\Omega-vk)\\ \frac{k^2}{2}-(\Omega-vk)\end{array}\right].
\end{equation}
Notice that we skip here normalization factors since these are non-normalizable solutions with complex wavevector.  Specifically, inside the lasing cavity ($|x|<L/2$), we have $4$ complex modes while, in each asymptotic subsonic region ($|x|>L/2$), we only have $2$ asymptotically bounded modes. Hence, in total, we have $8$ degrees of freedom, corresponding to the amplitudes of each of these modes. On the other hand, each matching at $x=\pm L/2$ imposes $4$ equations ($2$ for $z_{I}$ and $2$ for its derivative), so we have $8$ linear equations for $8$ degrees of freedom. Thus, in order to display nontrivial solutions, the $8\times 8$ matrix $A(\Omega)$ describing the homogeneous linear system of equations must satisfy
\begin{equation}\label{eq:BdGScat}
    \det A(\Omega)=0.
\end{equation}
The zeros of this equation yield the complex frequencies $\Omega$ of the lasing modes. The same process is repeated for $\Omega^*$, using the corresponding asymptotically bounded modes. Once obtained, the BdG spinors $z_I,z_S$ are determined up to a global factor. 

In the specific case of a degenerate lasing mode, $\Omega=-\Omega^*=i\Gamma$, we fix the normalization by the conditions
\begin{equation}
    z_I=\bar{z}_I,~z_S=\bar{z}_S,~(z_I|z_S)=i,
\end{equation}
and by $z_S(x)=-z^*_I(-x)$, where this latter relation is a specific feature of the FPBHL, resulting from its $PT$ invariance. 

\begin{figure*}[!t]
\begin{tabular}{@{}cc@{}}   
\stackinset{l}{0pt}{t}{0pt}{(a)}{\includegraphics[width=0.5\textwidth]{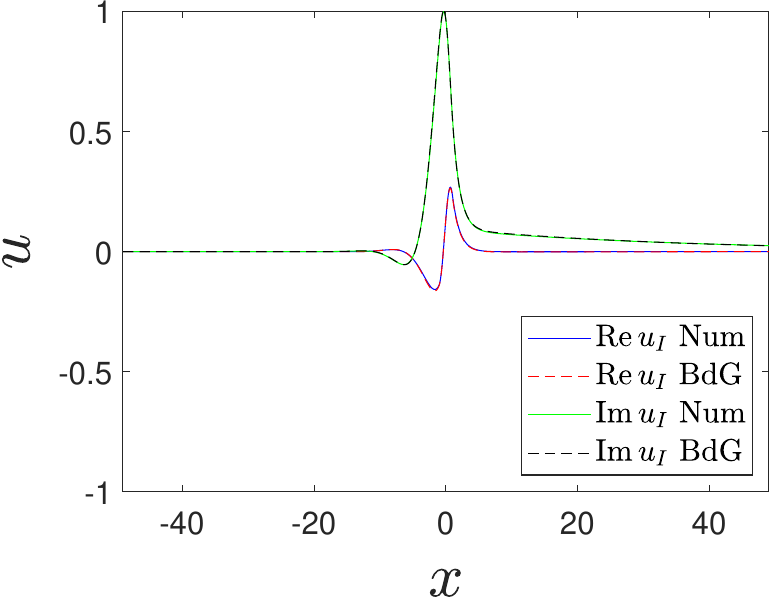}} 
& \stackinset{l}{0pt}{t}{0pt}{(b)}{\includegraphics[width=0.5\textwidth]{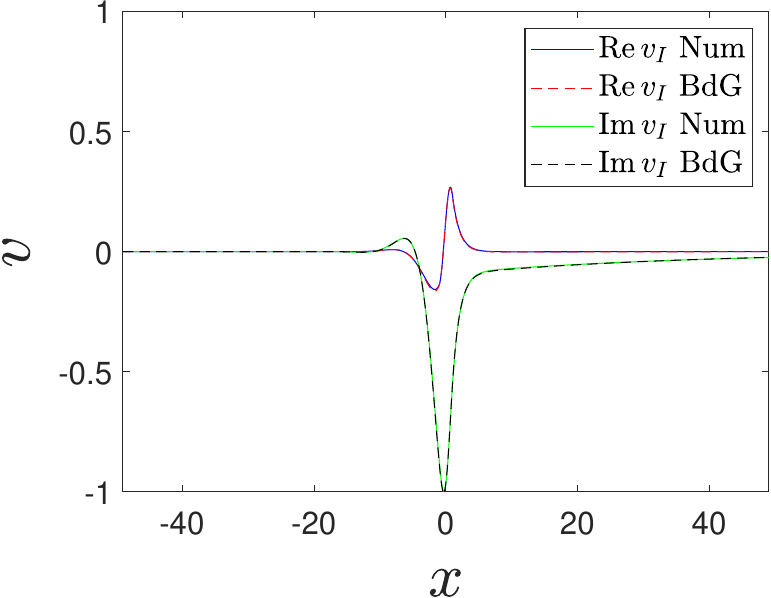}}\\
\stackinset{l}{0pt}{t}{0pt}{(c)}{\includegraphics[width=0.5\textwidth]{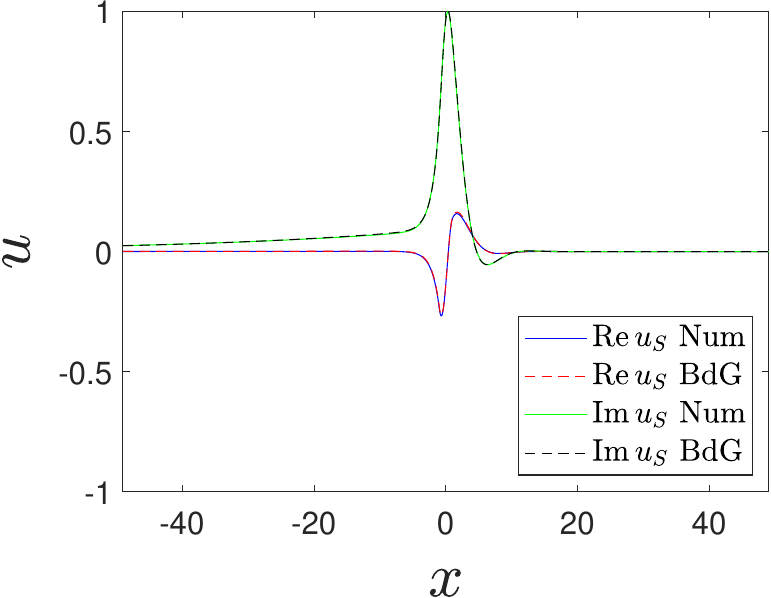}} 
& \stackinset{l}{0pt}{t}{0pt}{(d)}{\includegraphics[width=0.5\textwidth]{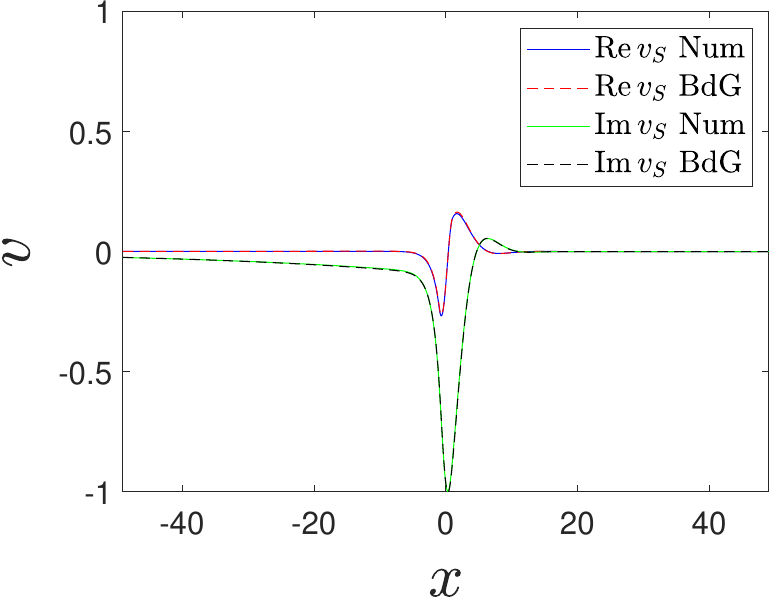}}
\end{tabular}
\caption{(a)-(d) Spatial profile of the real and imaginary part of the BdG components $u_I,v_I,u_S,v_S$ for a FPBHL with $v=0.95,c_2=0.4,L=2$. Solid lines are numerically obtained by finite methods and dashed lines are the analytical result from solving the BdG scattering problem [see Eq. (\ref{eq:BdGScat}) and ensuing discussion].}
\label{fig:LaserMode}
\end{figure*}

Alternatively, we have found here that $\Omega,\Omega^*$ and $z_I,z_S$ can be simply obtained by numerically diagonalizing the BdG operator $M_0$ using finite methods. A very good agreement is observed between both techniques, Fig. \ref{fig:LaserMode}, where the expected relations $u_I(x)=v^*_I(x)=-u^*_S(-x)=-v_S(-x)$ can be also seen.

\subsection{Truncated Wigner method}\label{SM:TW}

The Truncated Wigner (TW) method computes symmetric-ordered expectation values from ensemble averages of integrations of the Gross-Pitaevskii (GP) equation \cite{Sinatra2002,Carusotto2008}, which here reads 
\begin{eqnarray}\label{eq:GPEquationFPBHLTW}
i\partial_t\Psi_W(x,t)&=&\left[-\frac{1}{2}\partial_x^2+g(x)\left[|\Psi_W(x,t)|^2-1\right]+1\right]\Psi_W(x,t),
\end{eqnarray}
by using a stochastic initial condition 
\begin{equation}\label{eq:TWExpansionNC}
\Psi_W(x,0)=\left[1+A\delta\Psi_C(x)+\frac{\varphi_W(x)}{\sqrt{n_0}}\right]e^{ivx}\equiv \left[1+\frac{\delta\Psi_W(x)}{\sqrt{n_0}}\right]e^{ivx},~\delta\Psi_W(x)=\sqrt{n_0}A\delta\Psi_C(x)+\varphi_W(x),
\end{equation}
where we absorb the initial classical stimulation $A\delta\Psi_C(x)=A\cos(\pi x/L)\theta\left(x+L/2\right)\theta\left(L/2-x\right)$  into the definition of the field fluctuations. In turn, $\varphi_W(x)$ is expressed in terms of the BdG modes of the initial homogeneous condensate as
\begin{equation}\label{eq:TWBdG}
\left[\begin{array}{c}\varphi_W(x)\\ \varphi^*_W(x)\end{array}\right]=\frac{1}{\sqrt{L}}\sum_k \alpha_{k} s_ke^{ikx}+\alpha^*_k \bar{s}_ke^{-ikx}=\sum_k \alpha_{k} z_k(x)+\alpha^*_k \bar{z}_k(x),
\end{equation}
the amplitudes $\alpha_k, \alpha^*_k$ being here stochastic variables distributed according to the Wigner function of the initial state. Specifically, we make the usual assumption that the initial quantum state is a thermal state at temperature $T=1/\beta$ ($k_B=1$) in the comoving frame of the condensate \cite{Macher2009a,Busch2014}: 
\begin{equation}\label{eq:ThermalState}
    \hat{\rho}=\frac{e^{-\beta \hat{K}_C}}{Z},~Z=\textrm{Tr}\,e^{-\beta \hat{K}_C},~\hat{K}_C=\sum_k\Omega_k\hat{\alpha}^\dagger_k\hat{\alpha}_k,~\Omega_k=\sqrt{k^2+\frac{k^4}{4}}.
\end{equation}
The associated Wigner function is a Gaussian distribution characterized by the first and second-order momenta
\begin{eqnarray}
    \nonumber \braket{\alpha_k}&=&\braket{\hat{\alpha}_k}=0,~\braket{\alpha_{k'}\alpha_k}=\braket{\hat{\alpha}_{k'}\hat{\alpha}_k}=0,\\    \braket{\alpha^*_{k'}\alpha_k}&=&\frac{\braket{\hat{\alpha}^\dagger_{k'}\hat{\alpha}_k+\hat{\alpha}_k\hat{\alpha}^\dagger_{k'}}}{2}=\left(n_k+\frac{1}{2}\right)\delta_{kk'}=\frac{\coth\dfrac{\beta\Omega_k}{2}}{2}\delta_{kk'},~n_k=\frac{1}{e^{\beta\Omega_k}-1}.
\end{eqnarray}
By subtracting the constant factor $\delta_{kk'}/2$ in the last line, arising from the commutator between annihilation and creation operators, one retrieves the more usual normal-ordered expectation value $\braket{\hat{\alpha}_{k'}^\dagger\hat{\alpha}_{k}}$. The zero-temperature case considered in the main text amounts to setting $n_k=0$,
\begin{equation}
\braket{\alpha^*_{k'}\alpha_k}=\frac{\delta_{kk'}}{2}.
\end{equation}

For the TW simulations, we take the total length of the grid and particle number to be $L\approx 1885$ and $N=10^8$, with $n_0=N/L\approx 5.31\times 10^4\gg 1$ the condensate density. The number of modes included in the initial condition is $N_m=3000\ll N$, which amounts to implementing a momentum cutoff $|k|<k_C=5$. The numerical method employed to integrate the GP equation is explained in detail in Ref. \cite{deNova2016}.

Technically, one should use a number-conserving approximation in the computations, taking the number of particles in the condensate for each realization as $N_0=N-N_{\rm{NC}}$, with $N_{\rm{NC}}$ the number of non-condensed particles
\begin{eqnarray}
    N_{\rm{NC}}&=&\sum_k \left(|u_k|^2+|v_k|^2\right)\left(|\alpha_k|^2-\frac{1}{2}\right)+|v_k|^2.
\end{eqnarray}
This is translated into a modified version of Eq. (\ref{eq:TWExpansionNC}),
\begin{equation}\label{eq:TWExpansionConserving}
\Psi_W(x,0)=\left[\sqrt{1-\frac{N_{\rm{NC}}}{N}}+\frac{\delta\Psi_W(x)}{\sqrt{n_0}}\right]e^{ivx}.
\end{equation}
For the parameters considered, we have not found any significant effect resulting from the use of a non-conserving approximation.

Regarding the evaluation of observables, since averages over the TW ensemble are equivalent to symmetric-ordered expectation values, commutators of the form 
\begin{equation}
    [\hat{\Psi}(x,t),\hat{\Psi}^\dagger(x',t)]=\frac{1}{2\pi n_0}\int^{k_C}_{-k_C}\mathrm{d}k~e^{ik(x-x')}=\frac{\sin k_C(x-x')}{\pi n_0 (x-x')}
\end{equation}
arise in the symmetrization of observables evaluated at equal times, such as the density or the ETCF (the factor $n^{-1}_0$ results from our unit convention). However, their contribution becomes negligible when compared to that from the time-shift fluctuations at late times. Similarly, regarding the OTCF, we will face commutators of the form $[\hat{\Psi}(x,t),\hat{\Psi}^\dagger(x',t')]\sim n^{-1}_0\ll 1$, which are once more negligible when compared to the late time-shift fluctuations. As a result of neglecting these commutators, expectation values can be simply evaluated at late times by taking averages over the TW ensemble with the substitution $\hat{\Psi}(x,t)\to\Psi_W(x,t)$ and $\hat{\Psi}^\dagger(x,t)\to\Psi^*_W(x,t)$.

\begin{figure*}[!t]
\begin{tabular}{@{}cccc@{}}   
\stackinset{l}{0pt}{t}{0pt}{(a)}{\includegraphics[width=0.25\textwidth]{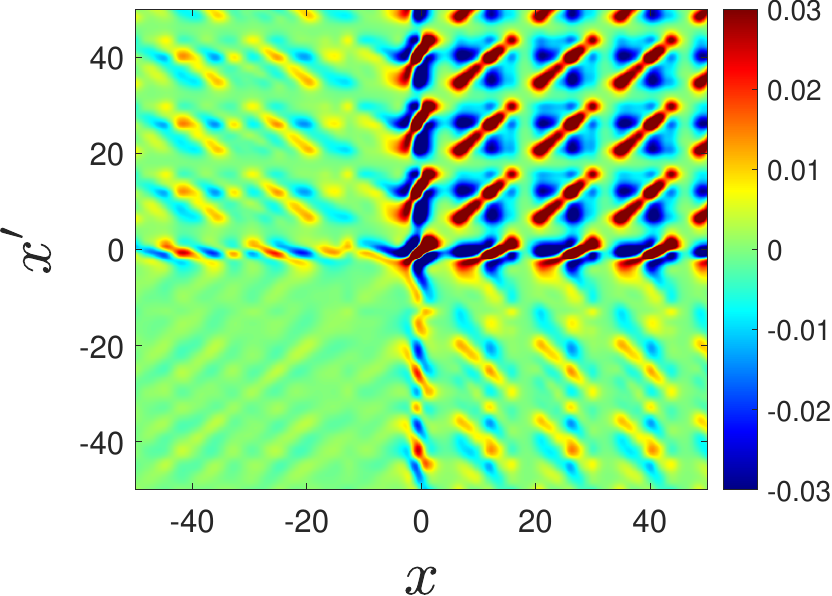}} 
& \stackinset{l}{0pt}{t}{0pt}{(b)}{\includegraphics[width=0.25\textwidth]{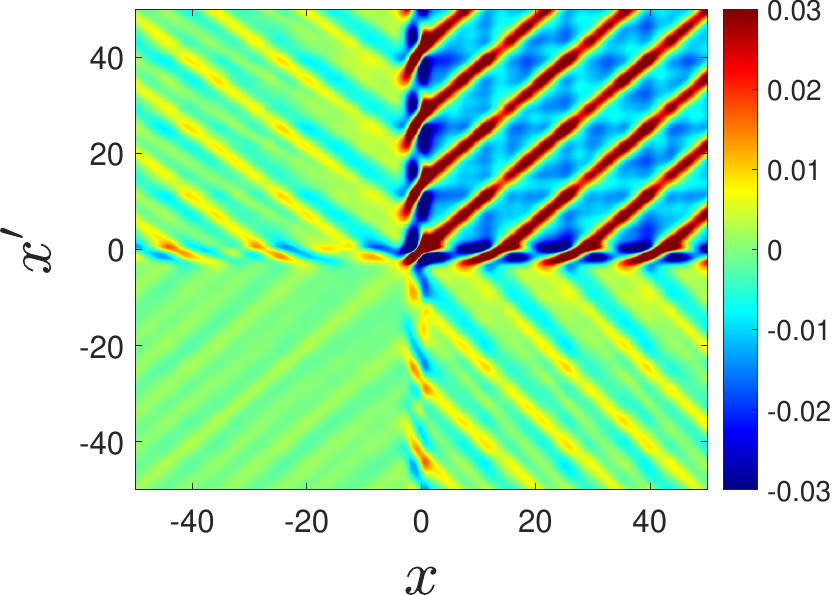}}& 
\stackinset{l}{0pt}{t}{0pt}{(c)}{\includegraphics[width=0.25\textwidth]{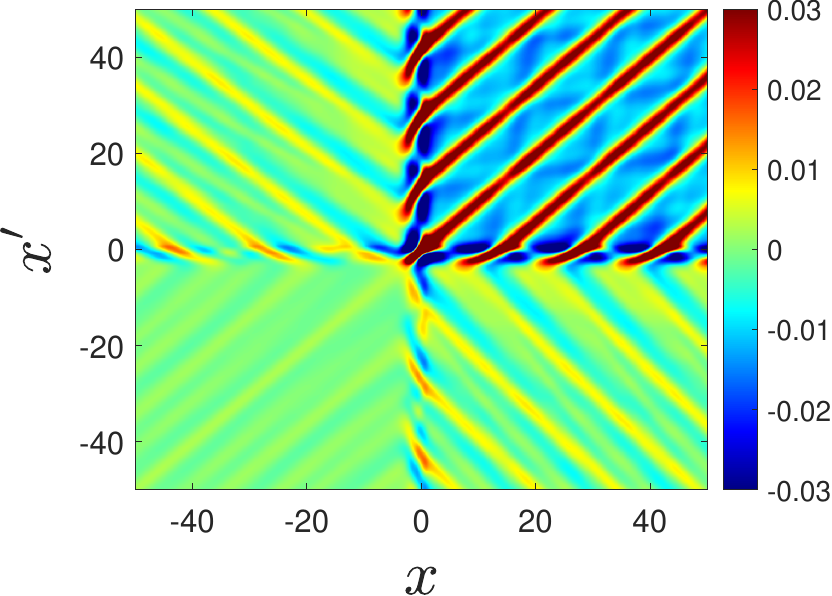}}& 
\stackinset{l}{0pt}{t}{0pt}{(d)}{\includegraphics[width=0.25\textwidth]{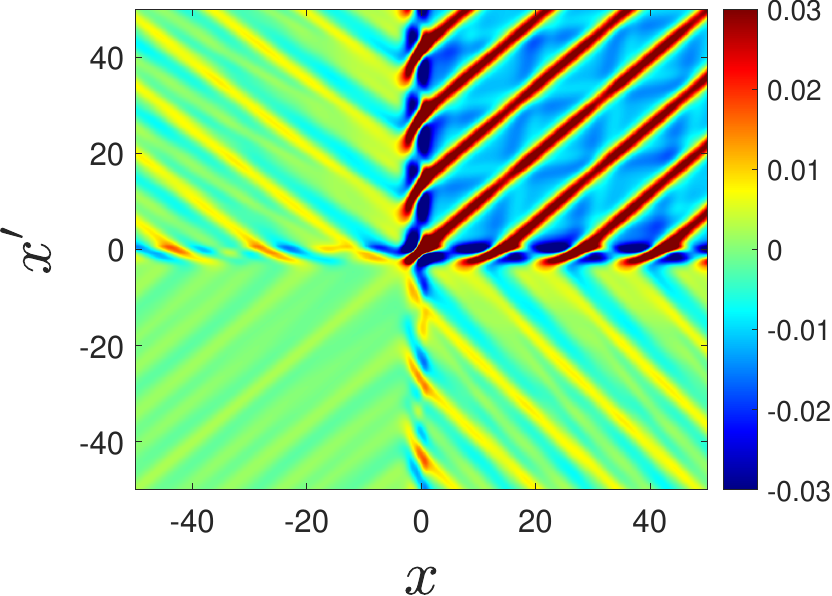}} 
\end{tabular}
\caption{TW computation of the ECTF $g^{(2)}(x,x',t)$ at $t=870$ for a purely quantum FPBHL with $v=0.95,c_2=0.4,L=2$ for different sizes $N_{\rm{stat}}$ of the TW ensemble. (a) $N_{\rm{stat}}=10$. (a) $N_{\rm{stat}}=100$. (c) $N_{\rm{stat}}=1000$. (d) Ideal HTC prediction $g_{\rm{HTC}}^{(2)}(x,x',\tau=0)$, Eq. (\ref{eq:HteoSM}).}
\label{fig:Stats}
\end{figure*}

Regarding statistics, expectation values are computed using ensembles of $N_{\rm{stat}}=1000$ simulations. We examine in Figs. \ref{fig:Stats}a-c the effect of finite statistics by considering TW ensembles of $N_{\rm{stat}}=10,100,1000$, respectively. We observe that $N_{\rm{stat}}=1000$ already displays a very good agreement with the ideal theoretical prediction, Fig. \ref{fig:Stats}d. Error bars in the histograms of Fig. \ref{fig:Histogram} of the main text represent the standard deviation of a binomial distribution with $N=N_{\rm{stat}}$ trials and probability $p$, where $p$ is the probability to find the random variable in the corresponding bin, accordingly computed from the expected theoretical distribution. For a uniform distribution (oscillation phase-shift $\phi_0$, upper row of Fig. \ref{fig:Histogram}), $p=1/N_{\rm{bins}}$, with $N_{\rm{bins}}$ the number of bins. For a Gaussian distribution (lasing amplitude $X$, lower row of Fig. \ref{fig:Histogram}), $p$ can be computed in terms of the error function. We have also checked that the main results are quite insensitive to the number of bins in general, and to the position $x_0$ and time of observation in the particular case of the $\phi_0$ distribution.

\subsection{Quantum distributions}\label{SM:QuantumDistribution} 

We further explain the theoretical model used for the quantum distributions of the initial lasing amplitude $X$ and the asymptotic phase-shift $\phi_0$, as well as their connection through their corresponding time operators.

\subsubsection{Initial state: Lasing distribution}

From the complex modes $z_I,z_S$, we construct a properly normalized mode 
\begin{equation}
    Z=\frac{z_I- iz_S}{\sqrt{2}}  \Longrightarrow (Z|Z)=1,
\end{equation}
whose quantum amplitude is 
\begin{equation}
    (Z|\hat{\Phi})=\frac{\hat{X}+ i\hat{P}}{\sqrt{2}}=\hat{a},~(\bar{Z}|\hat{\Phi})=-\hat{a}^\dagger. %,~\hat{\Phi}=\left[\begin{array}{c}\hat{\varphi}\\ \hat{\varphi}^{\dagger}\end{array}\right]
\end{equation}
The inverse relations read:
\begin{eqnarray}\label{eq:InverseLasingRelation}
     z_I&=&\frac{Z+\bar{Z}}{\sqrt{2}},~z_S=i\frac{Z-\bar{Z}}{\sqrt{2}},\\
     \nonumber \hat{X}&=&\frac{\hat{a}+\hat{a}^\dagger}{\sqrt{2}},~\hat{P}=-i\frac{\hat{a}-\hat{a}^\dagger}{\sqrt{2}}.
\end{eqnarray}
By expanding the field spinor at $t=0$ in terms of the quasiparticle modes and the classical field contribution as in Eq. (\ref{eq:InitialFieldSpinor}) of the main text, 
\begin{equation}
    \hat{\Phi}(x,0)=\sqrt{n_0}A\Phi_C(x)+\sum_k\hat{\alpha}_kz_k(x)+ \hat{\alpha}^\dagger_k\bar{z}_k(x),
\end{equation}
we find that
\begin{eqnarray}\label{eq:AniquilacionCero}
    \hat{a}&=&(Z|\hat{\Phi}(0))=\alpha_C+\frac{1}{\sqrt{L}}\sum_kc_k\hat{\alpha}_k+ d_k\hat{\alpha}^\dagger_k,\\
    \nonumber \alpha_C&=&(Z|\Phi_C)\sqrt{n_0}A,~c_k=\sqrt{L}(Z|z_k),~d_k=\sqrt{L}(Z|\bar{z}_k). %=-(\bar{z}|s_ke^{ikx})^*
\end{eqnarray}
Using this expansion, we compute the initial Wigner function which characterizes the distribution of the complex amplitude $\alpha$ associated with the annihilation operator $\hat{a}$ at $t=0$, 
\begin{eqnarray}    W_0(\alpha)&=&\int\frac{\mathrm{d}^2\eta}{\pi^2}e^{\eta^*\alpha-\eta\alpha^*}\braket{e^{\eta \hat{a}^\dagger-\eta^*\hat{a}}}=\frac{2}{\pi}\frac{e^{-2\frac{\lambda|\alpha-\alpha_C|^2-\Delta^*(\alpha-\alpha_C)^2-\Delta(\alpha^*-\alpha^*_C)^2}{\lambda^2-4|\Delta|^2}}}{\sqrt{\lambda^2-4|\Delta|^2}},
\end{eqnarray}
where the expectation value is evaluated here using the thermal state (\ref{eq:ThermalState}), $\braket{e^{\eta \hat{a}^\dagger-\eta^*\hat{a}}}=\textrm{Tr}[e^{\eta \hat{a}^\dagger-\eta^*\hat{a}}\hat{\rho}]$, and
\begin{eqnarray}
    \lambda&\equiv &\frac{1}{L}\sum_k(|c_k|^2+|d_k|^2)\coth\frac{\beta\Omega_k}{2}=\frac{1}{2\pi}\int\mathrm{d}k~(|c_k|^2+|d_k|^2)\coth\frac{\beta\Omega_k}{2},\\
    \nonumber \Delta&=&\Delta_x+i\Delta_y\equiv\frac{1}{L}\sum_kc_kd_k\coth\frac{\beta\Omega_k}{2}=\frac{1}{2\pi}\int\mathrm{d}k~c_kd_k\coth\frac{\beta\Omega_k}{2}.
\end{eqnarray}
In terms of phase-space operators, from Eqs. (\ref{eq:InverseLasingRelation})-(\ref{eq:AniquilacionCero}), we find
\begin{eqnarray}\label{eq:EspacioFaseCero}
    \hat{X}&=&i(z_S|\hat{\Phi}(0))=X_C+\frac{1}{\sqrt{L}}\sum_k \beta_k\hat{\alpha}_k+ \beta^*_k\hat{\alpha}^\dagger_k,~X_C=i(z_S|\Phi_C)\sqrt{n_0}A=\frac{\alpha_C+\alpha^*_C}{\sqrt{2}},~\beta_k=i\sqrt{L}(z_S|z_k)=\frac{c_k+d^*_k}{\sqrt{2}},\\
    \nonumber \hat{P}&=&-i(z_I|\hat{\Phi}(0))=P_C+\frac{1}{\sqrt{L}}\sum_k\delta_k\hat{\alpha}_k+ \delta^*_k\hat{\alpha}^\dagger_k,~P_C=-i(z_I|\Phi_C)\sqrt{n_0}A=-i\frac{\alpha_C-\alpha^*_C}{\sqrt{2}},~\delta_k=-i\sqrt{L}(z_I|z_k)=-i\frac{c_k-d^*_k}{\sqrt{2}}.
\end{eqnarray}
Regarding their Wigner distribution, by rewriting $\alpha$ in terms of the phase-space variables $X,P$,
\begin{equation}
    \alpha=\frac{X+iP}{\sqrt{2}},
\end{equation}
we get, after proper normalization and employing compact matrix notation, 
\begin{equation}
    W_0(X,P)=\frac{\sqrt{\det M}}{2\pi}e^{-\frac{1}{2}(\mathbf{X}-\mathbf{X}_C)^{\rm{T}}M(\mathbf{X}-\mathbf{X}_C)},~\mathbf{X}=\left[\begin{array}{cc} X \\ P\end{array}\right],~\mathbf{X}_C=\left[\begin{array}{cc} X_C \\ P_C\end{array}\right],~M=\frac{2}{\lambda^2-4|\Delta|^2}\left[\begin{array}{cc} \lambda-2\Delta_x & -2\Delta_y \\-2\Delta_y & \lambda+2\Delta_x\end{array}\right].
\end{equation}
From usual Gaussian theory, we have  $\braket{X}=X_C$ and $\braket{P}=P_C$, while the second-order momenta involving the uncertainties $\Delta X=X-\braket{X}$ and $\Delta P=P-\braket{P}$ are obtained from the correlation matrix
\begin{equation}
    C=\left[\begin{array}{cc} \braket{\Delta X^2} & \braket{\Delta X\Delta P} \\\braket{\Delta P\Delta X} & \braket{\Delta P^2}\end{array}\right]=M^{-1}=\left[\begin{array}{cc} \dfrac{\lambda}{2}+\Delta_x & \Delta_y \\\Delta_y & \dfrac{\lambda}{2}-\Delta_x\end{array}\right].
\end{equation}
Notice that $W_0(X,P)$ can be also derived directly from Eq. (\ref{eq:EspacioFaseCero}) after noticing Gaussianity, since any Gaussian distribution is determined by its first ($\braket{X}=\braket{\hat{X}}=X_C$,  $\braket{P}=\braket{\hat{P}}=P_C$) and second-order momenta,
% \begin{eqnarray}
%     \nonumber \braket{\Delta X^2}&=&\braket{\hat{X}^2}-\braket{\hat{X}}^2=\frac{1}{L}\sum_k|\beta_k|^2\braket{\hat{\alpha}_k\hat{\alpha}^\dagger_k+\hat{\alpha}^\dagger_k\hat{\alpha}_k}=\frac{1}{L}\sum_k|\beta_k|^2\coth\frac{\beta\Omega_k}{2}=\frac{1}{2\pi}\int\mathrm{d}k~|\beta_k|^2\coth\frac{\beta\Omega_k}{2}=\frac{\lambda}{2}+\Delta_x,
%     \\ \nonumber \braket{\Delta P^2}&=&\braket{\hat{P}^2}-\braket{\hat{P}}^2=\frac{1}{L}\sum_k|\delta_k|^2\braket{\hat{\alpha}_k\hat{\alpha}^\dagger_k+\hat{\alpha}^\dagger_k\hat{\alpha}_k}=\frac{1}{L}\sum_k|\delta_k|^2\coth\frac{\beta\Omega_k}{2}=\frac{1}{2\pi}\int\mathrm{d}k~|\delta_k|^2\coth\frac{\beta\Omega_k}{2}=\frac{\lambda}{2}-\Delta_x,\\
%     \nonumber \braket{\Delta X\Delta P}&=&\braket{\frac{\hat{X}\hat{P}+\hat{P}\hat{X}}{2}}-\braket{\hat{X}}\braket{\hat{P}}=\frac{1}{L}\sum_k\frac{\beta_k\delta^*_k+\beta^*_k\delta_k}{2}\braket{\hat{\alpha}_k\hat{\alpha}^\dagger_k+\hat{\alpha}^\dagger_k\hat{\alpha}_k}=\frac{1}{L}\sum_k\frac{\beta_k\delta^*_k+\beta^*_k\delta_k}{2}\coth\frac{\beta\Omega_k}{2}\\
%     &=&\frac{1}{2\pi}\int\mathrm{d}k~\frac{\beta_k\delta^*_k+\beta^*_k\delta_k}{2}\coth\frac{\beta\Omega_k}{2}=\Delta_y.
% \end{eqnarray}
\begin{eqnarray}
    \nonumber \braket{\Delta X^2}&=&\braket{\hat{X}^2}-\braket{\hat{X}}^2=\frac{1}{2\pi}\int\mathrm{d}k~|\beta_k|^2\coth\frac{\beta\Omega_k}{2}=\frac{\lambda}{2}+\Delta_x,
    \\ \braket{\Delta P^2}&=&\braket{\hat{P}^2}-\braket{\hat{P}}^2=\frac{1}{2\pi}\int\mathrm{d}k~|\delta_k|^2\coth\frac{\beta\Omega_k}{2}=\frac{\lambda}{2}-\Delta_x,\\
    \nonumber \braket{\Delta X\Delta P}&=&\frac{\braket{\hat{X}\hat{P}+\hat{P}\hat{X}}}{2}-\braket{\hat{X}}\braket{\hat{P}}=\frac{1}{2\pi}\int\mathrm{d}k~\frac{\beta_k\delta^*_k+\beta^*_k\delta_k}{2}\coth\frac{\beta\Omega_k}{2}=\Delta_y.
\end{eqnarray}
The marginal position, momentum distributions are also Gaussian,
\begin{equation}\label{eq:Marginal}
    W_0(X)=\frac{e^{-\frac{(X-X_C)^2}{2\Delta X^2}}}{\sqrt{2\pi\Delta X^2}},~W_0(P)=\frac{e^{-\frac{(P-P_C)^2}{2\Delta P^2}}}{\sqrt{2\pi\Delta P^2}},
\end{equation}
where, hereafter, for the sake of simplicity, we denote $\Delta X^2=\braket{\Delta X^2}$ and $\Delta P^2=\braket{\Delta P^2}$.

\begin{figure*}[!t]
\begin{tabular}{@{}cc@{}}   
\stackinset{l}{0pt}{t}{0pt}{(a)}{\includegraphics[width=0.5\textwidth]{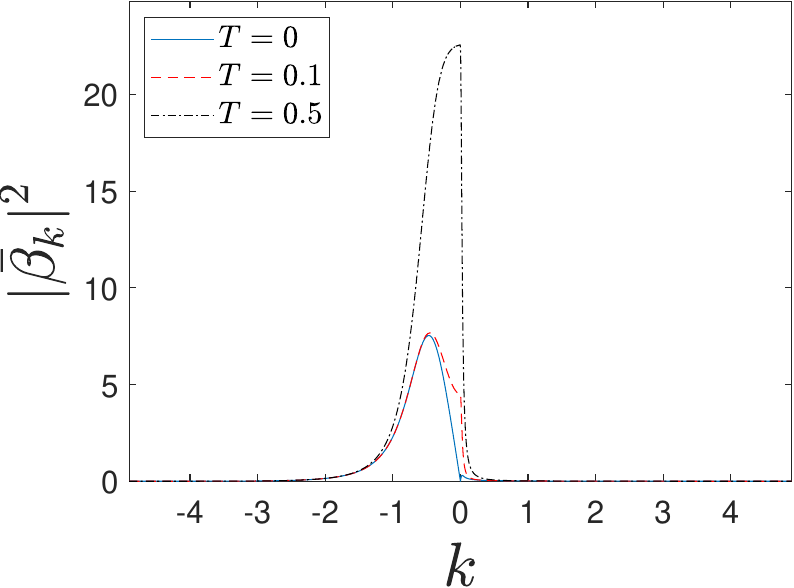}} 
& \stackinset{l}{0pt}{t}{0pt}{(b)}{\includegraphics[width=0.5\textwidth]{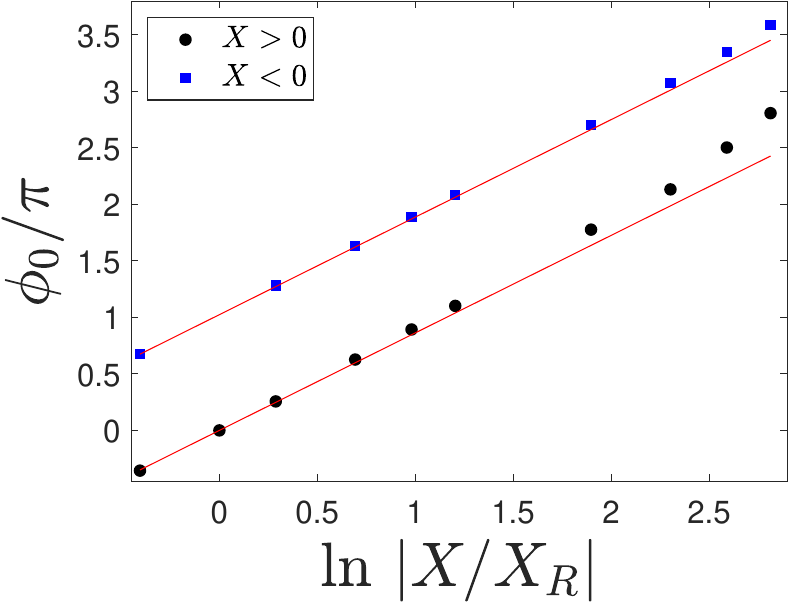}}\\
\stackinset{l}{0pt}{t}{0pt}{(c)}{\includegraphics[width=0.5\textwidth]{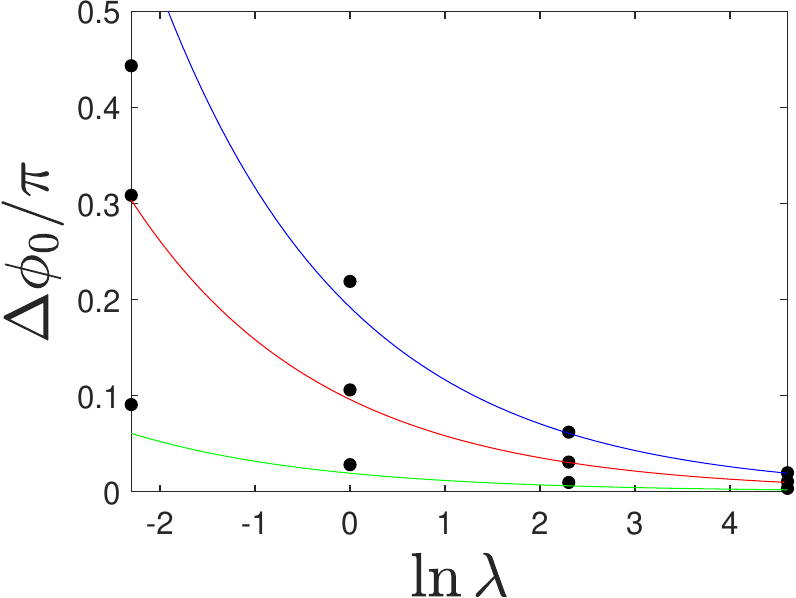}} 
& \stackinset{l}{0pt}{t}{0pt}{(d)}{\includegraphics[width=0.5\textwidth]{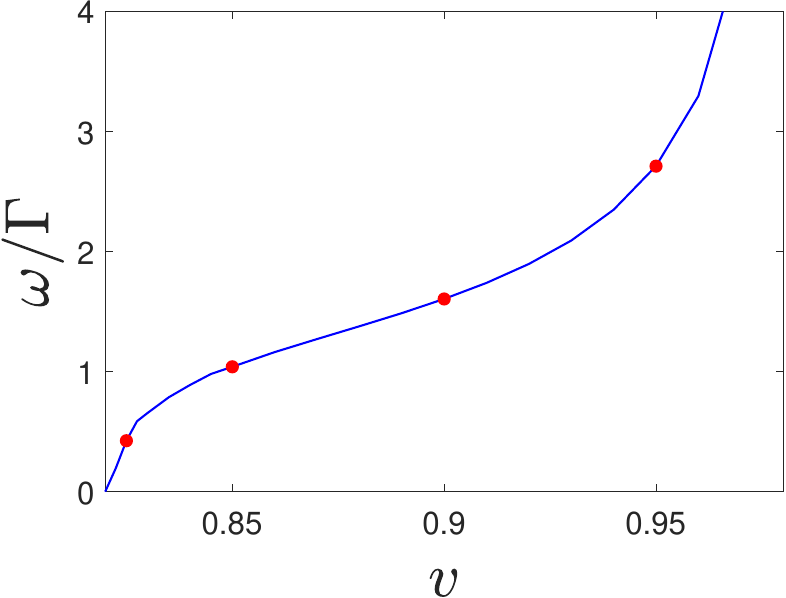}}
\end{tabular}
\caption{(a) The coefficient $|\bar{\beta}_k|^2\equiv|\beta_k|^2\coth\frac{\beta\Omega_k}{2}$ as a function of $k$ for temperatures $T=0$ (solid blue), $T=0.1$ (dashed red) and $T=0.5$ (dashed-dotted black). The background FPBHL parameters are $v=0.95,c_2=0.4,L=2$. (b) Asymptotic phase-shift $\phi_0$ as a function of the initial lasing amplitude $X$ for both $X>0$ (black dots) and $X<0$ (blue squares). Red lines represent the theoretical prediction from Eq. (\ref{eq:TimeShiftZ2}), where the reference amplitude is $X_R\approx 1.60$, corresponding to an initial classical amplitude $A=0.003$. The background FPBHL parameters are the same as in (a). (c) Phase-shift width $\Delta \phi_0$ as a function of the dimensionless parameter $\lambda$ for $A=0.005,0.01,0.05$ (black dots). Solid blue, red, green lines show the corresponding theoretical prediction from Eq. (\ref{eq:GaussianPhaseShift}). The background FPBHL parameters are $v=0.90,c_2=0.4,L=2$. (d) $\omega/\Gamma$ as a function of the initial flow velocity $v$ in the range $0.82\approx v_c<v<v_{1/2}\approx 0.98$ for fixed $c_2=0.4,L=2$.}
\label{fig:QuantumLasing}
\end{figure*}

Focusing on the lasing amplitude $X$, we separate pure quantum vacuum fluctuations from thermal fluctuations as
\begin{eqnarray}\label{eq:LasingFluctuations}
   \Delta X^2&=&\frac{1}{2\pi}\int\mathrm{d}k~|\beta_k|^2\coth\frac{\beta\Omega_k}{2}=\frac{1}{2\pi}\int\mathrm{d}k~|\beta_k|^2+\frac{1}{2\pi}\int\mathrm{d}k~|\beta_k|^2\frac{2}{e^{\beta\Omega_k}-1}\equiv\Delta X^2_Q+\Delta X^2_T.
\end{eqnarray}
In the main text, we work at zero temperature, so only quantum fluctuations contribute, $\Delta X^2=\Delta X^2_Q$. We represent $|\bar{\beta}_k|^2\equiv|\beta_k|^2\coth\frac{\beta\Omega_k}{2}$ in Fig. \ref{fig:QuantumLasing}a for different temperatures $T=0,0.1,0.5$. As it is clearly seen, $|\bar{\beta}_k|^2\simeq 0$ for wavevectors above our cutoff $k_C=5$.

The TW histograms for the distribution of $X$ shown in the main text are numerically computed using the initial TW condition, Eqs. (\ref{eq:TWExpansionNC}), (\ref{eq:TWBdG}):
\begin{equation}
   X=i(z_S|\Phi_W)=X_C+\frac{1}{\sqrt{L}}\sum_k\alpha_k\beta_k+ \alpha^*_k\beta^*_k,~\Phi_W=\left[\begin{array}{c}\delta\Psi_W\\ \delta\Psi^*_W\end{array}\right].
\end{equation}

\subsubsection{Early times: Time operator in a parametric amplifier}
Regarding dynamics, as explained in Eq. (\ref{eq:DPA}) of the main text, at early times when the
BdG approximation is still valid, the time evolution of the lasing mode is governed by the grand-canonical Hamiltonian 
\begin{equation}    
\hat{K}\simeq \hat{K}_{\rm{DPA}}\equiv i\Gamma\frac{(\hat{a}^\dagger)^2-\hat{a}^2}{2}=\Gamma\frac{\hat{X}\hat{P}+\hat{P}\hat{X}}{2},
\end{equation}
giving rise to the time-evolution operator $U(t)=e^{-i\hat{K} t}=S(-\Gamma t)$, with $ S(\varepsilon)=e^{\frac{\varepsilon^* \hat{a}^2-\varepsilon (\hat{a}^\dagger)^2}{2}}$ the usual squeezing operator. Thus, remarkably, a degenerate lasing mode represents a textbook realization of a degenerate parametric amplifier (PA).

In the Schr\"odinger picture, the eigenstates forming the continuous spectrum of the phase-space operators, $\hat{X}\ket{X}=X\ket{X}$ and $\hat{P}\ket{P}=P\ket{P}$, are squeezed as 
\begin{equation}
    U(t)\ket{X}=e^{\frac{\Gamma t}{2}}\ket{Xe^{\Gamma t}},~U(t)\ket{P}=e^{-\frac{\Gamma t}{2}}\ket{Pe^{-\Gamma t}},
\end{equation}
while the quantum state of the system evolves as $\hat{\rho}(t)=U(t)\hat{\rho}U^\dagger(t)$. In the Heisenberg picture, operators evolve as $O(t)=U^\dagger(t)OU(t)$, resulting in  
\begin{equation}
    \hat{a}(t)=\hat{a}\cosh\Gamma t+\hat{a}^\dagger\sinh\Gamma t,~\hat{X}(t)=\hat{X}e^{\Gamma t},~\hat{P}(t)=\hat{P}e^{-\Gamma t}.
\end{equation}
It is easy to show that this time evolution is translated into a time-dependent Wigner function for $\hat{\rho}(t)$,
\begin{equation}
W(\alpha,t)=W_0(\alpha(t))\equiv W_0(\alpha\cosh\Gamma t-\alpha^*\sinh\Gamma t).
\end{equation}
In terms of phase-space variables, this reads as
\begin{equation}
    W(X,P,t)=W_0(Xe^{-\Gamma t},Pe^{\Gamma t}).
\end{equation}
As a result, the marginal distribution of the lasing amplitude $W(X,t)=e^{-\Gamma t}W_0(Xe^{-\Gamma t})$ is also Gaussian, with exponentially increasing mean $X_C(t)=X_Ce^{\Gamma t}$ and width $\Delta X(t)=\Delta X e^{\Gamma t}$.

We note that $\hat{K}$ has not a unique associated vacuum, as any squeezed vacuum of the family $\ket{r}\equiv S(-r)\ket{0},~r\in\mathbb{R},$ is a valid choice, since $\hat{K}$ is invariant under these Bogoliubov transformations, $S(-r)\hat{K}S(r)=\hat{K}$. Specifically, 
\begin{equation}
   S(-r)\hat{K}S(r)=i\Gamma\frac{[\hat{a}^\dagger(r)]^2-\hat{a}^2(r)}{2}=i\Gamma\frac{(\hat{a}^\dagger)^2-\hat{a}^2}{2}=\hat{K},~\hat{a}(r)=S(-r)\hat{a}S(r)=\hat{a}\cosh r-\hat{a}^\dagger\sinh r, 
\end{equation}
so $\ket{r}$ is then the vacuum of $\hat{a}(r)$, $\hat{a}(r)\ket{r}=0$. Actually, $\hat{K}$ has a continuous spectrum of non-normalizable eigenstates, and $\ket{r=\Gamma t}$ is just the result of evolving $\ket{0}$ in time as $\ket{\Gamma t}=U(t)\ket{0}$. Consequently, the choice of a specific instantaneous vacuum of the family $\ket{r}$ at $t=0$ implies fixing a particular time origin $t_0$ as $r=\Gamma t_0$. This can be regarded as another form of spontaneous symmetry breaking, here induced by the quantization process \cite{Ribeiro2022}. Nonetheless, this vacuum ambiguity is broken in our work by the presence of a well-defined initial state at $t=0$. 

This conceptual connection permits us to define a time operator for a degenerate PA in terms of the position operator as
\begin{equation}
    \hat{t}\equiv-\frac{1}{\Gamma}\ln \left|\dfrac{\hat{X}}{X_R}\right|,
\end{equation}
with $X_R$ some reference amplitude. The operator $\hat{t}$ acts on any state $\ket{\Psi}$ in position representation, $\Psi(X)=\braket{X|\Psi}$, as
\begin{equation}
    \braket{X|\hat{t}|\Psi}=-\frac{1}{\Gamma}\ln \left|\dfrac{X}{X_R}\right|\braket{X|\Psi}=-\frac{1}{\Gamma}\ln \left|\dfrac{X}{X_R}\right|\Psi(X).
\end{equation}
By using this representation, and the fact that the momentum behaves as the usual differential operator $\braket{X|\hat{P}|\Psi}=-i\partial_X\Psi(X)$, it is immediate to show that $\hat{t}$ satisfies the expected commutation relation for a time operator (notice that $\hat{K}$ here plays the role of the Hamiltonian, generating the time evolution): 
\begin{equation}
    [\hat{t},\hat{K}]=-i.
\end{equation}
The spectrum of $\hat{t}$ is continuous and degenerate, $\hat{t}\ket{t}_\pm=t\ket{t}_\pm$, $t\in\mathbb{R}$, presenting two branches labeled as $\pm$,
\begin{equation}
    \ket{t}_\pm\equiv \sqrt{\left|\frac{dX_\pm}{dt}\right|}\ket{X_{\pm}(t)},~\hat{X}\ket{X_{\pm}(t)}=X_{\pm}(t)\ket{X_{\pm}(t)},~X_{\pm}(t)=\pm|X_R| e^{-\Gamma t}.
\end{equation}
The $\pm$ branches result from considering the time involution (i.e., the time-reversed evolution) of positive/negative position eigenstates, respectively, which are not connected between them through the amplification dynamics. Therefore, physically, the time operator measures the time needed for a given lasing amplitude $X$ to reach the reference values $\pm |X_R|$, which then fix a time origin. Different choices of $X_R$ amount to choose different time origins.

These eigenstates form a complete 
\begin{equation}
    \sum_{a=\pm}\int^\infty_{-\infty}\mathrm{d}t~\ket{t}_{a}\leftindex_{a}{\bra{t}}=\int^\infty_{0}\mathrm{d}X~\ket{X}\bra{X}+\int^0_{-\infty}\mathrm{d}X~\ket{X}\bra{X}=\int^\infty_{-\infty}\mathrm{d}X~\ket{X}\bra{X}=1
\end{equation}
and orthonormal 
\begin{equation}
\leftindex_{\pm}{\braket{t|t'}}_{\pm}=\delta(t-t'),~\leftindex_{\pm}{\braket{t|t'}}_{\mp}=0
\end{equation}
basis of the Hilbert space. Therefore, any quantum state can be expanded in temporal representation as
\begin{equation}
    \ket{\Psi}=\int^\infty_{-\infty}\mathrm{d}t~[\Psi_+(t)\ket{t}_+ + \Psi_-(t)\ket{t}_-]=\sum_{a=\pm}\int^\infty_{-\infty}\mathrm{d}t~\Psi_a(t)\ket{t}_a,~\Psi_a(t)\equiv \leftindex_a {\braket{t|\Psi}}.
\end{equation}

Using their explicit expression, it is immediate to show that these eigenstates can be generated through time translations as
\begin{equation}
    \ket{t}_\pm=\sqrt{\Gamma |X_R|}e^{-\frac{\Gamma t}{2}}\ket{\pm|X_R|e^{-\Gamma t}}=U(-t)\sqrt{\Gamma |X_R|}\ket{\pm|X_R|}\equiv U(-t)\ket{0}_{\pm}=e^{i\hat{K}t}\ket{0}_{\pm}.
\end{equation}
Thus, one only needs to characterize the dynamics for the two time origins $\ket{t=0}_\pm$, corresponding to initial lasing amplitudes $\ket{X=\pm X_R}$, since any quantum state can be written as a superposition of time translations of those states:
\begin{equation}
    \ket{\Psi}=\sum_{a=\pm}\int^\infty_{-\infty}\mathrm{d}t~\Psi_a(t)\ket{t}_a=\sum_{a=\pm}\int^\infty_{-\infty}\mathrm{d}t~\Psi_a(t)U(-t)\ket{0}_a=\sum_{a=\pm}\int^\infty_{-\infty}\mathrm{d}t~\Psi_a(t)e^{i\hat{K}t}\ket{0}_a.
\end{equation}
As a result, the time-evolution operator simply shifts the temporal wavefunction,
\begin{equation}
    U(t)\ket{\Psi}=\sum_{a=\pm}\int^\infty_{-\infty}\mathrm{d}t'~\Psi_a(t')\ket{t'-t}_a=\sum_{a=\pm}\int^\infty_{-\infty}\mathrm{d}t'~\Psi_a(t'+t)\ket{t'}_a.
\end{equation}
In general, as one would expect, $\hat{K}$ acts as $i\partial_t$ in temporal representation:
\begin{eqnarray}
\hat{K}\ket{\Psi}&=&\sum_{a=\pm}\int^\infty_{-\infty}\mathrm{d}t~\Psi_a(t)\hat{K}\ket{t}_a=\sum_{a=\pm}\int^\infty_{-\infty}\mathrm{d}t~\Psi_a(t)(-i\partial_t)(e^{i\hat{K}t}\ket{0}_a)=\sum_{a=\pm}\int^\infty_{-\infty}\mathrm{d}t~i\partial_t \Psi_a(t)\ket{t}_a.
\end{eqnarray}
The above results apply to any density matrix $\hat{\rho}$, which can be accordingly expanded in four temporal branches as
\begin{equation}
    \hat{\rho}=\sum_{a,b=\pm}\int^\infty_{-\infty}\int^\infty_{-\infty}\mathrm{d}t\,\mathrm{d}t'~\rho_{ab}(t,t')\ket{t}_{a}\leftindex_{b}{\bra{t'}},~\rho_{ab}(t,t')\equiv\leftindex_{a}{\braket{t|\hat{\rho}|t'}}_b.
\end{equation}
The existence of such time operator $\hat{t}$ evades the celebrated no-go theorem by Pauli because the PA Hamiltonian is not bounded. It can be shown that a non-degenerate PA, which describes the dynamics of non-degenerate lasing modes, $\textrm{Re}\,\Omega\neq 0$ \cite{Finazzi2010,deNova2024}, also has an associated time operator; a detailed study of the properties of the time operator emerging in parametric amplifiers is left for future work.

\subsubsection{Late times: Phase-shift distribution}

At the BdG level, any initial lasing amplitude $X$ is exponentially amplified in time as $Xe^{\Gamma t}$, eventually dominating the dynamics. At some point, the lasing instability saturates when it becomes of the order of the condensate itself, $Xe^{\Gamma t}\sim \sqrt{n_0}\gg 1$. Once in this regime, the dynamics is described by the full GP equation. For sufficiently long times, the system eventually approaches the CES state, where we can define another time operator, $\hat{t}_0$, arising from the quantum amplitude of the temporal Floquet-Nambu-Goldstone mode and describing the fluctuations of the global time origin \cite{deNova2025}. Since the CES state is periodic, this time-shift can be expressed as an oscillation phase-shift, $\hat{\phi}_0=-\omega \hat{t}_0$. 

We now connect both time operators through the dynamics using a simple model which only involves the amplification of the lasing mode. The results of the previous section show that we only need to determine the evolution of the reference lasing amplitudes $\pm X_R$, since the evolution of any other initial amplitude $X$ is obtained by adding a time-shift 
\begin{equation}
    t_0=-\frac{1}{\Gamma}\ln\left|\frac{X}{X_R}\right|.
\end{equation}
We assume that the reference amplitudes asymptotically yield the same CES state but with a different oscillation phase-shift that only depends on $X_R$, $\phi_0(X_R)=\phi_R$ and $\delta\equiv \phi_0(-X_R)-\phi_0(X_R)$. Without loss of generality, we take $\phi_R=0$ as late-time origin. Hence, any initial lasing amplitude $X$ will asymptotically give rise to a CES state with a phase-shift $\phi_0(X)$ given by
\begin{eqnarray}\label{eq:TimeShiftZ2}
    \phi_0(X)&=&-\omega t_0=\frac{\omega}{\Gamma}\ln\left|\frac{X}{X_R}\right|,~X>0,\\
    \nonumber\phi_0(X)&=&\delta-\omega t_0=\delta+\frac{\omega}{\Gamma}\ln\left|\frac{X}{X_R}\right|,~X<0.
\end{eqnarray}
We note that the additional phase-shift $\delta$ reflects the $\mathbb{Z}_2$ symmetry-breaking of a BHL at the nonlinear level \cite{Michel2013,Michel2015}. For computational convenience, we are not yet restricting $\phi_0$ to its first Brillouin zone, $\phi_0\in[-\pi,\pi)$, since the time-shifts arising from the initial PA time operator are not bounded.

We check the above relation by simulating classical GP trajectories where the initial noise $\varphi_W$ is removed from Eq. (\ref{eq:TWExpansionNC}). This imprints a lasing amplitude $X\propto A$, which is translated into a late phase-shift $\phi_0$, from where the relation $\phi_0(X)$ can be extracted. The results are shown in Fig. \ref{fig:QuantumLasing}b, in good agreement with Eq. (\ref{eq:TimeShiftZ2}). The deviations for the largest values of $X$ can be attributed to the onset of nonlinear corrections.

In the main text, within each TW ensemble we take as reference the initial classical lasing amplitude, $X_R=X_C>0$, so the histograms are centered around $\phi_0=0$. In addition, we assume there that the fluctuations are moderate enough so that the values $X<0$ can be neglected and $\phi_0(X)$ is an injective relation. This is for instance the case of small lasing fluctuations, $\Delta X\ll X_C$, where we find
\begin{equation}
    \phi_0\simeq \frac{\omega}{\Gamma}\frac{X-X_C}{X_C}.
\end{equation}
As a result, $\phi_0$ is also Gaussian, with
\begin{equation}\label{eq:GaussianPhaseShift}
\braket{\phi_0}=0,~\Delta\phi_0=\sqrt{\braket{\phi^2_0}}=\frac{\omega}{\Gamma}\frac{\Delta X}{X_C}\ll 1.
\end{equation}
We check this relation by introducing a dimensionless parameter $\lambda$ that rescales the density as $n_0\to \lambda n_0$ while leaving unchanged the rest of relevant parameters. This leads to the same mean-field evolution and same width $\Delta X$, while $X_C$ scales as $X_C\to\lambda^{1/2} X_C$, which implies  $\Delta X/X_C\to \lambda^{-1/2}\Delta X/X_C$. The corresponding TW results for the phase-shift width $\Delta \phi_0$ are shown in Fig. \ref{fig:QuantumLasing}c as a function of $\lambda$ for different classical amplitudes $A=0.005,0.01,0.05$; the corresponding theoretical predictions (\ref{eq:GaussianPhaseShift}) are depicted as solid lines. As expected, the results converge  in the limit of large $\lambda$, i.e., for small $\Delta X/X_C$.

In general, the phase-shift distribution is computed from the initial lasing distribution as 
\begin{equation}\label{eq:PhaseShiftDistribution}
    W(\phi_0)=\int^{\infty}_{-\infty}\mathrm{d}X~\delta(\phi_0-\phi_0(X))W_0(X),
\end{equation}
since the relation $\phi_0(X)$ is not necessarily injective. Moreover, one also has to include the negative branch, $X<0$. 

This is the case of the strong fluctuating scenario $X_C\ll\Delta X$, which in particular includes the case of a purely quantum BHL, $X_C=0$. In this limit, we can take $X_C\simeq 0$ and work with the normalized variable 
\begin{equation}\label{eq:WignerTimeNormal}
    z\equiv\frac{X}{\sqrt{2}\Delta X},~W_0(z)=\frac{e^{-z^2}}{\sqrt{\pi}},
\end{equation}
while setting an arbitrary reference amplitude $X_R$ to replace the choice $X_R=X_C$, since $X_C=0$ does not lead to any mean-field dynamics. Equation (\ref{eq:TimeShiftZ2}) is then rewritten as
\begin{eqnarray}\label{eq:Phiz}
    \phi_0(z)&=&\phi_R+\frac{\omega}{\Gamma}\ln|z|,~z>0,\\
    \nonumber \phi_0(z)&=&\delta+\phi_R+\frac{\omega}{\Gamma}\ln|z|,~z<0,
\end{eqnarray}
with $\phi_R\equiv\dfrac{\omega}{\Gamma}\ln\dfrac{\sqrt{2}\Delta X}{|X_R|}$. The corresponding mean and variance are
\begin{eqnarray}
    \braket{\phi_0}&=&\phi_R+\frac{\delta}{2}+\frac{\omega}{\Gamma}\braket{\ln|z|}=\phi_R+\frac{\delta}{2}+\frac{\psi\left(\frac{1}{2}\right)}{2}\frac{\omega}{\Gamma}\approx \phi_R+\frac{\delta}{2}-0.982\frac{\omega}{\Gamma},\\
    \nonumber \Delta\phi^2_0&=&\braket{\phi^2_0}-\braket{\phi_0}^2=\left(\frac{\delta}{2}\right)^2+\frac{\omega^2}{\Gamma^2}(\braket{\ln^2|z|}-\braket{\ln|z|}^2)=\left(\frac{\delta}{2}\right)^2+\frac{\psi'\left(\frac{1}{2}\right)}{4}\frac{\omega^2}{\Gamma^2}\approx \left(\frac{\delta}{2}\right)^2+ 1.233\dfrac{\omega^2}{\Gamma^2},
\end{eqnarray}
where we have used $W_0(z)=W_0(-z)$ and 
\begin{equation}
    \braket{\ln^n |z|}=\int^\infty_{-\infty}\mathrm{d}z~W_0(z)\ln^n|z| =\frac{1}{2^n\sqrt{\pi}}\int^\infty_{0}\mathrm{d}t~\frac{\ln^n t}{\sqrt{t}} e^{-t}=\frac{\Gamma^{(n)}\left(\frac{1}{2}\right)}{2^n\Gamma\left(\frac{1}{2}\right)},
\end{equation}
$\psi(z)$ being the digamma function, i.e., the logarithmic derivative of the Euler gamma function $\Gamma(z)$,
\begin{equation}
    \psi(z)\equiv\frac{d\ln\Gamma(z)}{dz}=\frac{\Gamma'(z)}{\Gamma(z)}.
\end{equation}
We stress that $W(\phi_0)$ only depends on the global phase-shift $\phi_R$, the $\mathbb{Z}_2$-symmetry breaking phase-shift $\delta$, and the dimensionless parameter $\omega/\Gamma$, the latter measuring the ratio between the CES frequency and the lasing growth rate. Notice that $\delta$ only biases the distribution in such a way that, if $\omega/\Gamma=0$, half of the events present $\phi_0=\phi_R$ and half of the events present $\phi_0=\delta/2+\phi_R$. On the other hand, the reference amplitude $X_R$ only introduces a trivial phase-shift $\phi_R$ but does not affect the width $\Delta\phi_0$. Remarkably, $\Delta\phi_0$ does not even depend on $\Delta X$ in the limit $X_C\ll \Delta X$. 

Therefore, $\omega/\Gamma$ is the critical parameter controlling the uniformity of the late phase-shift distribution. In Figure \ref{fig:QuantumLasing}d, we represent $\omega/\Gamma$ as a function of the initial flow velocity $v$ in the range $v_c<v<v_{1/2}$ for fixed $c_2=0.4,L=2$. It increases from the critical velocity $v_c\approx 0.82$, where $\omega=0$, eventually diverging at $v_{1/2}\approx 0.98$, where $\Gamma=0$. 

\begin{figure*}[!t]
\begin{tabular}{@{}cccc@{}}   
\stackinset{l}{0pt}{t}{0pt}{(a)}{\includegraphics[width=0.25\textwidth]{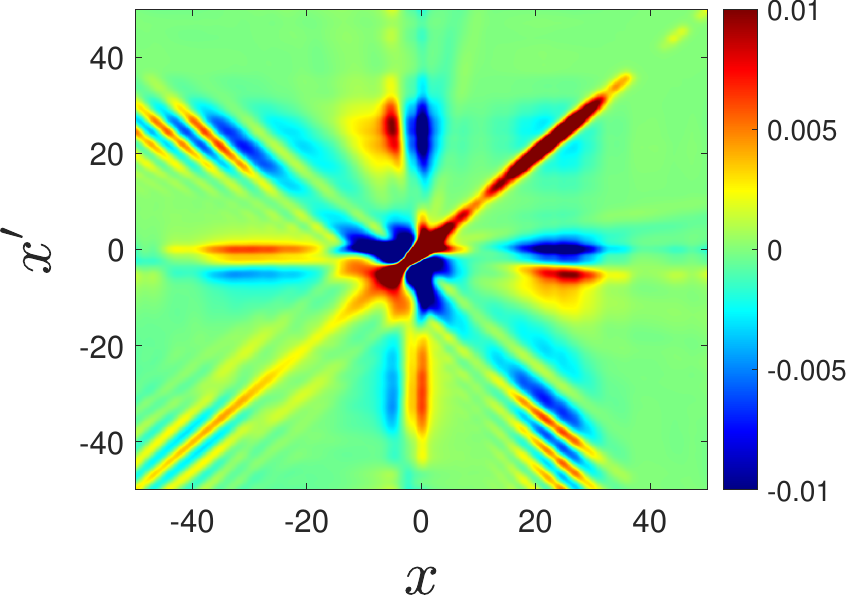}} 
& \stackinset{l}{0pt}{t}{0pt}{(b)}{\includegraphics[width=0.25\textwidth]{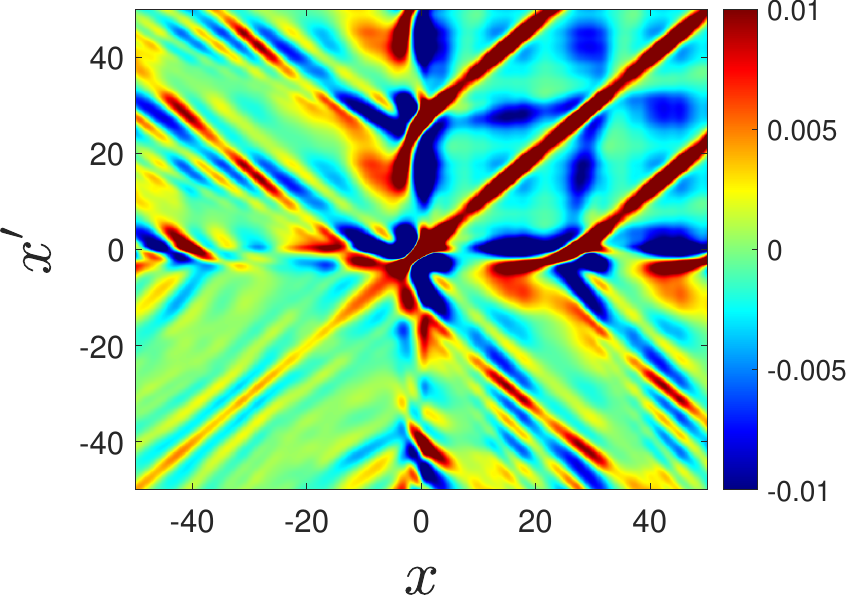}}& 
\stackinset{l}{0pt}{t}{0pt}{(c)}{\includegraphics[width=0.25\textwidth]{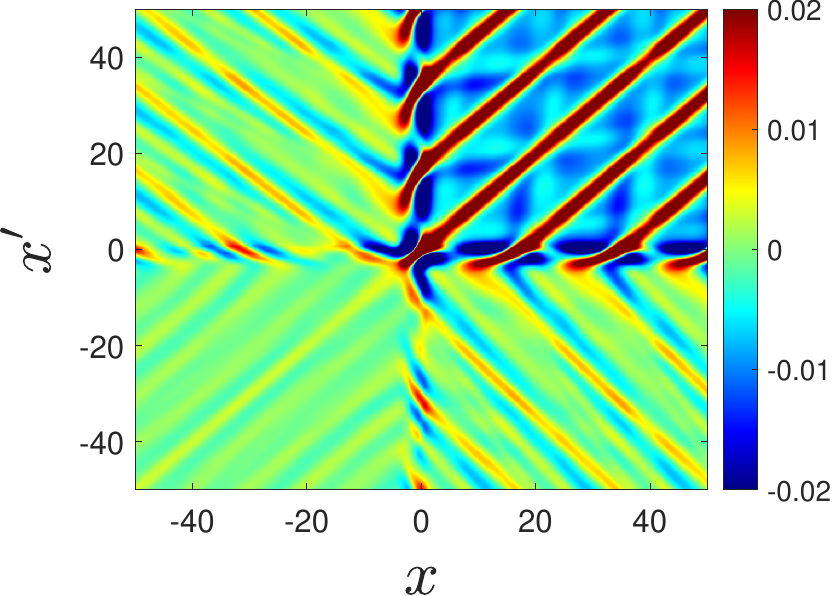}}& 
\stackinset{l}{0pt}{t}{0pt}{(d)}{\includegraphics[width=0.25\textwidth]{FigureSM/G295}}  
\\
\stackinset{l}{0pt}{t}{0pt}{(e)}{\includegraphics[width=0.25\textwidth]{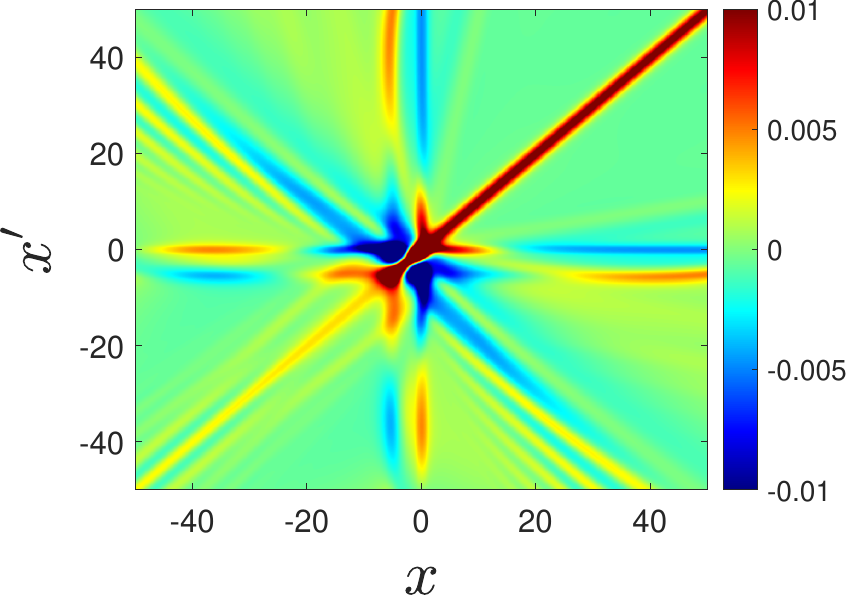}} 
& \stackinset{l}{0pt}{t}{0pt}{(f)}{\includegraphics[width=0.25\textwidth]{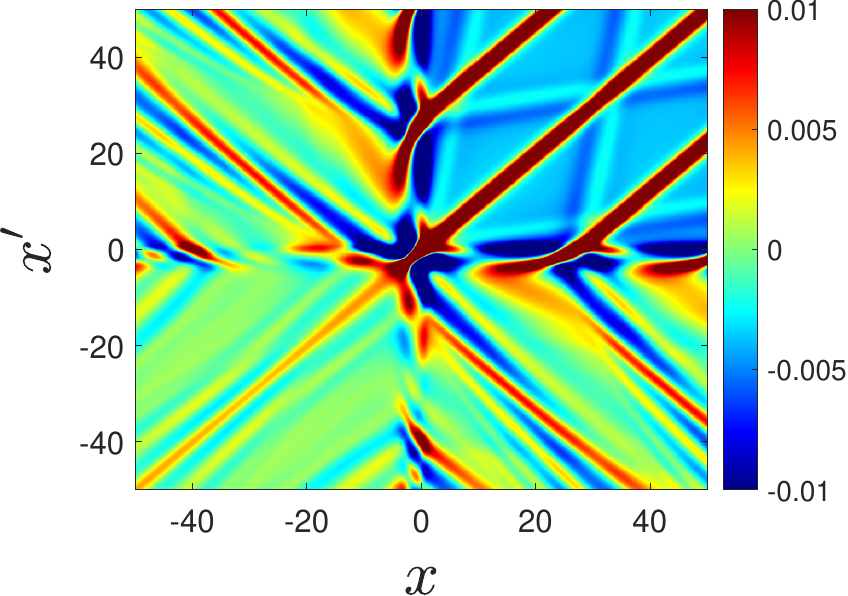}}& 
\stackinset{l}{0pt}{t}{0pt}{(g)}{\includegraphics[width=0.25\textwidth]{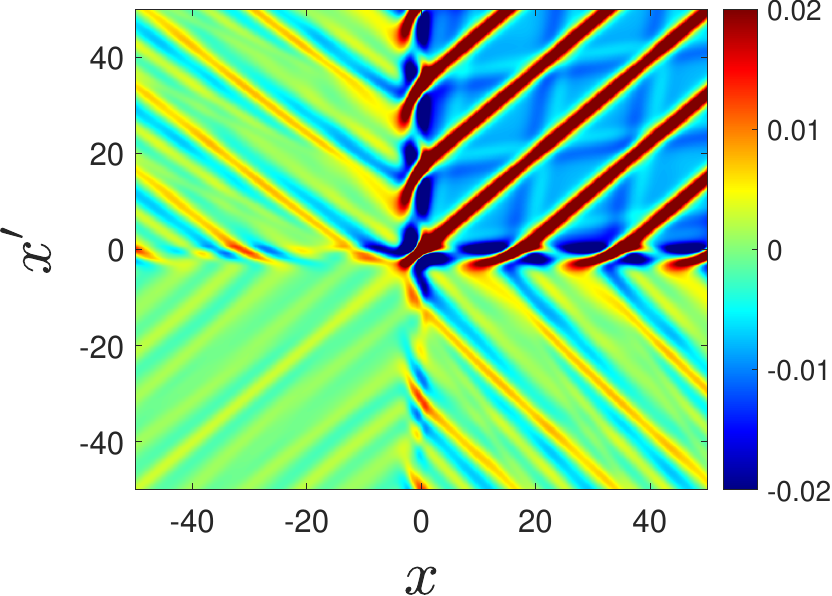}}& 
\stackinset{l}{0pt}{t}{0pt}{(h)}{\includegraphics[width=0.25\textwidth]{FigureSM/G295Teo}} \\
\stackinset{l}{0pt}{t}{0pt}{(i)}{\includegraphics[width=0.25\textwidth]{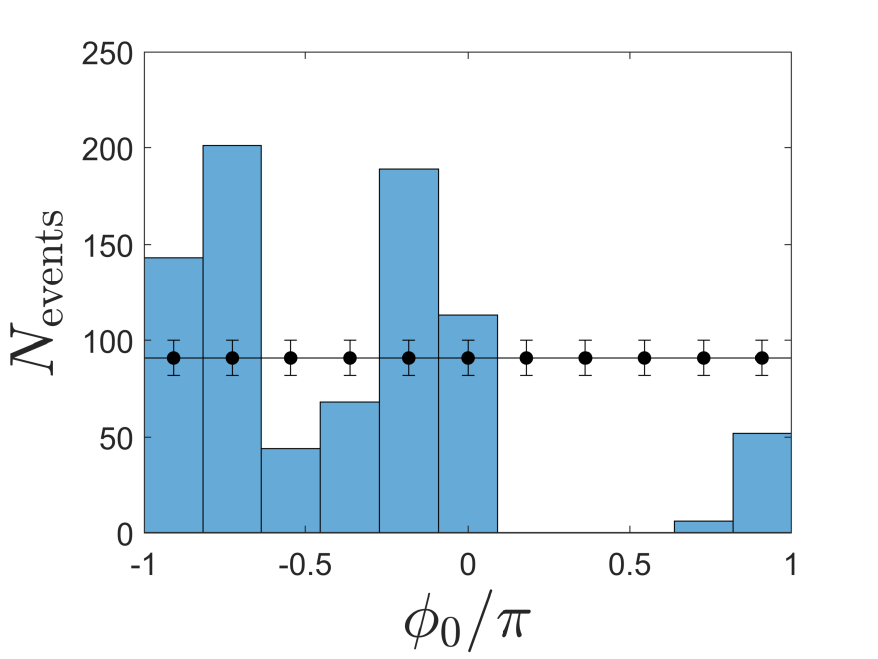}} 
& \stackinset{l}{0pt}{t}{0pt}{(j)}{\includegraphics[width=0.25\textwidth]{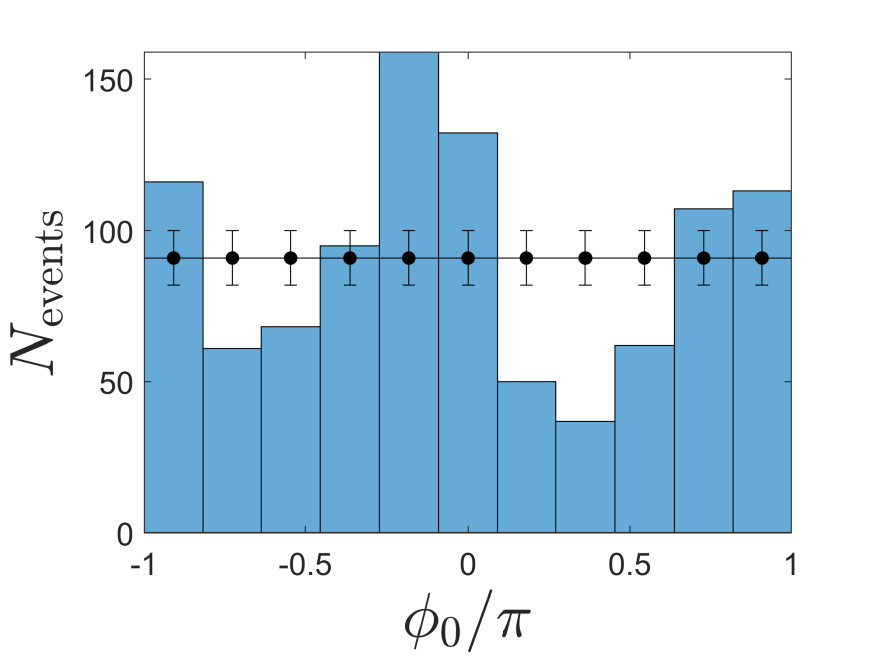}}& 
\stackinset{l}{0pt}{t}{0pt}{(k)}{\includegraphics[width=0.25\textwidth]{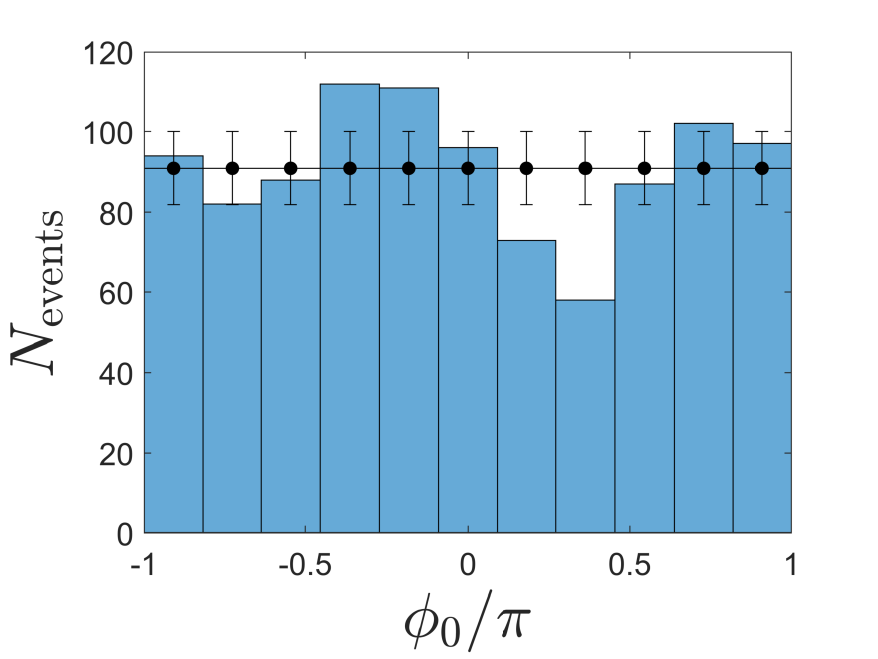}}
& 
\stackinset{l}{0pt}{t}{0pt}{(l)}{\includegraphics[width=0.25\textwidth]{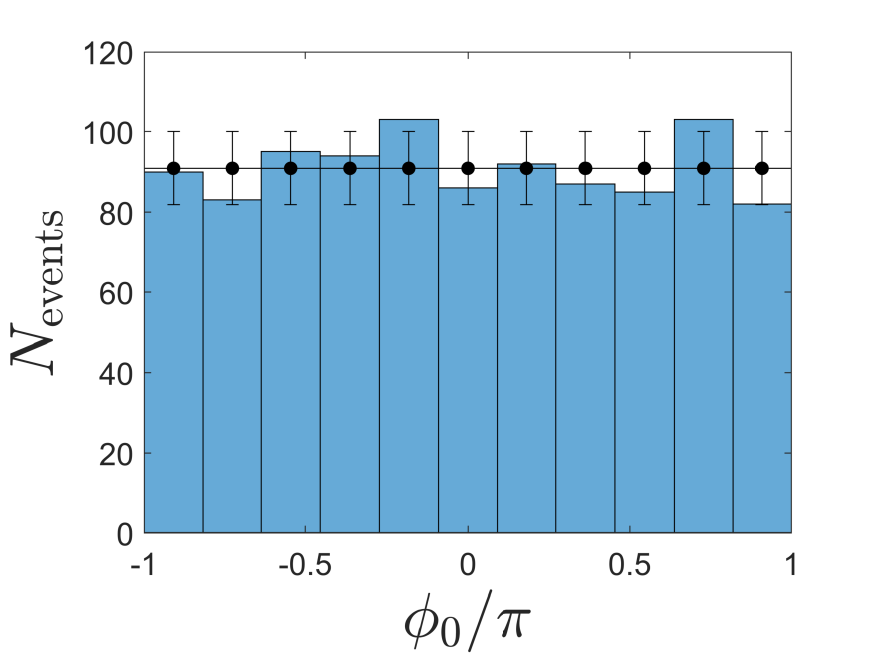}}
\\ 
\stackinset{l}{-3pt}{t}{0pt}{(m)}{\includegraphics[width=0.25\textwidth]{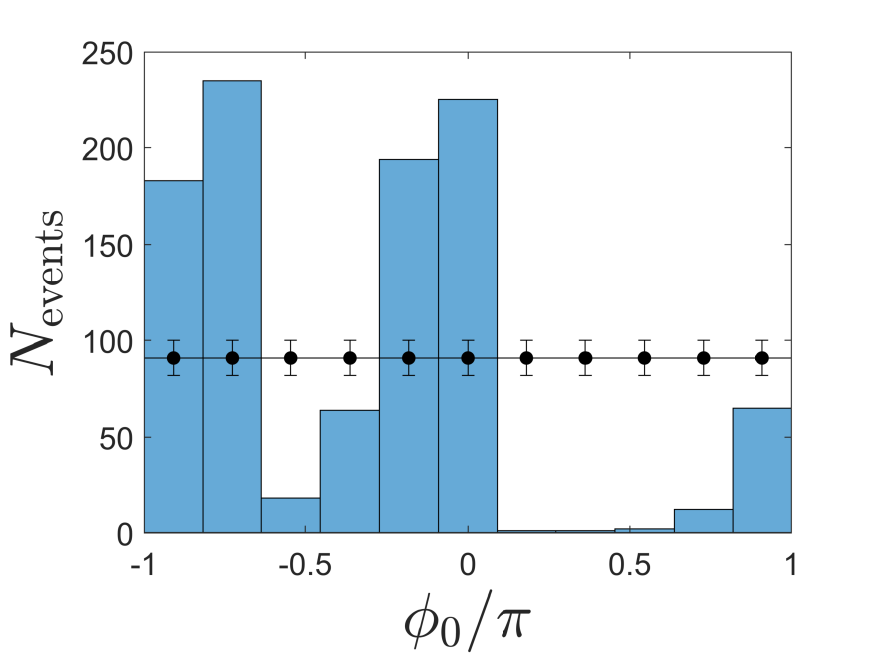}} 
& \stackinset{l}{0pt}{t}{0pt}{(n)}{\includegraphics[width=0.25\textwidth]{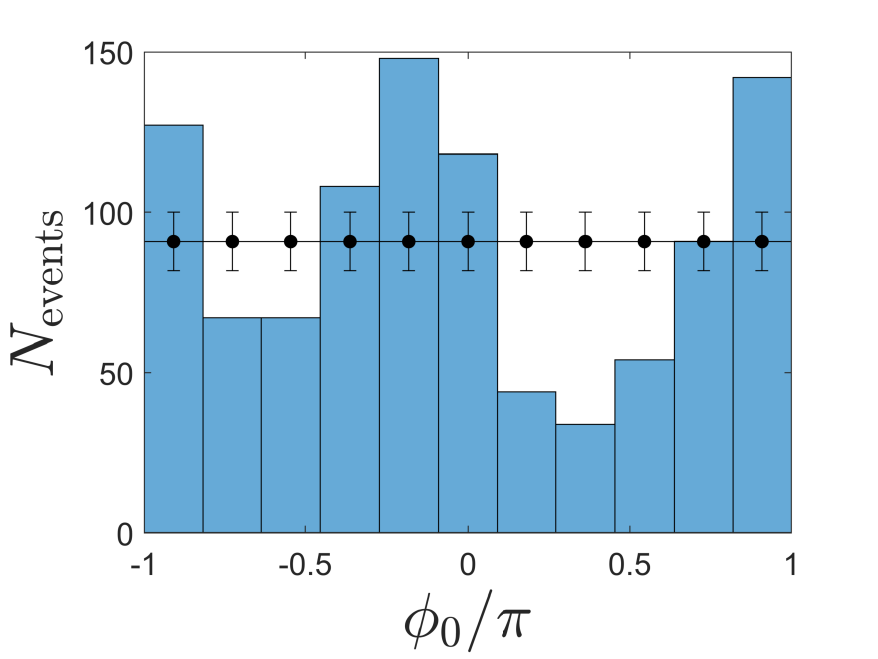}}& 
\stackinset{l}{0pt}{t}{0pt}{(o)}{\includegraphics[width=0.25\textwidth]{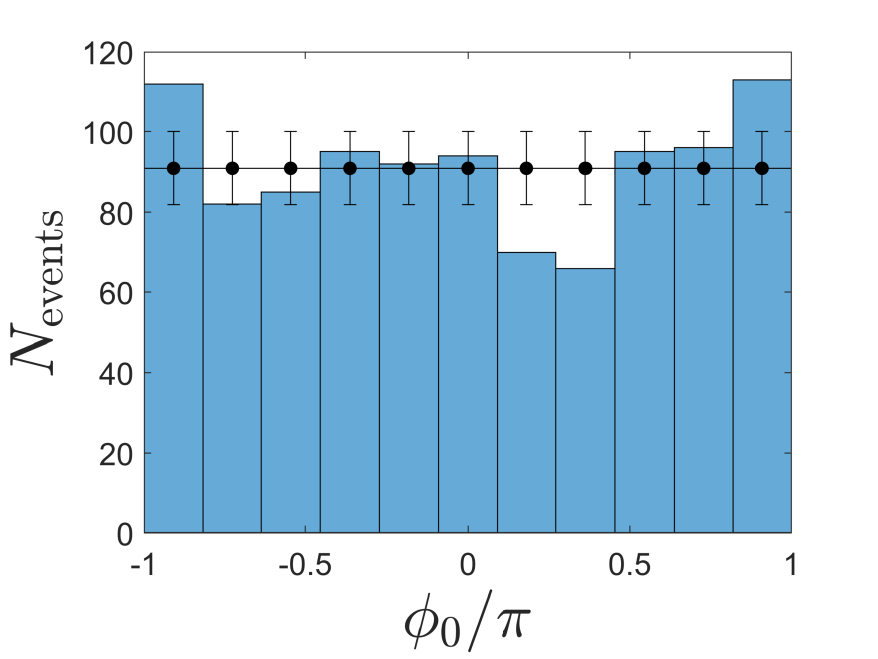}}
& 
\stackinset{l}{0pt}{t}{0pt}{(p)}{\includegraphics[width=0.25\textwidth]{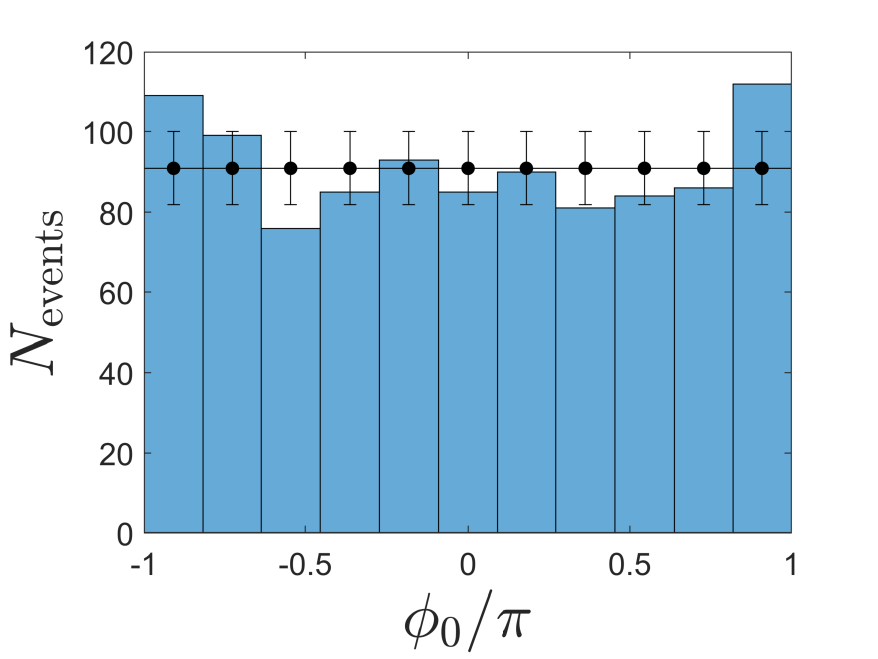}}
\end{tabular}
\caption{(a)-(d) TW computation of the ECTF $g^{(2)}(x,x',t)$ at $t=870$ for a purely quantum FPBHL with fixed $c_2=0.4,L=2$ and $v=0.825,0.850,0.900,0.950$, respectively. (e)-(h) Ideal HTC prediction (\ref{eq:HteoSM}) for (a)-(d). (i)-(l) TW histogram of $\phi_0$ for (a)-(d). Horizontal line with error bars: uniform distribution and its statistical uncertainty. (m)-(p) Histograms resulting from sampling $1000$ random events using Eqs. (\ref{eq:WignerTimeNormal}), (\ref{eq:Phiz}).}
\label{fig:CriticalTradeoff}
\end{figure*}

The crucial role played by $\omega/\Gamma$ is observed in Fig. \ref{fig:CriticalTradeoff}, where we consider a purely quantum BHL with increasing $v=0.825,0.850,0.900,0.950$ (marked as red dots in Fig. \ref{fig:QuantumLasing}d). Specifically, in the first row, we show the TW results for the late-time ETCF. In the second row, we show the theoretical HTC prediction (\ref{eq:HteoSM}), resulting from uniformly averaging over all possible phase-shifts. As we can see, the agreement worsens close to the phase transition, where $\omega/\Gamma$ vanishes, indicating the absence of uniformity in the phase-shift distribution. This conclusion is further supported by the corresponding TW histograms of $\phi_0$ in third row of Fig. \ref{fig:CriticalTradeoff}, where we observe unevenly distributed blanks. The simple model from Eqs. (\ref{eq:WignerTimeNormal}), (\ref{eq:Phiz}) is able to reproduce this behavior, last row of Fig. \ref{fig:CriticalTradeoff}, where we show the histograms resulting from generating $1000$ random Gaussian events for the dimensionless variable $z$ and then evaluating $\phi_0(z)$; hereafter, we confine $\phi_0$ to its first Brillouin zone $\phi_0\in[-\pi,\pi)$ since we deal with density-based observables, which are periodic in $\phi_0$.

Hence, in order to achieve a genuine spontaneous symmetry breaking realizing all phases $\phi_0\in[-\pi,\pi)$, it is required $\omega/\Gamma\gtrsim 1$, even in the purely quantum limit. On the other hand, $\omega/\Gamma$ cannot be too large since $\Gamma$ close to zero implies large lasing times for the initial instability to be amplified up to the saturation regime. Thus, there is a tradeoff between lasing and spontaneous symmetry breaking. However, in practice, this tradeoff does not require a delicate fine-tuning. For instance, $v=0.85$ (second column in Fig. \ref{fig:CriticalTradeoff}) already displays a broad phase-shift distribution, and $v=0.9$, $v=0.95$ (third, fourth columns in Fig. \ref{fig:CriticalTradeoff}) are close to an ideal uniform distribution (horizontal line with error bars). In particular, the rightmost column corresponds to the simulation analyzed in the main text.

\subsection{HTC correlations}\label{SM:HTCorr}

\begin{figure}[!t]
\includegraphics[width=\textwidth]{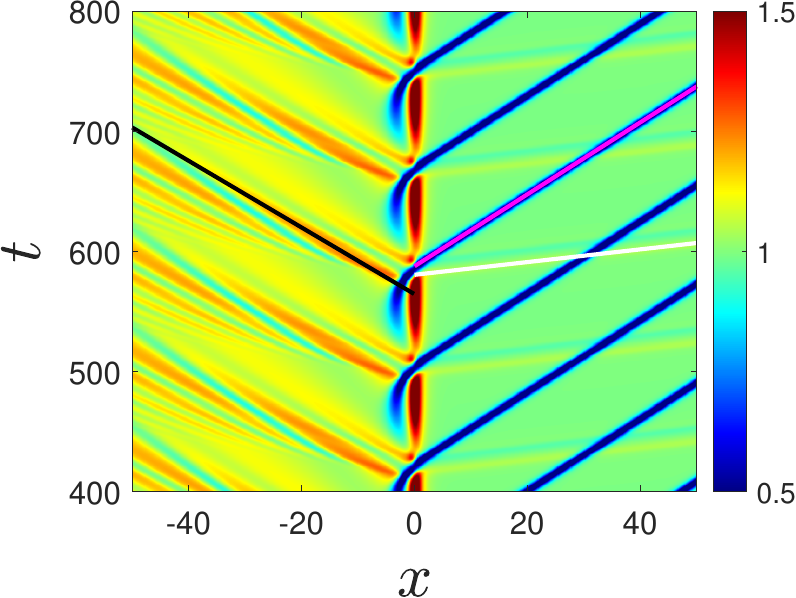}
\caption{CES density $n_0(x,t)$ for a mean-field simulation with $v=0.85,c_2=0.4,L=2$. Solid lines are ballistic fits of the trajectory of the upstream wave (black), soliton (magenta), and downstream wave (white).}
\label{fig:Track}
\end{figure}

At late times, expectation values for density-based observables can be computed using the distribution $W(\phi_0)$ from Eq. (\ref{eq:PhaseShiftDistribution}),
\begin{equation}
    \braket{O}\simeq\int^\pi_{-\pi}\mathrm{d} \phi_0~W(\phi_0)O[\phi_0],
\end{equation}
where $O[\phi_0]$ results from evaluating the observable $O$ after the substitution $\hat{\Psi}(x,t)= \tilde{\Psi}_0(x,\phi_0+\omega t)$. For the density and second-order correlation function, this implies:
\begin{eqnarray}\label{eq:HTCWignerSM}
n(x,t)&=&\int^\pi_{-\pi}\mathrm{d}\phi_0~W(\phi_0)\tilde{n}_0(x,\phi_0+\omega t),\\
\nonumber g^{(2)}(x,x',t,t')&=&\int^\pi_{-\pi}\mathrm{d}\phi_0~W(\phi_0)\tilde{n}_0(x,\phi_0+\omega t)\tilde{n}_0(x',\phi_0+\omega t')-n(x,t)n(x',t'),
\end{eqnarray}
with $\tilde{n}_0(x,\phi_0+\omega t)\equiv |\tilde{\Psi}_0(x,\phi_0+\omega t)|^2$. For an ideal HTC, $W(\phi_0)$ can be approximated by a uniform distribution, 
\begin{equation}
    W(\phi_0)\simeq \dfrac{1}{2\pi}.
\end{equation}
By combining this result with the time-periodicity of $n_0(x,t_0+t)\equiv \tilde{n}_0(x,\phi_0+\omega t)$ we recover the theoretical HTC predictions of the main text:
\begin{eqnarray}\label{eq:HteoSM}
n_{\rm{HTC}}(x)&=&\frac{1}{T}\int^{T}_{0}\mathrm{d}t_0~n_0(x,t_0)=n^{(0)}_0(x),\\
\nonumber g_{\rm{HTC}}^{(2)}(x,x',\tau)&=&\frac{1}{T}\int^{T}_{0}\mathrm{d}t_0~n_0(x,t_0)n_0(x',t_0+\tau)-n_{\rm{HTC}}(x)n_{\rm{HTC}}(x')=\sideset{}{'}\sum^\infty_{m=-\infty}n^{(-m)}_0(x)n^{(m)}_0(x')e^{-im\omega \tau},
\end{eqnarray}
where we have used the temporal Fourier expansion of the CES density,
\begin{equation}
    n_{0}(x,t)=\sum^\infty_{m=-\infty}n^{(m)}_0(x)e^{-im\omega t},
\end{equation}
and the $'$ denotes that the sum excludes the term $m=0$. Hence, remarkably, the OTCF of an ideal HTC is purely oscillatory as it does not have zero Fourier component.

The spatial correlation patterns predicted in the main text are obtained by assuming that the CES density $n_0(x,t)$ is composed of traveling features in the upstream ($x<0$) and downstream ($x>0$) regions with trajectories $x_{i}(t)=x^{(0)}_{i}+v_{i}t$, where $i=u,w,d$ label the upstream wave, the downstream wave, and the downstream soliton, respectively. Specifically, $v_u<0$ and $v_{w}>v_d>0$, whose values are fitted from the mean-field CES wavefunction. Figure \ref{fig:Track} shows a detailed fit of each traveling feature, where we also see that the emitted waves have a positive amplitude over the background density, while the solitons carry a strong density depletion.

Consequently, we can approximate the density in the upstream/downstream region as 
\begin{eqnarray}\label{eq:Cynthia}
    n_0(x,t)&\simeq & n_u\left(t-\frac{x}{v_u}\right),~x<0,\\
    \nonumber n_0(x,t)&\simeq & n_d\left(t-\frac{x}{v_d}\right)+n_w\left(t-\frac{x}{v_w}\right),~x>0,
\end{eqnarray}
where $n_{i}(t+T)=n_{i}(t)$ are periodic functions localized around $t=-x^{(0)}_i/v_i+nT,~n\in\mathbb{Z}$, and $n_w(t)\ll n_d(t)$ as the downstream region is dominated by the soliton emission. Their Fourier expansion reads 
\begin{equation}
    n_{i}(t)=\sum^\infty_{m=-\infty}n^{(m)}_ie^{-im\omega t}.
\end{equation}
Therefore, we expect homogeneous and time-independent values for the average density in the asymptotic regions:
\begin{eqnarray}
n_{\rm{HTC}}(x)&\simeq&\frac{1}{T}\int^{T}_{0}\mathrm{d}t_0~n_u(t_0)= n^{(0)}_u,~x<0,\\
\nonumber n_{\rm{HTC}}(x)&\simeq &\frac{1}{T}\int^{T}_{0}\mathrm{d}t_0~[n_d(t_0)+n_w(t_0)]= n^{(0)}_d+n^{(0)}_w,~x>0,
\end{eqnarray}
as observed in Fig. \ref{fig:ClassicalQuantum}d of main text. Regarding the second-order correlation function, we define
\begin{eqnarray}\label{eq:CorrelationBands}
    \nonumber g_{ij}(x,x',\tau)&\equiv& g_{ij}(\tau_{ij})\equiv \frac{1}{T}\int^{T}_{0}\mathrm{d}t_0~n_i\left(t_0\right)n_j\left(t_0+\tau_{ij}\right)-n^{(0)}_in^{(0)}_j=\sideset{}{'}\sum^\infty_{m=-\infty}n^{(-m)}_in^{(m)}_je^{-im\omega\tau_{ij}},\\
    \tau_{ij}&\equiv&\tau+\frac{x}{v_i}-\frac{x'}{v_j}.
\end{eqnarray}
It is immediate to see that $g_{ij}(\tau_{ij})$ is localized around  $\tau_{ij}=\dfrac{x^{(0)}_i}{v_i}-\dfrac{x^{(0)}_j}{v_j}+nT,~n\in\mathbb{Z}$, which implies that $g_{ij}(x,x',\tau)$ is localized in the $(x,x')$ plane around the equispaced, parallel bands
\begin{equation}
    \dfrac{x'}{v_j}-\dfrac{x}{v_i}=\dfrac{x^{(0)}_j}{v_j}-\dfrac{x^{(0)}_i}{v_i}+\tau+nT.
\end{equation}
This yields Eqs. (\ref{eq:HawkingParallelOOT}), (\ref{eq:NonLinearAndreedHawkingOOT}) in the main text. Specifically, 
\begin{eqnarray}\label{eq:AHteo}
g_{\rm{HTC}}^{(2)}(x,x',\tau)&\simeq&g_{uu}(x,x',\tau),~x,x'<0,\\
\nonumber g_{\rm{HTC}}^{(2)}(x,x',\tau)&\simeq&g_{ud}(x,x',\tau),~x<0,~x'>0,\\
\nonumber g_{\rm{HTC}}^{(2)}(x,x',\tau)&\simeq&g_{du}(x,x',\tau),~x>0,~x'<0,\\
\nonumber g_{\rm{HTC}}^{(2)}(x,x',\tau)&\simeq&g_{dd}(x,x',\tau)+g_{dw}(x,x',\tau)+g_{wd}(x,x',\tau),~x,x'>0.
\end{eqnarray}
The self-correlations of the upstream waves and the downstream solitons are described by the $g_{uu}$ and $g_{dd}$ functions, while the analog of the Hawking and Andreev correlations are encapsulated in the $g_{ud},g_{du}$ and $g_{wd},g_{dw}$ functions, respectively. The correlations $g_{uw},g_{wu},g_{ww}$ are neglected as they are too weak to be observed, in analogy to the subdominant character of the normal-normal $u-d1$ correlations from the Andreev-Hawking effect \cite{Recati2009}.

\begin{figure*}[!t]
\begin{tabular}{@{}ccc@{}}   
\stackinset{l}{0pt}{t}{0pt}{(a)}{\includegraphics[width=0.33\textwidth]{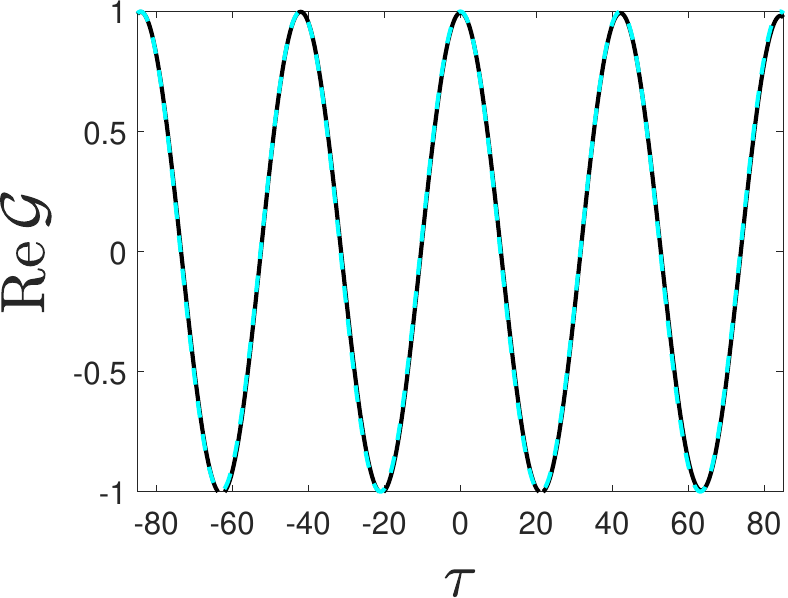}} 
& \stackinset{l}{0pt}{t}{0pt}{(b)}{\includegraphics[width=0.33\textwidth]{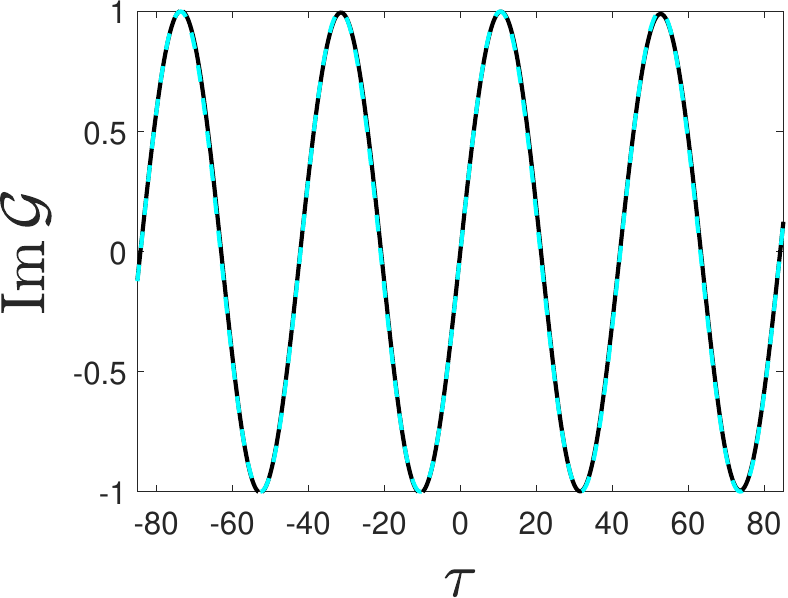}}& 
\stackinset{l}{0pt}{t}{0pt}{(c)}{\includegraphics[width=0.33\textwidth]{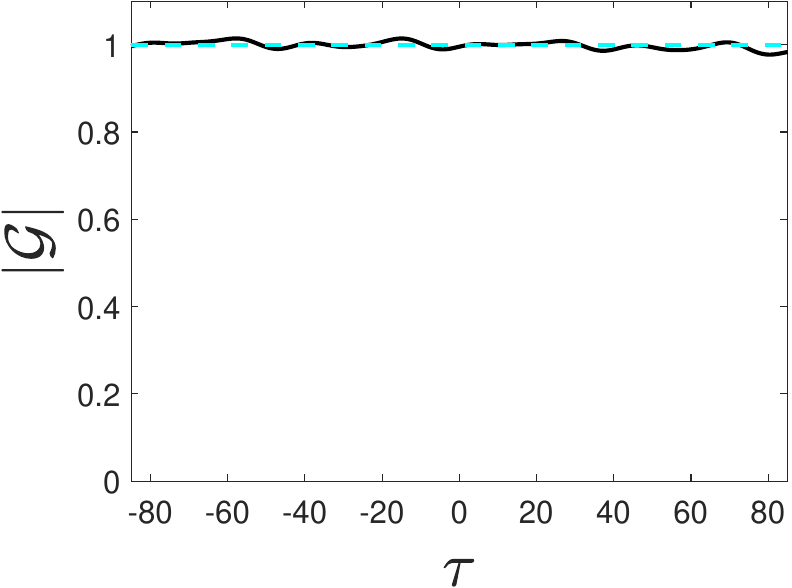}} 
\end{tabular}
\caption{Figure of merit $\mathcal{G}(\tau)$ for a purely quantum FPBHL with $v=0.95,c_2=0.4,L=2$ (see Fig. \ref{fig:ClassicalQuantum}k of main text), computed via TW method (solid black) versus theoretical prediction $\mathcal{G}_{\rm{HTC}}(\tau)=e^{i\omega\tau}$ (dashed cyan). (a) Real part $\textrm{Re}\,\mathcal{G}(\tau)$. (b) Imaginary part $\textrm{Im}\,\mathcal{G}(\tau)$. (c) Absolute value $|\mathcal{G}(\tau)|$.}
\label{fig:MeritoSM}
\end{figure*}

We also evaluate the spatial Fourier transform of the OTCF within a certain region $\mathcal{R}\in\mathbb{R}^2$,   
\begin{equation}
g^{(2)}(k,k',t,t')=\int_{\mathcal{R}}\mathrm{d}x\mathrm{d}x'~g^{(2)}(x,x',t,t')e^{-ikx}e^{-ik'x'}.
\end{equation}
In practice, we take the region $\mathcal{R}$ as one of the four quadrants of the Cartesian plane. Consequently, from Eq. (\ref{eq:AHteo}), we only need to compute the Fourier transform of $g_{ij}(x,x',\tau)$, which is expected to exhibit a discrete peak structure from Eq. (\ref{eq:CorrelationBands}),
\begin{equation}
    g_{ij}(x,x',\tau)=\sideset{}{'}\sum^\infty_{m=-\infty}g^{(m)}_{ij}(\tau)e^{-im\frac{\omega}{v_i}x}e^{im\frac{\omega}{v_j}x'},~g^{(m)}_{ij}(\tau)=n^{(-m)}_in^{(m)}_je^{-im\omega\tau}.
\end{equation}
In the main text, we focus on the downstream-upstream region $\mathcal{R}=(0,\infty)\times (-\infty,0)$, where the analog of the Hawking correlations are exhibited between the upstream waves and the downstream solitons. In particular, we evaluate $g^{(2)}(x,x',t,t')$ at the peak of the upstream/downstream spectrum, $k_{u,d}=\dfrac{\omega}{|v_{u,d}|}$. For an HTC, we obtain
\begin{equation}
     g^{(2)}_{\rm{HTC}}(k_d,k_u,\tau)\simeq g^{(-1)}_{du}(\tau)=n^{(1)}_dn^{(-1)}_ue^{i\omega \tau}.
\end{equation}
Thus, the HTC prediction for the figure of merit in the main text is just
\begin{equation}
    \mathcal{G}_{\rm{HTC}}(\tau)=\frac{g^{(2)}_{\rm{HTC}}(k_d,k_u,\tau)}{g^{(2)}_{\rm{HTC}}(k_d,k_u,\tau=0)}=e^{i\omega\tau}.
\end{equation}
In Fig. \ref{fig:MeritoSM}, we compare this prediction with the TW simulation in Fig. \ref{fig:ClassicalQuantum}k of the main text in more detail, finding an excellent agreement. 

We remark that the approximate nature of these expressions, which can be traced back to Eq. (\ref{eq:Cynthia}), stems solely from the dispersion of the upstream and downstream waves (especially in the upstream case), while the ballistic character of the soliton is instead \textit{exact}. 

\subsection{Experimental effects}\label{SM:Exp}

We now explicitly simulate the inclusion of realistic experimental effects, such as finite temperature and experimental noise. In actual experiments, the main source of noise are run-to-run fluctuations in background parameters such as the particle number or the confining potential, whose main effect is to add a stochastic component to the amplitude of the deterministic background Bogoliubov-Cherenkov-Landau (BCL) radiation \cite{Wang2016,Wang2017,Kolobov2021,Steinhauer2022}. 

We can model the effect of BCL fluctuations by adding a stochastic component to the classical amplitude $A$ in the initial TW condition (\ref{eq:TWExpansionNC}), $A\to A+\delta A$, with $\delta A$ a random variable with zero mean, $\braket{\delta A}=0$, and statistical deviation $\sqrt{\braket{\delta A^2}}=\Delta A$. These BCL fluctuations further increase the width $\Delta X$ of the lasing amplitude. Specifically, from Eq. (\ref{eq:EspacioFaseCero}), the lasing amplitude now reads
\begin{equation}
     \hat{X}=X_C+\delta X_C+ \frac{1}{\sqrt{L}}\sum_k \beta_k\hat{\alpha}_k+ \beta^*_k\hat{\alpha}^\dagger_k,~\delta X_C\equiv i(z_S|\Phi_C) \sqrt{n_0}\delta A.
\end{equation}
As a result, we still have $\braket{X}=X_C$ but now [see Eq. (\ref{eq:LasingFluctuations})]
\begin{equation}
     \Delta X^2=\Delta X^2_C+\Delta X^2_T+\Delta X^2_Q,~\Delta X^2_C=|(z_S|\Phi_C)|^2n_0\Delta A^2.
\end{equation}
Hence, the main effect of including finite temperature and experimental noise is to enhance the fluctuations of the lasing mode, which further randomizes the asymptotic phase-shift $\phi_0$. Moreover, if we assume that the BCL noise is Gaussian, the Wigner distribution for $X$ will be still a Gaussian of the form (\ref{eq:Marginal}), with $\Delta X$ given by the equation above. Therefore, all our previous results still apply  even when including experimental effects.

\begin{figure*}[!t]
\begin{tabular}{@{}ccc@{}}   
\stackinset{l}{0pt}{t}{0pt}{(a)}{\includegraphics[width=0.33\textwidth]{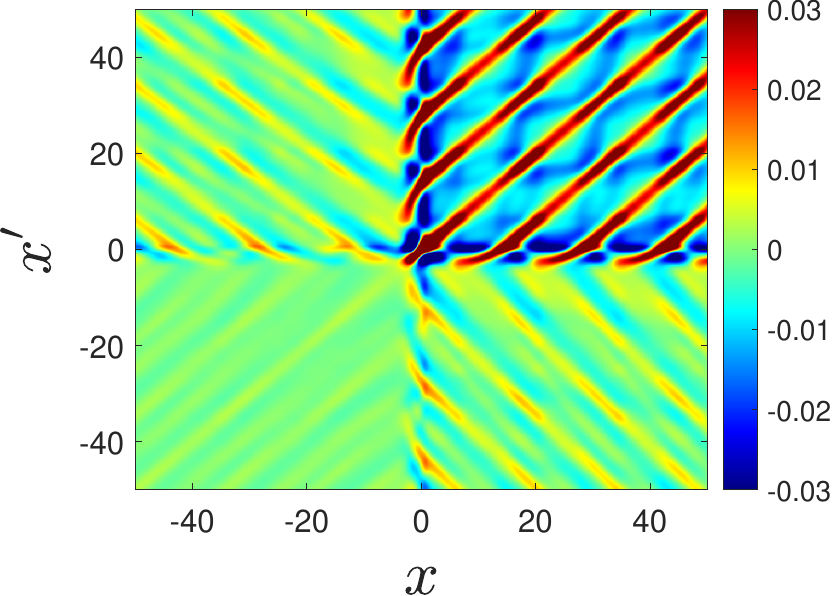}} 
& \stackinset{l}{0pt}{t}{0pt}{(b)}{\includegraphics[width=0.33\textwidth]{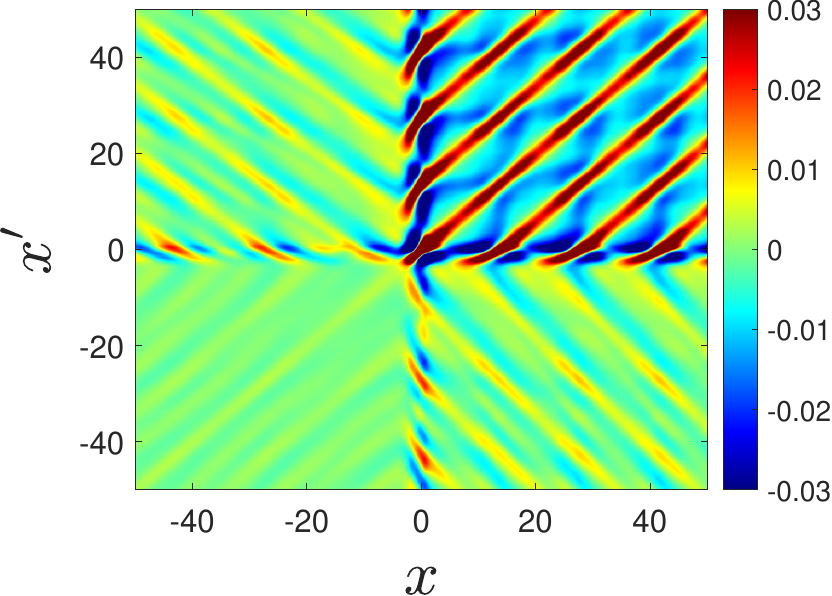}}& 
\stackinset{l}{0pt}{t}{0pt}{(c)}{\includegraphics[width=0.33\textwidth]{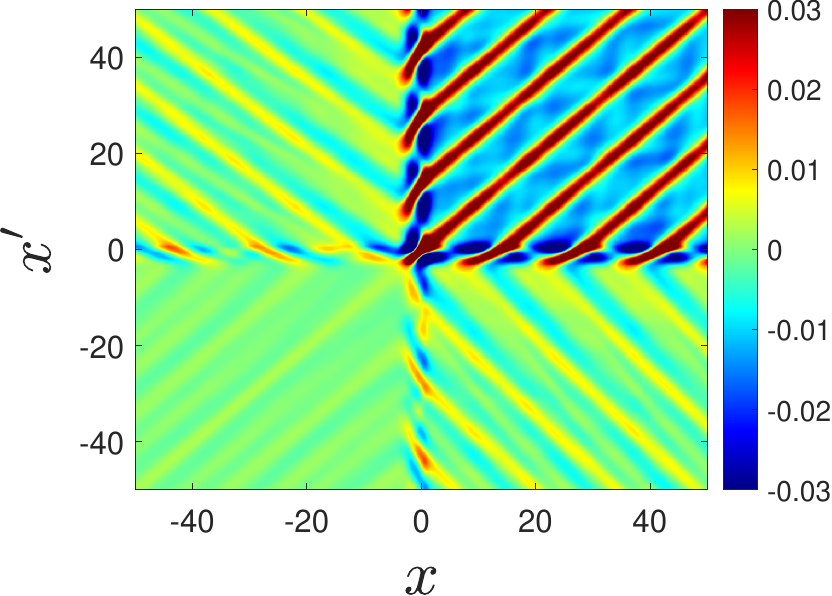}}\\ 
{\includegraphics[width=0.33\textwidth]{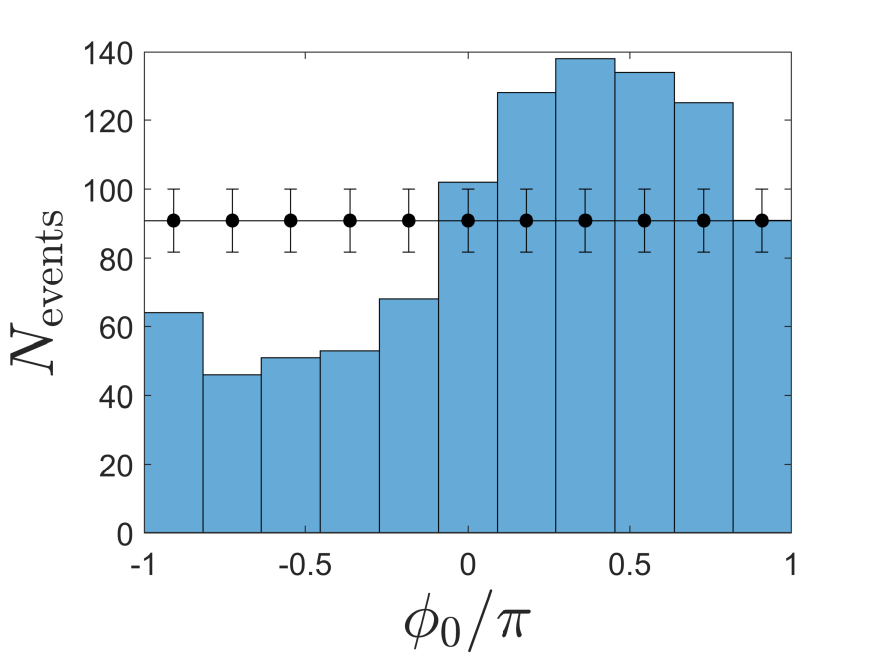}} 
& \stackinset{l}{0pt}{t}{0pt}{(b)}{\includegraphics[width=0.33\textwidth]{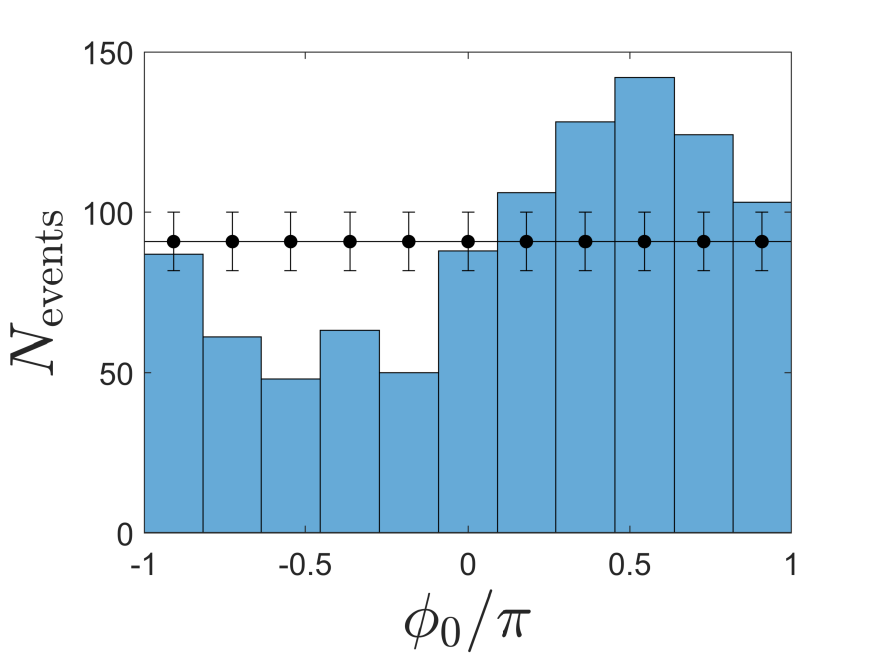}}& 
\stackinset{l}{0pt}{t}{0pt}{(c)}{\includegraphics[width=0.33\textwidth]{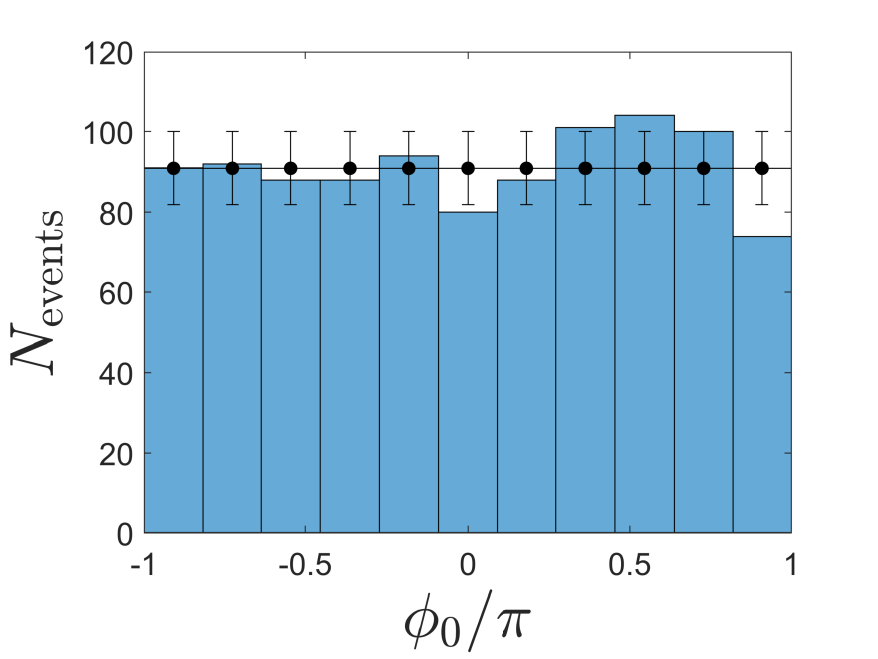}}
\end{tabular}
\caption{(a)-(c) Realistic TW computation of the ECTF $g^{(2)}(x,x',t)$ at $t=870$ for a FPBHL with $v=0.95,c_2=0.4,L=2$, including  finite temperature $T$ and experimental noise in the classical amplitude. The corresponding values are $T=0.5,0.5,0$, $A=0.01,0.005,0$, and $\Delta A=0.005,0.002,0.01$. (d)-(f) TW histogram of $\phi_0$ for (a)-(c). For the simulation in (c), where $A=0$, the time reference is set by a mean-field simulation with $A=0.003$.}
\label{fig:Experimental}
\end{figure*}

Figure \ref{fig:Experimental} shows the results of including finite temperature and experimental fluctuations in the TW method. In the first two columns, we show the effect of a finite temperature $T=0.5$ combined with some BCL noise $\Delta A=0.005,0.002$ around a classical amplitude $A=0.01,0.005$, respectively. We observe that the ETCF in the first row are close to the ideal case, as can be also inferred from the broad histograms in the second row. These can be compared with the results of including only quantum fluctuations shown in the main text, giving rise to much narrower distributions.

The third column represents the more realistic simulation, where $T=0$ and only pure BCL noise is added to the initial quantum fluctuations, $A=0$ and $\Delta A=0.01$. This is the type of model that successfully reproduced the experimental results in Ref. \cite{Kolobov2021}, where the effect of temperature was found to be negligible and BCL fluctuations were sufficiently strong to stochastically stimulate Hawking radiation. We observe that the agreement with an ideal HTC is still excellent, and experimental fluctuations do not destroy the HTC. 

Indeed, the main impediment to observe the BHL effect in current experiments is the strong background of fluctuating BCL radiation, which overshadows the lasing amplification in its early stages. This would correspond to set $\Delta A\sim 1$ in the above discussed simulation, so the initial state is no longer close to a BHL configuration. Nevertheless, a detailed analysis of experimental effects is beyond the scope of the present work since that will depend on the specific setup proposed for the achievement of the BHL, which remains one of the biggest present challenges in the field of analog gravity.

\end{appendices}

\end{document}